\PassOptionsToClass{floatfix}{revtex4-2}
\documentclass[aip,jcp,reprint,amsmath,amssymb]{revtex4-2}

\usepackage{graphicx}
\graphicspath{{./}{assets/}}
\usepackage{dcolumn}
\usepackage{bm}
\usepackage{booktabs}
\usepackage{placeins}
\providecommand{\alttext}[1]{}
\usepackage[colorlinks=true,linkcolor=blue,citecolor=blue,urlcolor=blue]{hyperref}

\newcommand{\ee}{\mathrm{ee}}

\begin{document}

\title{Symmetry-Induced Conservation Laws, Three-Channel Attribution, and Solid-Phase Recycling in Prebiotic Enantiomeric Excess Amplification}

\author{Zeyu Jia}
\email{jiazeyu@tju.edu.cn}
\affiliation{School of Biomedical Engineering and Technology, Tianjin Medical University, Tianjin 300070, China}

\author{Anlin Gong}
\altaffiliation{A. Gong and B. Wu contributed equally to this work.}
\affiliation{School of Pharmacy, Tianjin Medical University, Tianjin 300070, China}

\author{Bingmei Wu}
\altaffiliation{A. Gong and B. Wu contributed equally to this work.}
\affiliation{School of Pharmacy, Tianjin Medical University, Tianjin 300070, China}

\date{\today}

\begin{abstract}
Enantiomeric excess is the primary observable used to quantify chiral amplification across prebiotic chemistry, asymmetric autocatalysis, and multiphase crystallization. Because enantiomeric excess divides net chiral excess by total chiral mass, identical increases can arise either from parity-breaking synthesis of chiral centers or from symmetric phase and oligomer partitioning that depletes soluble mass. We establish a pathwise conservation law, a model-free three-channel attribution identity, and a continuous path-covariant Aumann--Shapley attribution for mirror-symmetric reaction networks. In any closed network lacking autocatalytic feedback or epimerization, net chiral excess remains invariant along every stochastic trajectory, locking liquid-phase amplification inversely to surviving chiral mass while Shear--Kramers--Lorentz relative-entropy balance bounds dissipation. We re-analyze seven experimental benchmark systems and show that standalone phase removal loses $91.7\%$--$98.9\%$ of chiral solute into racemic sediment in closed batches and washes out under open racemic inflow. We show that coupling eutectic partitioning with racemization recycling achieves $96.5\%$ homochiral crystal recovery, whereas evaporative autocatalysis crosses the open-flow bifurcation threshold with $81.9\%$ homochiral yield. Finally, we extend our non-crossing parity invariant across a fifty-nine-point multicomponent peptide ligation and seeded co-precipitation benchmark, evaluate a one-hundred-twelve-row calcite adsorption dataset, and resolve mineral-occlusion non-identifiability using three-block antipodal crystal pairs.
\end{abstract}

\maketitle

\section{Introduction}
\label{sec:intro}

The emergence of homochirality from nearly racemic prebiotic precursors is a foundational problem in chemical physics and origin-of-life research.\cite{Frank1953,Kondepudi1983,Blackmond2004,Saito2013,Sallembien2022,Ribo2025ChemPhysChem} Motivated by a question raised during a freshman undergraduate chemistry lecture on how a slight initial enantiomeric majority can be amplified across distinct reaction networks (see Acknowledgments), we examine what scalar reporting of enantiomeric excess,
\begin{equation}
\ee(t) = \frac{c_L(t) - c_D(t)}{c_L(t) + c_D(t)} = \frac{\Delta(t)}{M(t)},
\label{eq:ee_def}
\end{equation}
conceals by projecting a two-dimensional chiral inventory---net chiral excess $\Delta = c_L - c_D$ in the numerator and total chiral mass $M = c_L + c_D$ in the denominator---onto a single coordinate.

Across experimental studies, chiral amplification has been demonstrated via nonlinear catalyst dimer reservoirs and autocatalytic alkylation in the Soai reaction,\cite{Guillaneux1994,Girard1998,Soai1995,Blackmond2001,Athavale2020} attrition-enhanced Ostwald ripening and temperature-cycling deracemization,\cite{Viedma2005,Noorduin2008,Suwannasang2013,Deck2024,Deck2025} eutectic solid--liquid fractionation and self-disproportionation of enantiomers (SDE),\cite{Morowitz1969,Soloshonok2006,Klussmann2006,Breslow2006,Blackmond2010} chiral mineral adsorption,\cite{Hazen2001,Kim2016} and peptide ligation or crystallization-induced dynamic resolution.\cite{Deng2024,VanDongen2025,VanDongen2026Size} Because these protocols report $\ee(t)$, identical $\ee(t)$ curves can arise from distinct mechanisms with opposite implications for yield, open-flow persistence, and dissipation.\cite{Plasson2007,Bissette2013,Seifert2012,Rao2016}

Here we formulate a conservation, attribution, and identifiability framework for chiral networks:
\begin{enumerate}
\item We prove a pathwise $\mathbb{Z}_2$ conservation law showing that under $1{:}1$ heterochiral pairings and achiral transport without autocatalysis or epimerization, $\Delta(t)$ is invariant along every trajectory ($M_f / M_0 = \ee_0 / \ee_f$).
\item We derive ordered, discrete Shapley, and continuous Aumann--Shapley attribution identities decomposing $\Delta\ee$ into phase removal (Channel R), subpool fractionation (Channel S), and chiral-excess shift (Channel B), together with Shear--Kramers--Lorentz dissipation bounds.
\item We re-attribute seven experimental systems, establish a $1 \times K$ non-crossing parity invariant for peptide ligation,\cite{Deng2024} and show that while standalone Channel R loses $91.7\%$--$98.9\%$ of solute and washes out under open inflow, coupling Channel R with racemization recycling (Channel B)\cite{Deck2024,Deck2025} or evaporative autocatalysis (Channels R + S + B)\cite{Ribo2025ChemPhysChem} achieves $81.9\%$--$96.5\%$ homochiral recovery.
\item We derive an analytical wet--dry law, resolve a 1D mineral-occlusion ridge,\cite{Kim2016} and give an algebraic experimental recipe.
\end{enumerate}

\section{Symmetry-Induced Conservation and Three-Channel Attribution}
\label{sec:theory}

\subsection{Stoichiometric Decomposition and Pathwise Conservation}
\label{sec:conservation}

Consider chiral monomers $(L_i, D_i)$, homochiral oligomers $(P_{LL}, P_{DD})$, heterochiral duplexes $P_{LD}$, and solid phases. Each species $k$ carries chiral charge $q_k \in \mathbb{Z}$ ($q_{L_i} = +1$, $q_{D_i} = -1$, $q_{P_{LL}} = +2$, $q_{P_{DD}} = -2$, $q_{P_{LD}} = 0$) and chiral-center mass $m_k \in \mathbb{Z}_{\ge 0}$ ($m_{L_i} = m_{D_i} = 1$, $m_{P_{LL}} = m_{P_{DD}} = m_{P_{LD}} = 2$). For any active phase $\mathcal{A}$ with copy numbers $\bm{n}(t)$ in volume $\Omega$, total chiral inventory and signed chiral excess are
\begin{equation}
M(t) = \sum_{k \in \mathcal{A}} m_k n_k(t), \qquad \Delta(t) = \sum_{k \in \mathcal{A}} q_k n_k(t).
\label{eq:M_Delta_def}
\end{equation}
Each reaction $\rho$ with stoichiometric vector $\bm{\nu}_\rho$ induces increments $\Delta M_\rho = \bm{m}^\top \bm{\nu}_\rho$ and $\Delta q_\rho = \bm{q}^\top \bm{\nu}_\rho$. Under $\mathbb{Z}_2$ inversion symmetry $S$ ($L \leftrightarrow D$) without autocatalysis or epimerization, internal ligations ($L + L \rightleftharpoons P_{LL}$, $D + D \rightleftharpoons P_{DD}$, $L + D \rightleftharpoons P_{LD}$) conserve chiral centers within $\mathcal{A}$ ($\Delta M_\rho = 0$, $\Delta q_\rho = 0$), while $1{:}1$ heterochiral precipitation ($L + D \to P_{LD,\mathrm{solid}}$) removes two chiral centers ($\Delta M_\rho = -2$) with $\Delta q_\rho = -(+1) - (-1) = 0$. Along every trajectory with $K(t)$ firings,
\begin{equation}
\begin{aligned}
\Delta(t) &= \Delta(0) + \sum_{\xi=1}^{K(t)} \Delta q_{\rho_\xi} = \Delta(0), \\
\ee(t) &= \frac{\Delta(0)}{M(t)} = \ee_0 \frac{M_0}{M(t)}.
\end{aligned}
\label{eq:pathwise_lock}
\end{equation}
\noindent\textbf{Proposition 1 (Pathwise $\mathbb{Z}_2$ Conservation and Open-Flow Washout).} Equation~\eqref{eq:pathwise_lock} holds pathwise prior to ensemble averaging: whenever $\Delta q_\rho = 0$ for all reactions removing mass from $\mathcal{A}$, amplifying $\ee(t)$ in $\mathcal{A}$ requires inversely proportional mass depletion $M(t) / M_0 = \ee_0 / \ee(t)$.

In an open reactor with inflow $F_{\mathrm{in}}$ at $\ee_{\mathrm{in}}$ and outflow $F_{\mathrm{out}}$ at $\ee_{\mathrm{out}}$, differentiating $\ee = \Delta / M$ gives
\begin{equation}
\begin{aligned}
\frac{\mathrm{d}\ee}{\mathrm{d}t} &= -\frac{\ee}{M}\left(F_{\mathrm{in}} - F_{\mathrm{out}} + J_M\right) \\
&\quad + \frac{1}{M}\left(\ee_{\mathrm{in}} F_{\mathrm{in}} - \ee_{\mathrm{out}} F_{\mathrm{out}} + J_\Delta\right),
\end{aligned}
\label{eq:open_balance}
\end{equation}
where $J_M$ and $J_\Delta$ are internal production rates of $M$ and $\Delta$. For well-mixed outflow ($\ee_{\mathrm{out}} = \ee$) and racemic inflow ($\ee_{\mathrm{in}} = 0$), outflow terms cancel:
\begin{equation}
\frac{\mathrm{d}\ee}{\mathrm{d}t} = -\frac{F_{\mathrm{in}}}{M}\ee - \frac{J_M}{M}\ee + \frac{J_\Delta}{M}.
\label{eq:open_washout_law}
\end{equation}
If internal reactions cannot generate net chiral excess ($J_\Delta = 0$) and racemic feed balances mass loss ($F_{\mathrm{in}} = -J_M + F_{\mathrm{out}}$ at steady $M = M_0$), then $\mathrm{d}\ee / \mathrm{d}t = -(F_{\mathrm{out}} / M_0)\ee < 0$, driving washout $\ee(t) \to 0$. Sustaining homochirality under open racemic re-feeding requires $J_\Delta > \ee\left(F_{\mathrm{in}} + J_M\right)$ (Channel B), which equals $\ee F_{\mathrm{out}}$ at stationary active mass ($M = M_0$) and reduces to $J_\Delta > \ee F_{\mathrm{in}}$ when $J_M = 0$.

\subsection{Three-Channel Attribution Identity}
\label{sec:attribution}

Between initial state $(M_0, \Delta_0, \ee_0)$ and final state $(M_f, \Delta_f, \ee_f)$ of $\mathcal{A}$, alongside a target subpool (e.g., monomer with $M_{\mathrm{sub},f}$, $\Delta_{\mathrm{sub},f}$, and $\ee_{\mathrm{sub},f} = \Delta_{\mathrm{sub},f} / M_{\mathrm{sub},f}$), the shift $\Delta\ee_{\mathrm{sub}} = \ee_{\mathrm{sub},f} - \ee_0$ decomposes into three channels:
\begin{itemize}
\item \textbf{Channel R (Phase Removal):} Removal of $1{:}1$ $L{:}D$ pairs into a precipitate, sublimate, or sink ($\Delta_f = \Delta_0$, $M_f < M_0$), generalizing Morowitz's eutectic mass-balance law\cite{Morowitz1969,Klussmann2006} and Soloshonok's self-disproportionation of enantiomers (SDE).\cite{Soloshonok2006}
\item \textbf{Channel S (Subpool Fractionation):} Internal partitioning between monomer and oligomer/dimer subpools at fixed $M_f$ and $\Delta_f$ ($\ee_{\mathrm{sub},f} \neq \ee_f$), generalizing the Horeau--Langenbeck and Kagan $\mathrm{ML}_2$ homochiral/heterochiral dimer reservoir effect.\cite{Guillaneux1994,Girard1998,Blackmond2001}
\item \textbf{Channel B (Net Chiral-Excess Shift):} Creation or interconversion of signed chiral excess via autocatalysis or epimerization ($\Delta_f \neq \Delta_0$).\cite{Frank1953,Soai1995,Viedma2005}
\end{itemize}

\noindent\textbf{Proposition 2 (Ordered and Shapley Three-Channel Attribution).} Using the conserved-$\Delta$ intermediate $\ee_R^* = \Delta_0 / M_f = \ee_0 (M_0 / M_f)$ yields the ordered telescoping sequence $\ee_0 \xrightarrow{+A_R} \ee_R^* = \Delta_0/M_f \xrightarrow{+A_B} \ee_f = \Delta_f/M_f \xrightarrow{+A_S} \ee_{\mathrm{sub},f}$:
\begin{align}
\ee_{\mathrm{sub},f} - \ee_0 &= A_R + A_S + A_B, \nonumber\\
A_R &= \frac{\Delta_0}{M_f} - \ee_0 = \ee_0\left(\frac{M_0}{M_f} - 1\right), \nonumber\\
A_S &= \ee_{\mathrm{sub},f} - \ee_f = \frac{\Delta_{\mathrm{sub},f}}{M_{\mathrm{sub},f}} - \frac{\Delta_f}{M_f}, \nonumber\\
A_B &= \ee_f - \frac{\Delta_0}{M_f} = \frac{\Delta_f - \Delta_0}{M_f}.
\label{eq:ordered_attribution}
\end{align}
When phase-level inventory $(M_f, \Delta_f, \ee_f)$ is tracked ($A_S = 0$), averaging over the two factor-replacement orderings $(M_0 \to M_f, \Delta_0 \to \Delta_f)$ gives the symmetric Shapley attribution:
\begin{align}
A_R^{\mathrm{Sh}} &= \frac{1}{2}\left(\Delta_0 + \Delta_f\right)\left(\frac{1}{M_f} - \frac{1}{M_0}\right), \nonumber\\
A_B^{\mathrm{Sh}} &= \frac{1}{2}\left(\Delta_f - \Delta_0\right)\left(\frac{1}{M_0} + \frac{1}{M_f}\right),
\label{eq:shapley_attribution}
\end{align}
satisfying $A_R^{\mathrm{Sh}} + A_B^{\mathrm{Sh}} = \ee_f - \ee_0$ and isolating covariance $\delta_{\mathrm{int}} = A_B - A_B^{\mathrm{Sh}} = \frac{1}{2}(\Delta_f - \Delta_0)(M_0 - M_f)/(M_0 M_f)$. When time-resolved trajectories $\gamma : t \in [0,T] \mapsto \bm{x}(t)$ with $\dot{\bm{x}} = \sum_{\alpha \in \mathcal{C}} \bm{f}_\alpha(\bm{x},t)$ are observed, two-endpoint replacement becomes order-sensitive near $\ee_0 = 0$; integrating along $\gamma$ defines the continuous path-covariant Aumann--Shapley attribution\cite{BelahcenMussard2026} $\Phi_\alpha[\gamma] = \int_0^T \nabla_{\bm{x}} F(\bm{x}(t))\cdot \bm{f}_\alpha(\bm{x}(t))\,\mathrm{d}t$, which admits a closed-form rational chordal antiderivative on $F(M,\Delta) = \Delta/M$ (completeness residual $\le 2.7\times 10^{-14}$ across $m \in \{1,\dots,64\}$ chords; Sec.~S2.B and Table~S8). On crystallization dynamic resolution from $\ee_0 = 10^{-6} \to 0^+$,\cite{Deck2024,VanDongen2025} $\Phi_R[\gamma]$ resolves path ordering that discrete Shapley assigns a fixed $0.2500$ share, giving $\Phi_R = 8.2\times 10^{-5} < 10^{-4}$ ($\Phi_B = 0.9999$) under early dissolution versus $\Phi_R = \Phi_B = 0.5000$ under early deracemization (Table~S8).

\begin{figure*}[!t]
\centering
\includegraphics[width=0.85\textwidth]{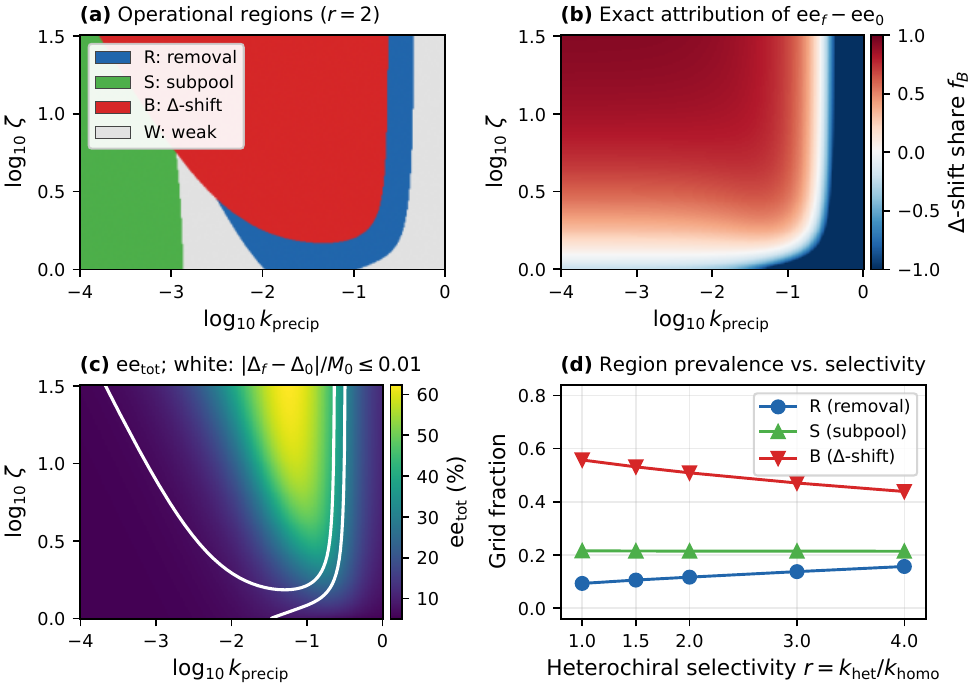}
\caption{Three-channel map across precipitation rate $k_{\mathrm{precip}}$ and autocatalytic feedback $\zeta$ at selectivity $r = k_{\mathrm{het}}/k_{\mathrm{homo}} = 2$. (a) Dominant channel (Channel R: blue; Channel S: green; Channel B: red; washout/weak: gray). (b) Share $f_B = A_B / (\ee_f - \ee_0)$. (c) Final active $\ee_{\mathrm{tot}}$, with white contour $|\Delta_f - \Delta_0|/M_0 \le 0.01$. (d) Regime prevalence across $r \in \{1.0, 1.5, 2.0, 3.0, 4.0\}$.}
\label{fig:taxonomy}
\end{figure*}

\begin{figure*}[!t]
\centering
\includegraphics[width=0.85\textwidth]{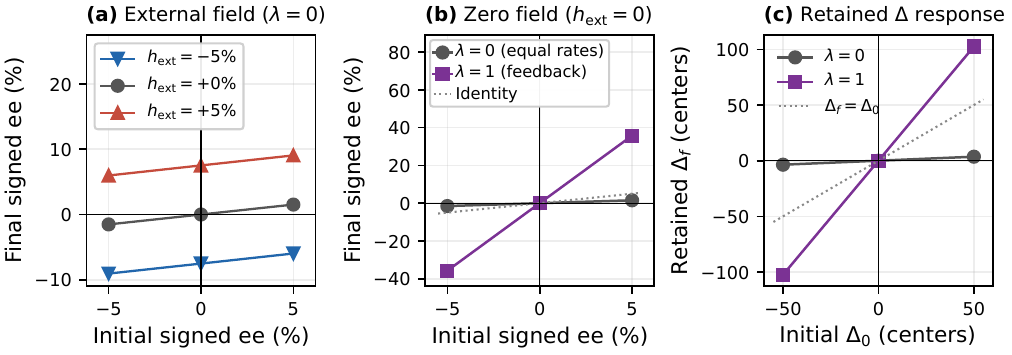}
\caption{Symmetry diagnostics separating external chiral fields ($h_{\mathrm{ext}} \neq 0$) from Channel B bifurcation ($\lambda > 0$, $h_{\mathrm{ext}} = 0$). (a) Bias $h_{\mathrm{ext}} = \pm 5\%$ ($\lambda = 0$) shifts trajectories vertically with non-zero intercept at $\ee_0 = 0$. (b) At $h_{\mathrm{ext}} = 0$, Channel B ($\lambda = 1$) preserves odd parity ($\ee_f(-\ee_0) = -\ee_f(\ee_0)$) with slope $> 1$. (c) Retained $\Delta_f$ vs.\ $\Delta_0$, showing $|\Delta_f| > |\Delta_0|$ only when Channel B is active.}
\label{fig:covariance}
\end{figure*}

We map these regimes across $512\,000$ ODE integrations on $320 \times 320$ grids spanning five heterochiral selectivities $r \in \{1.0, 1.5, 2.0, 3.0, 4.0\}$ (Fig.~\ref{fig:taxonomy} and Table~S1). Channel R dominates $9.4\%$--$15.7\%$ of the grid at high $k_{\mathrm{precip}}$ and low $\zeta$, where $|\Delta_f - \Delta_0|/M_0 \le 0.01$; Channel S accounts for $21.5\%$--$21.6\%$ at low $k_{\mathrm{precip}}$ and low $\zeta$; and Channel B governs $43.9\%$--$55.7\%$ of the domain, including $50.9\%$ at $r = 2.0$. Figure~\ref{fig:covariance} separates an external chiral bias $h_{\mathrm{ext}} \neq 0$, which breaks odd parity with $\ee_f(0) \neq 0$, from spontaneous Channel B feedback at $h_{\mathrm{ext}} = 0$ and $\lambda > 0$, which preserves odd parity while amplifying $|\Delta_f| > |\Delta_0|$.

\section{Stochastic Parity, Microdroplet Fluctuations, and Thermodynamic Dissipation}
\label{sec:stochastic_thermo}

\subsection{Chemical Master Equation Parity and Microdroplet Fluctuations}
\label{sec:cme}

To verify conservation and parity in finite compartments without continuum approximations, we formulate the Chemical Master Equation (CME) on the 5-species state space $\bm{n} = (n_L, n_D, n_{LL}, n_{DD}, n_{LD}) \in \mathbb{Z}_{\ge 0}^5$\cite{Kim2020JCP} for a closed compartment of chiral capacity
\begin{equation}
N_{\mathrm{total}} = n_L + n_D + 2\left(n_{LL} + n_{DD} + n_{LD}\right).
\label{eq:cme_lattice_constraint}
\end{equation}
Indexing dimer count $k = n_{LL} + n_{DD} + n_{LD} \in \{0, \dots, \lfloor N_{\mathrm{total}}/2 \rfloor\}$, $(n_{LL}, n_{DD}, n_{LD})$ admit $\binom{k+2}{2} = (k+1)(k+2)/2$ integer compositions and $(n_L, n_D)$ admit $N_{\mathrm{total}} - 2k + 1$, giving full 5-species cardinality (valid for all even and odd $N_{\mathrm{total}} \in \mathbb{Z}_{\ge 0}$) and reduced 2-DOF active monomer-simplex cardinality $\mathcal{S}_{\mathrm{red}}(N_{\mathrm{total}}) = \{(n_L, n_D) \in \mathbb{Z}_{\ge 0}^2 : n_L + n_D \le N_{\mathrm{total}}\}$:
\begin{align}
|\mathcal{S}_5(N_{\mathrm{total}})| &= \sum_{k=0}^{\lfloor N_{\mathrm{total}}/2 \rfloor} \binom{k+2}{2}\left(N_{\mathrm{total}} - 2k + 1\right) \nonumber\\
&= \binom{\lfloor N_{\mathrm{total}}/2\rfloor + 4}{4} + \binom{\lceil N_{\mathrm{total}}/2\rceil + 3}{4}, \nonumber\\
|\mathcal{S}_{\mathrm{red}}(N_{\mathrm{total}})| &= \binom{N_{\mathrm{total}} + 2}{2}.
\label{eq:cme_dimension}
\end{align}
\noindent\textbf{Proposition 3 (CME Parity Commutator and Finite-Volume Invariance).} At $N_{\mathrm{total}} = 70$, Eq.~\eqref{eq:cme_dimension} gives $|\mathcal{S}_5(70)| = \binom{39}{4} + \binom{38}{4} = 156\,066$ five-species microstates and $|\mathcal{S}_{\mathrm{red}}(70)| = \binom{72}{2} = 2556$ monomer-simplex states. When $k_{\mathrm{hyd}} = 0$, the monomer-projected generator $W_{\mathrm{red}}$ is autonomous with $2553$ relaxation modes and $3$ absorbing monomer states $(0,0), (1,0), (0,1)$ (Sec.~S3.A). The simplex $\mathcal{S}_{\mathrm{red}}(70)$ splits into $36^2 = 1296$ even-sum states absorbing into $(0,0)$ and $36\times 35 = 1260$ odd-sum states absorbing into $(1,0)$ or $(0,1)$, whereas for $k_{\mathrm{hyd}} > 0$ Markovian closure holds on $W_5$. With permutation matrix $S$ exchanging $(n_L, n_{LL}) \leftrightarrow (n_D, n_{DD})$ at $\lambda = 0$, sparse evaluation gives $\|W - S W S\|_{\max} \le 4.4 \times 10^{-16}$ on both $W_{\mathrm{red}}$ and $W_5$. Setting $\ee = 0$ on $n_L + n_D = 0$, any symmetric initial distribution $P(\bm{n}, 0) = S P(\bm{n}, 0)$ remains $S$-invariant at all $t \ge 0$, enforcing $\mathbb{E}[\Delta(t)] = 0$ and $\mathbb{E}[\ee(t)] = 0$ by odd parity $\ee(S\bm{n}) = -\ee(\bm{n})$.

Individual microdroplets exhibit finite-count mass-depletion fluctuations while conserving integer chiral excess. Across $7\,500$ Gillespie stochastic trajectories over $N = 100$ wet--dry cycles spanning $N_0 \in \{50, 100, 250, 500, 1000\}$ and $\ee_0 \in \{0.0\%, 1.0\%, 5.0\%\}$ with $n_{L,D}(0) = (N_0 \pm \Delta_0)/2$, pathwise conservation holds across every run under Channels R and S, giving $\max |\Delta_f - \Delta_0| = 0$ on all $2\,500$ racemic paths and $\Lambda = \Delta_f/\Delta_0 = 1$ on all $5\,000$ scalemic paths (Sec.~S3.B and Table~S2). Because $\Delta_f = \Delta_0$ is fixed on each trajectory while surviving count $M_f$ fluctuates, endpoint dispersion $\mathrm{SD}(\ee_f) = |\Delta_0|\,\mathrm{SD}(1/M_f)$ at $\ee_0 = 1.0\%$ falls from $6.1\%$ at $N_0 = 50$ to $3.3\%$ at $N_0 = 1000$ with log--log slope $-0.17$. At nominal $\ee_0 = 5.0\%$, integer parity rounding maps $N_0 = 50$ to $\Delta_0 = 2$ ($\ee_{0,\mathrm{real}} = 4.0\%$, $\mathrm{SD}(\ee_f) = 7.7\%$) and $N_0 = 100$ to $\Delta_0 = 6$ ($\ee_{0,\mathrm{real}} = 6.0\%$, $\mathrm{SD}(\ee_f) = 10.7\%$), above which $\mathrm{SD}(\ee_f)$ declines monotonically to $7.0\%$ at $N_0 = 1000$ with slope $-0.19$ over $[100, 1000]$ and $-0.038$ over $[50, 1000]$.

\subsection{Reaction-Resolved Entropy Production and Amplification Efficiency}
\label{sec:thermo}

Amplifying $\ee$ away from racemic equilibrium requires free-energy dissipation.\cite{Seifert2012,Rao2016,RaoEsposito2018,Remlein2026} For reversible reactions $\rho$ satisfying local detailed balance with fluxes $J_\rho^\pm(t)$, the entropy production rate is
\begin{equation}
\sigma(t) = R \sum_{\rho} \left(J_\rho^+(t) - J_\rho^-(t)\right) \ln\frac{J_\rho^+(t)}{J_\rho^-(t)} \ge 0,
\label{eq:sigma_rate}
\end{equation}
with $R = 8.3145\times 10^{-3}\text{ kJ mol}^{-1}\text{ K}^{-1}$, satisfying the exact Shear--Kramers--Lorentz relative-entropy balance $\mathrm{d}\mathcal{F}_{\mathrm{tot}}/\mathrm{d}t = RT\,\mathrm{d}\mathcal{L}(\bm{Z}\|\bm{Z}_{\mathrm{eq}})/\mathrm{d}t = -T\sigma(t) \le 0$ for any closed detailed-balanced network (numerical residual $\le 5.5\times 10^{-13}$; Sec.~S3.D).\cite{RaoEsposito2018,Remlein2026} Along any monotonic amplification trajectory ($\mathrm{d}\ee/\mathrm{d}t > 0$), normalizing $T\sigma(t)$ by surviving product inventory $M_{\mathrm{prod}}(t)$ and changing variables to $u = \ee(t)$ defines specific free-energy dissipation per mole of product and dimensionless efficiency $\eta_G(\ee)$:
\begin{align}
\Delta G_{\mathrm{diss}}(\ee) &= \int_0^{t(\ee)} \frac{T\,\sigma(s)}{M_{\mathrm{prod}}(s)}\,\mathrm{d}s = RT \int_{\ee_0}^{\ee} c(u)\,\mathrm{d}u, \label{eq:G_diss_def}\\
\eta_G(\ee) &= \frac{\ee - \ee_0}{\Delta G_{\mathrm{diss}}(\ee) / (RT)},
\label{eq:eta_G_def}
\end{align}
where $c(u) = [\sigma(u)/R] / [M_{\mathrm{prod}}(u)(\mathrm{d}u/\mathrm{d}t)] > 0$ is marginal specific dissipation ($RT$ per unit $\ee$ gain per mole of product), so $\eta_G(\ee) = [(\ee - \ee_0)^{-1}\int_{\ee_0}^{\ee} c(u)\,\mathrm{d}u]^{-1}$ is dimensionless and measures $\ee$ gain per $RT$ dissipated in the product pool.

\begin{figure*}[!t]
\centering
\includegraphics[width=0.85\textwidth]{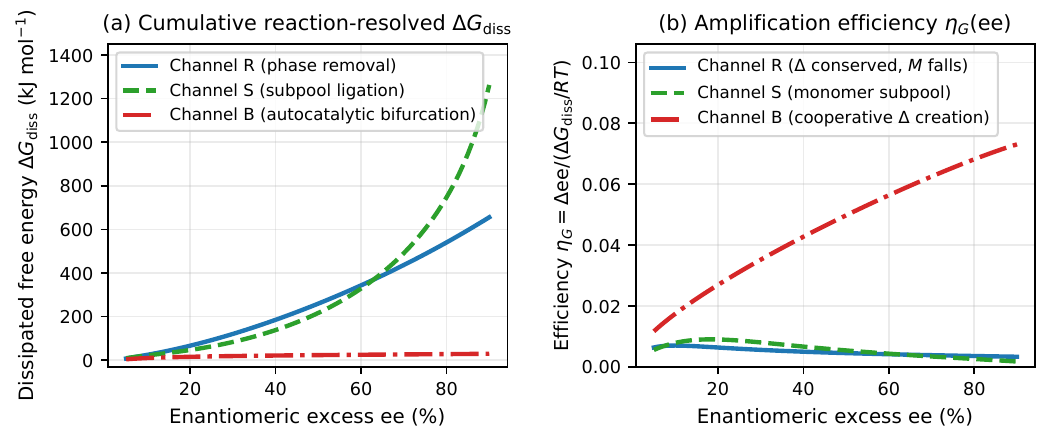}
\caption{Reaction-resolved thermodynamics from $\ee_0 = 3\%$ at $T = 298.15\text{ K}$. (a) Specific dissipated free energy $\Delta G_{\mathrm{diss}}(\ee)$ ($\text{kJ mol}^{-1}$ of product) from Eq.~\eqref{eq:G_diss_def} for Channel R (solid blue), Channel S (dashed green), and Channel B (dash-dotted red). (b) Dimensionless efficiency $\eta_G(\ee)$ from Eq.~\eqref{eq:eta_G_def}.}
\label{fig:thermo}
\end{figure*}

We integrate $\Delta G_{\mathrm{diss}}(\ee) = RT\,g_\alpha(\ee)$ and $\eta_G(\ee)$ across $\ee \in [5\%, 90\%]$ from autonomous reversible mass-action networks\cite{Hochberg2019Life,Ribo2024ACSPhysChemAu} using dimensionless accumulators $g_\alpha(t) = \Delta G_{\mathrm{diss},\alpha}(t)/(RT)$ ($\alpha \in \{R, S, B\}$) seeded at $\ee_0 = 3\%$ and $M_0 = 1.0$ at $T = 298.15\text{ K}$ ($RT = 2.48\text{ kJ mol}^{-1}$; Sec.~S3.C and Table~S3):
\begin{enumerate}
\item \textbf{Channel R:} Two-step heterochiral crystallization ($L + D \rightleftharpoons S_1 \rightleftharpoons S_2$ with pre-nucleation buffer $B_0 = S_1(0) = 6.0 M_0$, Sec.~S3.C)\cite{Deck2024,Wu2026CDTA} conserves $\Delta = c_L - c_D = \Delta_0$, so surviving liquid mass $M_{\mathrm{prod}}(t) = c_L(t) + c_D(t)$ falls as $M_{\mathrm{prod}}(u)/M_0 = \ee_0 / u$. Because a shrinking supernatant bears cumulative nucleation and crystal-maturation dissipation, with $70\%$ of monomer mass precipitating between $3\%$ and $10\%$ $\ee$, $\Delta G_{\mathrm{diss},R}(\ee)$ rises convexly from $24.9$ to $260$ and $487\text{ kJ mol}^{-1}$ at $\ee = 10\%, 50\%,$ and $75\%$ [Fig.~\ref{fig:thermo}(a)]. Consequently, after a slight initial rise from $6.3\times 10^{-3}$ at $5\%$ $\ee$ to a peak of $7.0\times 10^{-3}$ near $10\%$ $\ee$, $\eta_{G,R}(\ee)$ decreases monotonically across $10\%$--$90\%$ $\ee$, falling to $4.5\times 10^{-3}$ at $50\%$ and $3.7\times 10^{-3}$ at $75\%$ [Fig.~\ref{fig:thermo}(b); Table~S3].
\item \textbf{Channel S:} Reversible homochiral and heterochiral dimerization ($2L \rightleftharpoons P_{LL}$, $2D \rightleftharpoons P_{DD}$, $L + D \rightleftharpoons P_{LD}$ at selectivity $r_{\mathrm{sel}} = 10$) coupled with secondary sequestration ($P_{ij} \rightleftharpoons Q_{ij}$ at heterodimer buffer $B_0 = P_{LD}(0) = 5.0 M_0$, Sec.~S3.C)\cite{Deng2024,Higgs2025,HagelschuerFeringa2026} depletes free monomer $M_{\mathrm{mon}}(t) = c_L(t) + c_D(t)$ while homochiral side-flux grows with $\ee(t)$. Specific dissipation $\Delta G_{\mathrm{diss},S}(\ee)$ rises convexly from $22.1$ to $217$ and $603\text{ kJ mol}^{-1}$ at $10\%, 50\%,$ and $75\%$ $\ee$. Correspondingly, $\eta_{G,S}(\ee)$ exhibits a brief initial rise below $\sim 18\%$ $\ee$ to a shallow peak of $9.0\times 10^{-3}$ before declining monotonically through $90\%$ $\ee$, dropping across the $10\%, 50\%,$ and $75\%$ benchmarks from $7.8\times 10^{-3}$ to $5.4\times 10^{-3}$ and $3.0\times 10^{-3}$ [Fig.~\ref{fig:thermo}(b); Table~S3].
\item \textbf{Channel B:} Under chemostatted precursor driving ($A_{\mathrm{chem}} \rightleftharpoons L, D$), cooperative autocatalysis ($A_{\mathrm{chem}} + 2i \rightleftharpoons 3i$), epimerization ($L \rightleftharpoons D$), and heterochiral antagonism ($L + D \rightleftharpoons \mathrm{LD}_{\mathrm{bath}}$) at effective affinity $\Delta\mu_{\mathrm{eff}} = RT\ln 24 \approx 3.18\,RT$ (Sec.~S3.C),\cite{Hochberg2019Life,Ribo2024ACSPhysChemAu,PengHong2026Entropy,GhosalBisker2026PCCP} near-racemic turnover dissipates background free energy while $\mathrm{d}u/\mathrm{d}t \propto u(1-u^2)$ remains small. As autocatalytic feedback accelerates $\mathrm{d}u/\mathrm{d}t$ and heterochiral antagonism subsides as $(1-u^2)$, marginal cost satisfies $c_B'(u) < 0$, keeping $\Delta G_{\mathrm{diss},B}$ modest at $10.0$, $23.4$, and $27.3\text{ kJ mol}^{-1}$ across $10\%, 50\%,$ and $75\%$ $\ee$ while $\eta_{G,B}(\ee)$ increases monotonically across the entire $5\%$--$90\%$ domain from $1.7\times 10^{-2}$ ($10\%$) to $5.0\times 10^{-2}$ ($50\%$) and $6.5\times 10^{-2}$ ($75\%$) [Fig.~\ref{fig:thermo}(b)]. Sweeping initial buffers $\Omega = B_0/M_0 \in \{0, 0.5, 2.0, 6.0\}$\cite{HeinFeringa2025Supramolecular,BilancioniEsposito2025Gears,GhosalBisker2026PCCP} establishes that even at zero initial buffer ($\Omega = 0$), reaching $75\%$ $\ee$ from $3\%$ requires $\Delta G_{\mathrm{diss}} = 66.90\text{ kJ mol}^{-1}$ ($\eta_G = 0.0267$) for Channel R and $98.60\text{ kJ mol}^{-1}$ ($\eta_G = 0.0181$) for Channel S versus $27.29\text{ kJ mol}^{-1}$ ($\eta_G = 0.0654$) for Channel B, while $\Omega = 6.0$ inflates Channels R and S by $7.28\times$ and $7.21\times$ ($487.15$ and $710.99\text{ kJ mol}^{-1}$; or $7.79\times$ and $7.76\times$ from $61.30$ and $95.03\text{ kJ mol}^{-1}$ vs.\ $23.24\text{ kJ mol}^{-1}$ under isometric $\Delta\mu = RT\ln 24$; Secs.~S3.C--S3.D and Tables~S3 and S9).
\end{enumerate}

\section{Unified Re-Attribution of Experimental Systems}
\label{sec:experiments}

\begin{table*}[!t]
\caption{Three-channel re-attribution of seven experimental systems and multiphase recycling regimes. Here $M_{\mathrm{ret}}/M_0$ is retained active-phase mass fraction, $(A_R, A_S, A_B)$ are ordered shares from Eq.~\eqref{eq:ordered_attribution}, and the $\ee_f$ column reports the target subpool $\ee_{\mathrm{sub},f}$ when $A_S \neq 0$ (so $A_R + A_S + A_B = \ee_f - \ee_0$). Entries marked (---) indicate that $\ee$ was reported without phase-resolved inventories $(M_{\mathrm{sol}}, M_{\mathrm{solid}})$ needed to separate Channels R/S from Channel B (with $A_B$ reporting the net shift $\ee_f - \ee_0$ conditional on unmeasured phase split).}
\label{tab:attribution}
\footnotesize
\begin{ruledtabular}
\begin{tabular}{llcccccc}
System / Regime & Physical Mechanism & $\ee_0$ (\%) & $\ee_f$ (\%) & $M_{\mathrm{ret}}/M_0$ & $A_R$ (\%) & $A_S$ (\%) & $A_B$ (\%) \\
\colrule
Soai (1995)\cite{Soai1995} & Autocatalytic alkanol alkylation & $0.5$ & $99.5$ & $1.000$ & $0.0$ & $+6.3$ & $+92.7$ \\
Blackmond (2001)\cite{Blackmond2001} & Catalyst dimer sequestration & $10.0$ & $68.0$ & $1.000$ & $0.0$ & $+58.0$ & $0.0$ \\
Klussmann (2006)\cite{Klussmann2006} & Serine eutectic partitioning & $2.0$ & $99.0$ & $0.0202$ & $+97.0$ & $0.0$ & $0.0$ \\
Deng (2024)\cite{Deng2024} & Peptide ligation + epimerization & $3.2$ & $42.2$\textsuperscript{e} & --- & --- & $+24.6$\textsuperscript{e} & $+14.4$\textsuperscript{e} \\
van Dongen (2025)\cite{VanDongen2025} & Crystallization dynamic resolution & $0.0$ & $91.0$ & $0.460$ & $0.0$\textsuperscript{a} & $0.0$ & $+91.0$\textsuperscript{a} \\
Viedma (2005)\cite{Viedma2005} & Attrition deracemization & $5.0$ & $99.5$ & $1.000$\textsuperscript{b} & --- & --- & $+94.5$\textsuperscript{b} \\
Noorduin (2008)\cite{Noorduin2008} & CPL + Viedma ripening & $0.0$ & $99.0$ & $1.000$\textsuperscript{b} & --- & --- & $+99.0$\textsuperscript{b} \\
\colrule
Regime I: Pure R (5-state, cycle 80) & Closed wet--dry precipitation & $5.0$ & $60.2$ & $0.0831$ & $+55.2$ & $0.0$ & $0.0$ \\
Regime I: Pure R (8-species, cycle 100) & Closed 8-species precipitation & $5.0$ & $57.14$ & $0.0110$\textsuperscript{d} & $+52.14$ & $0.0$ & $0.0$ \\
Regime II: Coupled R + B\cite{Deck2024,Deck2025} & Eutectic + solid racemization & $20.0$ & $96.50$\textsuperscript{b} & $1.000$\textsuperscript{b} & $0.0$\textsuperscript{b} & $0.0$ & $+76.50$\textsuperscript{b} \\
Regime III: Coupled R + S + B\cite{Ribo2025ChemPhysChem} & Evaporative open autocatalysis & $20.0$ & $92.87$ & $0.9467$\textsuperscript{c} & $+1.13$ & $+6.36$ & $+65.38$ \\
\end{tabular}
\end{ruledtabular}
\raggedright\small
\textsuperscript{a}Ordered attribution from $\ee_0 = 0$; Eq.~\eqref{eq:shapley_attribution} gives $A_R^{\mathrm{Sh}} = +24.57\%$ and $A_B^{\mathrm{Sh}} = +66.43\%$.
\textsuperscript{b}Two-phase inventory ($M_{\mathrm{tot}}/M_0 = 1.000$); Regime~II solid ($M_{\mathrm{solid}}/M_0 = 0.9650$) reaches $\ee_{\mathrm{solid}} = 100.00\%$ ($A_{R,\mathrm{solid}} = +0.73\%$, $A_{B,\mathrm{solid}} = +79.27\%$; Shapley $+2.11\% / +77.89\%$).
\textsuperscript{c}Soluble $M_{\mathrm{sol}}/M_0 = 0.9467$ ($M_{\mathrm{chiral}}/M_0 = 0.8818$ at $\ee_{\mathrm{chiral}} = 92.87\%$, $\ee_{\mathrm{sol}} = 86.51\%$, Shapley $+2.87\% / +63.64\%$, $Y_{\mathrm{hom}} = 81.90\%$; $2^3$ hierarchy in Table~S6 and Sec.~S6.B).
\textsuperscript{d}Species-1 retained monomer fraction $M_{1,f}/M_{1,0} = 0.0110$ ($\Delta_{1,f}/\Delta_{1,0} = 0.126$), whereas $(A_R, A_S, A_B)$ evaluate the pooled 8-species inventory ($M_{\mathrm{tot},f}/M_{\mathrm{tot},0} = 0.0875$; cross-precipitation decomposition in Table~S4).
\textsuperscript{e}Reported $\ee_f$ is $\ee_{\mathrm{sub},f}$ when $A_S \neq 0$ (pooled $\ee_{\mathrm{tot},f} = 93.20\%$, $10.00\%$, and $17.60\%$ for Soai, Blackmond, and Deng; Table~S4); (---) denotes unmeasured phase splits.
\end{table*}

\subsection{Re-Attribution Across Seven Landmark Experimental Systems}
\label{sec:seven_systems}

Re-analyzing seven experimental benchmarks shows that identical scalar $\ee$ gains conceal order-of-magnitude differences in mass retention and channel origin (Table~\ref{tab:attribution} and Table~S4; entries marked (---) lacked phase-resolved inventories $(M_{\mathrm{sol}}, M_{\mathrm{solid}})$). Two contrasts stand out:
\begin{enumerate}
\item \textbf{Channel B + Channel S in the Soai Reaction:\cite{Guillaneux1994,Girard1998,Soai1995,Blackmond2001,Athavale2020,Mohler2025SoaiKinetics}} Autocatalytic alkylation creates net chiral centers via Channel B ($A_B = +92.7\%$), while zinc alkoxide $\mathrm{ML}_2$ aggregation\cite{Guillaneux1994,Girard1998,Blackmond2001,Athavale2020,Mohler2025SoaiKinetics} enriches the active catalyst via Channel S ($A_S = +6.3\%$). Catalyst dimerization without turnover\cite{Blackmond2001} gives Channel S alone ($A_S = +58.0\%$, $A_B = 0$).
\item \textbf{Eutectic Partitioning vs.\ Dynamic Resolution:\cite{Morowitz1969,Soloshonok2006,Klussmann2006,VanDongen2025}} Serine eutectic equilibration\cite{Morowitz1969,Klussmann2006} amplifies supernatant $\ee$ ($2.0\% \to 99.0\%$) solely via Channel R ($A_R = +97.0\%$, $A_B = 0$), leaving $98.0\%$ of solute in racemic solid ($M_{\mathrm{ret}}/M_0 = 0.020$; Table~\ref{tab:attribution}). By contrast, crystallization dynamic resolution\cite{VanDongen2025} from $\ee_0 = 0$ gives ordered $A_R = 0\%$ and $A_B = +91.0\%$, whereas Eq.~\eqref{eq:shapley_attribution} resolves $\Delta_f = 0.42 M_0$ and $M_f = 0.46 M_0$ into $A_B^{\mathrm{Sh}} = +66.4\%$ and $A_R^{\mathrm{Sh}} = +24.6\%$ (Table~S4).
\end{enumerate}

\subsection{Non-Crossing Parity Invariant and Mechanistic Decoupling in Dual-Monomer Ligation}
\label{sec:deng}

\begin{figure*}[!t]
\centering
\includegraphics[width=0.85\textwidth]{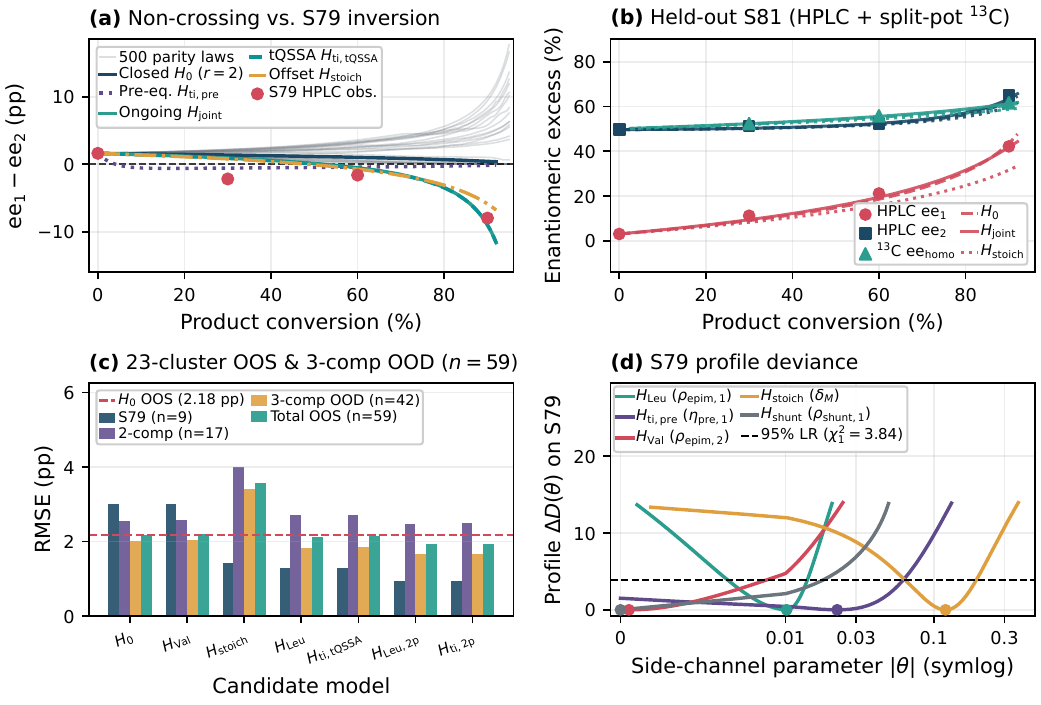}
\caption{Non-crossing parity invariant and multicomponent validation of peptide ligation\cite{Deng2024} (S79--S86). (a) Relative excess $\ee_1 - \ee_2$ vs.\ conversion: $500$ randomized Channel S rate laws (gray) and closed $H_0$ ($r=2$, dark blue) obey Eq.~\eqref{eq:noncrossing_invariant} without crossing zero, whereas S79 HPLC data (red circles) invert from $+1.7\text{ pp}$ to $-8.0\text{ pp}$, captured by ongoing thioimidate epimerization $H_{\mathrm{joint}}$ and $H_{\mathrm{ti,tQSSA}}$ (teal). (b) Held-out S81 HPLC and $^{13}\text{C}$-NMR trajectories. (c) Zero-shot out-of-sample RMSE across S79 ($n=9$), binary S80--S82 ($n=17$), three-component OOD S83--S86 ($n=42$), and total OOS ($n=59$, $23$ clusters) for $H_0$, $H_{\mathrm{Val}}$, $H_{\mathrm{stoich}}$, $H_{\mathrm{Leu}}$, $H_{\mathrm{ti,tQSSA}}$, $H_{\mathrm{Leu},2\mathrm{p}}$, and $H_{\mathrm{ti},2\mathrm{p}}$ vs.\ $H_0$ OOS ($2.18\text{ pp}$). (d) S79 profile deviance $\Delta D(\theta)$ vs.\ $|\theta|$ (symlog, $\mathrm{linthresh}=0.01$), including Leu shunt $H_{\mathrm{shunt}}$ ($\rho_{\mathrm{shunt},1}$).}
\label{fig:deng}
\end{figure*}

Dual-monomer peptide ligation experiments by Deng \emph{et al.}\cite{Deng2024} reveal a relative-excess sign inversion that rules out equimolar Channel S fractionation alone. In their mixed Val + Leu reactor (Table~S79 of Ref.~\onlinecite{Deng2024}), monomer excesses start at $\ee_{\mathrm{Leu}}(0) - \ee_{\mathrm{Val}}(0) = +1.7\text{ pp}$ at $0\%$ conversion, but invert sign to $-2.2\text{ pp}$ at $30\%$, $-1.7\text{ pp}$ at $60\%$, and $-8.0\text{ pp}$ at $90\%$ conversion [red circles in Fig.~\ref{fig:deng}(a)].

\noindent\textbf{Proposition 4 (Dual-Monomer Non-Crossing Parity Invariant).} In any closed dual-monomer network with monomers $(L_1, D_1)$ and $(L_2, D_2)$ undergoing parity-symmetric homochiral ($k_{\mathrm{homo}}$) and heterochiral ($k_{\mathrm{het}} = r k_{\mathrm{homo}}$) couplings without epimerization ($J_{\Delta_i} = 0$), writing $\Delta_i = L_i - D_i$, $M_i = L_i + D_i$, $\Delta = \Delta_1 + \Delta_2$, and $M = M_1 + M_2$ gives $\mathrm{d}(\Delta_1 - \Delta_2)/\mathrm{d}t = -\frac{1}{2}(k_{\mathrm{homo}}+k_{\mathrm{het}})M(\Delta_1 - \Delta_2) - \frac{1}{2}(k_{\mathrm{homo}}-k_{\mathrm{het}})\Delta(M_1 - M_2)$ and intensive difference $\mathrm{d}(\ee_1 - \ee_2)/\mathrm{d}t = -\frac{1}{2}(k_{\mathrm{het}}-k_{\mathrm{homo}})\Delta(\ee_1 + \ee_2)(\ee_1 - \ee_2)$ under symmetric self- plus cross-ligation, whereas under selective electrophile--nucleophile cross-ligation between substrate 1 (Ac-Leu-CN) and substrate 2 (Val-OH), $\mathrm{d}(\Delta_1 - \Delta_2)/\mathrm{d}t = -k_{\mathrm{het}}(t)[M_2(t)\Delta_1(t) - M_1(t)\Delta_2(t)]$. At equimolar stoichiometry ($\delta_M = (M_{1,0}-M_{2,0})/M_{2,0} = 0$, so $M_1(t) = M_2(t) = M_c(t) = M(t)/2$), this yields
\begin{equation}
\begin{aligned}
\frac{\mathrm{d}(\Delta_1 - \Delta_2)}{\mathrm{d}t} &= -k_{\mathrm{het}}(t)\,M_c(t)\left(\Delta_1 - \Delta_2\right), \\
\frac{\mathrm{d}(\ee_1 - \ee_2)}{\mathrm{d}t} &= -\frac{k_{\mathrm{het}}(t) - k_{\mathrm{homo}}(t)}{2}\,M_c(t) \\
&\quad \times \left(1 + \ee_1\ee_2\right)\left(\ee_1 - \ee_2\right),
\end{aligned}
\label{eq:noncrossing_invariant}
\end{equation}
which integrates to
\begin{equation}
\begin{aligned}
\Delta_1(t) - \Delta_2(t) &= \left[\Delta_1(0) - \Delta_2(0)\right] \\
&\quad \times \exp\left(-\int_0^t k_{\mathrm{het}}(s) M_c(s)\,\mathrm{d}s\right).
\end{aligned}
\label{eq:noncrossing_solution}
\end{equation}
Because the exponential factor in Eq.~\eqref{eq:noncrossing_solution} is positive, $\Delta_1(t) - \Delta_2(t)$ and $\ee_1(t) - \ee_2(t) = [\Delta_1(t)-\Delta_2(t)]/M_c(t)$ \emph{cannot change sign or cross zero} under any parity-symmetric equimolar Channel S law. Across $500$ randomly sampled Channel S rate laws ($2.5\times 10^5$ conversion points up to $\xi = 0.99$), zero trajectories cross zero (maximum integral-invariant residual $2.7\times 10^{-12}$), while reference $H_0$ ($r=2$) retains $[\ee_1(90\%)-\ee_2(90\%)] / [\ee_1(0)-\ee_2(0)] = 0.27 > 0$ ($+0.43\text{ pp}$ vs.\ $+1.61\text{ pp}$). Extending Proposition~4 to $1 \times K$ competitive ligation between electrophile $1$ and $K$ nucleophiles\cite{Deng2024,Barat2024} gives $\mathrm{d}(\Delta_1 - \sum_{k=1}^K \Delta_{2,k})/\mathrm{d}t = -\sum_{k=1}^K k_{\mathrm{het},k}(t,\bm{x})(M_{2,k}\Delta_1 - M_1\Delta_{2,k})$ alongside global chiral balance $\Delta_1 + \sum_{k=1}^K \Delta_{2,k} + 2\sum_{k=1}^K \Delta_{P,\mathrm{homo},k} = \mathrm{const}$ ($0/240$ sign flips, residual $6.66\times 10^{-16}$; Sec.~S5.C).

We evaluate competing mechanisms on S79 first against binary S80--S82 ($n=17$, $11$ clusters; Fig.~\ref{fig:deng}(b), Sec.~S5.B, and Table~S5) and then zero-shot across the $59$-point ($23$-cluster) S80--S86 benchmark including three-component OOD Tables~S83--S86 ($n=42$, using single-amino-acid selectivities $r_{\mathrm{Val}} = 2.0$, $r_{\mathrm{Leu}} = 1.5152$, Table~S49 $K_{\mathrm{eq}} = 0.8279\text{ M}^{-1}$, and $^{13}\text{C}$ stock $\ee_1(0) = 3.05\%$; Fig.~\ref{fig:deng}(c), Sec.~S5.C, and Table~S10). Stoichiometric offset $H_{\mathrm{stoich}}$ degrades accuracy relative to $H_0$ on binary S81 HPLC ($5.4\text{ pp}$; $3.8\text{ pp}$ pooled, $p = 0.045$) and across all $59$ OOS points ($3.58$ vs.\ $2.18\text{ pp}$ for $H_0$, $p = 1.09\times 10^{-4}$ vs.\ $H_{\mathrm{ti,tQSSA}}$). Reversible thioimidate $\alpha$-CH epimerization ($L_1 \rightleftharpoons TI_1 \rightleftharpoons D_1$, Channel B)\cite{Deng2024,Barat2024} captures S79 ($\mathrm{RMSE} = 1.0$--$1.3\text{ pp}$ vs.\ $3.0\text{ pp}$ for $H_0$) and binary S81 ($1.2$--$1.3\text{ pp}$ HPLC and $0.27$--$0.30\text{ pp}$ $^{13}\text{C}$-NMR, outperforming $H_{\mathrm{stoich}}$ on $7/11$ clusters at one-sided sign-flip $p = 0.031$--$0.033$ and $H_0$ on $8/11$ split-pot clusters at $p = 0.032$--$0.033$; Table~S5). On the $59$-point OOS set, two-parameter $H_{\mathrm{Leu},2\mathrm{p}}$ and $H_{\mathrm{ti},2\mathrm{p}}$ ($r_{\mathrm{Val}} = 2.13$) lower RMSE to $1.92$ and $1.94\text{ pp}$ ($1.65$ and $1.66\text{ pp}$ on OOD S83--S86, $p = 0.0127$ vs.\ $H_0$; Table~S10). Because conversion rescaling cancels $f_{\mathrm{TI}}([\mathrm{RSH}])$ at fixed thiol via $k_{\mathrm{epim,eff}}(\xi) = k_{\mathrm{epim,TI}} + k_{\mathrm{epim,free}}/(K_{\mathrm{TI}}[\mathrm{RSH}](\xi))$ (practical rank $2/3$ on S79), diluting $[\mathrm{RSH}]_0$ from $400$ to $50\text{ mM}$ breaks tQSSA invariance\cite{Byun2024} with a $3.13\text{ pp}$ Briggs--Haldane shift and a $21.77\text{ pp}$ free-nitrile vs.\ adduct divergence at $\xi = 0.90$, while Tables~S50 and S52 $^{1}\text{H}$-NMR kinetics\cite{Deng2024,Byun2024} cut $[\mathrm{Adduct}](t)$ RMSE from $5.82$ to $1.14\text{ mM}$ and from $2.55$ to $0.62\text{ mM}$ and quenched RPKA across $\{50, 150, 400\}\text{ mM}$ restores $4/4$ FIM rank ($\lambda_{\min} = 9.6$ at $48$ scalars, $15.1$ at $72$ scalars).

Coupling heterochiral peptide or azapeptide ligation with diastereoselective phase separation\cite{Deng2024,Yan2024} governs the $5$-day seeded dipeptide crystallization series\cite{Ahn2022,Nahas2026} (Tables~S90 and S93--S97; Sec.~S5.D and Table~S11). Unseeded solutions below $c_{\mathrm{nuc,LD,tot}} = 53.75\text{ mM}$ yield $0.0\text{ mM}$ solid (S93 and S97), whereas Hill--Sheng partitioning ($k_{p,\mathrm{Ala}} = 0.184$, $k_{p,\mathrm{Leu}} = 0.758$) and Guggenheim solid solutions\cite{Ma2024,Sun2023} ($a_{0,\mathrm{Leu}} = -1.38$) lower held-out Day-5 binary S94+S96 $\ee$ RMSE from $12.58\text{ pp}$ ($k_p = 0$) to $2.20$ and $3.85\text{ pp}$ ($11.06$ to $3.24$ and $2.40\text{ pp}$ zero-refit on S96).

\section{Multiphase Wet--Dry Coupling, Solid-Phase Recycling, and Identifiability}
\label{sec:multiphase}

\subsection{Integrable Wet--Dry Law and Solid-Phase Racemization Recycling}
\label{sec:coupled_recycling}

Consider a cyclic wet--dry reactor in which each dry phase precipitates a fraction $\mu_{\mathrm{het}}(M)$ of $1{:}1$ $L{:}D$ pairs ($\Delta q = 0$) and a fraction $\mu_{\mathrm{homo}}(M)$ of homochiral solute. Each cycle updates $\Delta_{n+1} = [1 - \mu_{\mathrm{homo}}(M)]\Delta_n$ and $M_{n+1} = [1 - \mu_{\mathrm{het}}(M) - \mu_{\mathrm{homo}}(M) + \mu_{\mathrm{het}}(M)\ee_n^2]M_n$. In the small-step continuum limit ($\mu_{\mathrm{het}}, \mu_{\mathrm{homo}} \ll 1$, Sec.~S6.A), the per-cycle step reduces to
\begin{equation}
\begin{aligned}
\frac{\mathrm{d}\ee}{\mathrm{d}n} &= C(M)\,\ee\left(1 - \ee^2\right), \\
C(M) &= \frac{\mu_{\mathrm{het}}(M)}{1 - \mu_{\mathrm{het}}(M) - \mu_{\mathrm{homo}}(M)}.
\end{aligned}
\label{eq:logistic_wetdry}
\end{equation}
For constant $C(M) = C_{\mathrm{const}} = 0.017$, Eq.~\eqref{eq:logistic_wetdry} integrates to the logistic law
\begin{equation}
\ee(N) = \frac{\ee_0}{\sqrt{\ee_0^2 + \left(1 - \ee_0^2\right)\mathrm{e}^{-2 C_{\mathrm{const}} N}}},
\label{eq:logistic_solution}
\end{equation}
whereas in the pre-saturation Channel R regime ($\mu_{\mathrm{homo}} = 0$, so $\Delta = \Delta_0$ and $M = M_0\ee_0/\ee$), bimolecular saturation gives $C(M)(1-\ee^2) = [C_0/(1 + K/M)]\,g(\ee)$ with $g(\ee) = r(1-\ee^2)/[(r+2)-(r-2)\ee^2]$ and $\kappa = K/M_0$ (Sec.~S6.A). Separating variables in $N(\ee_f) = \int_{\ee_0}^{\ee_f} \mathrm{d}u/[u(1-u^2)C(\Delta_0/u)]$ yields the closed-form law ($r = 40$, $C_0 = 0.043$, $\kappa = 0.142$, median error $7.4\%$):
\begin{equation}
\begin{aligned}
N(\ee_f) &= \frac{1}{C_0}\Bigg\{\ln\frac{\ee_f}{\ee_0} + \frac{2}{r}\ln\!\left[\frac{\ee_f\left(1-\ee_0^2\right)}{\ee_0\left(1-\ee_f^2\right)}\right] \\
&\quad + \frac{\kappa}{r\,\ee_0}\Bigl[\left(r-2\right)\left(\ee_f - \ee_0\right) \\
&\quad + 4\left(\operatorname{atanh}\ee_f - \operatorname{atanh}\ee_0\right)\Bigr]\Bigg\}.
\end{aligned}
\label{eq:quadrature_wetdry}
\end{equation}

\begin{figure}[!tbp]
\centering
\includegraphics[width=0.92\columnwidth]{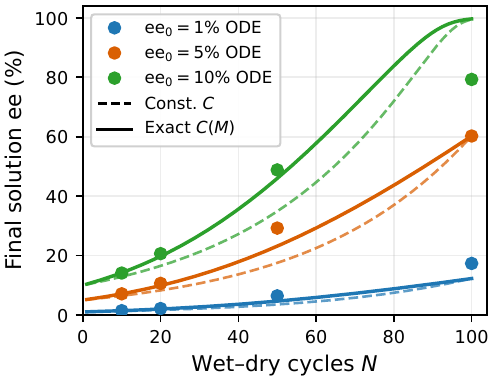}
\caption{Analytical wet--dry law vs.\ ODE integrations over $N = 100$ cycles for $\ee_0 \in \{1\%, 5\%, 10\%\}$ (circles: ODE; dashed: constant-$C$ Eq.~\eqref{eq:logistic_solution}; solid: quadrature Eq.~\eqref{eq:quadrature_wetdry}). At $\ee_0 = 10\%$ and $N > 60$, unclipped $C(M)$ overshoots the ODE points because the solution reaches the homochiral solubility boundary ($\ee \to 80\%$), where minority exhaustion and homochiral co-precipitation compress the per-cycle step.}
\label{fig:analytic_law}
\end{figure}

Figure~\ref{fig:analytic_law} shows that Eq.~\eqref{eq:quadrature_wetdry} tracks the ODE points through $N \le 60$ for all seeds and $N = 100$ for $\ee_0 = 5\%$. At $\ee_0 = 10\%$ and $N > 60$, unclipped $C(M)$ exceeds the ODE points ($\sim 99\%$ vs.\ $79.4\%$ at $N = 100$) due to \emph{boundary saturation near the homochiral wall}: once $\ee \to 80\%$ and $M/M_0 \to 12.5\%$, minority $D$ is nearly exhausted and homochiral $L$-solid co-precipitation ($\mu_{\mathrm{homo}} > 0$) caps per-cycle enrichment.

\begin{figure*}[!t]
\centering
\includegraphics[width=0.85\textwidth]{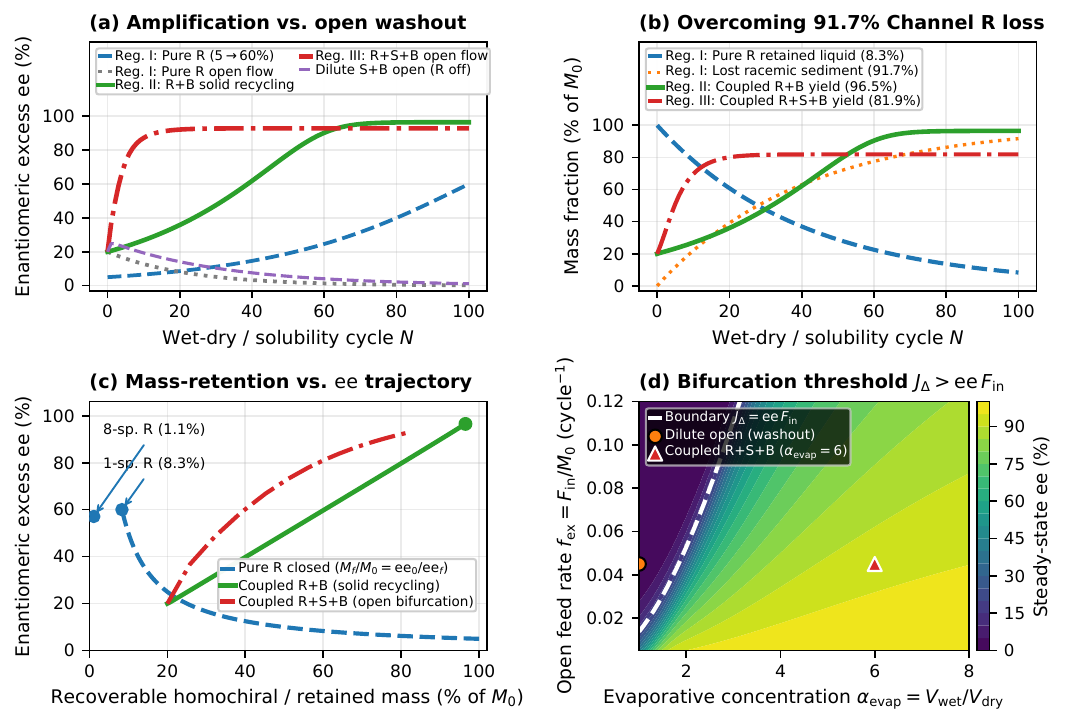}
\caption{Overcoming the Channel R mass-loss ceiling and open-flow washout via multiphase coupling (Regimes I--III). (a) $\ee(N)$ over $N = 100$ cycles for Regime I closed Channel R ($5\% \to 60.0\%$, blue dashed), Regime I open flow ($20\% \to 0.22\%$, gray dotted), Regime II coupled Channel R + B solid recycling\cite{Deck2024,Deck2025} ($20\% \to 96.50\%$ total, $100.00\%$ solid, green solid), Regime III coupled Channel R + S + B open evaporative bifurcation\cite{Ribo2025ChemPhysChem} ($\alpha_{\mathrm{evap}} = 6.0$, $20\% \to 92.87\%$, red dash-dotted), and dilute flow without Channel R ($\alpha_{\mathrm{evap}} = 1.0$, $20\% \to 1.16\%$, purple dashed). (b) Recoverable homochiral mass (\% of $M_0$), contrasting $91.7\%$ loss of pure Channel R ($5.0\%$ homochiral yield out of $8.3\%$ retained solute) against $96.5\%$ crystal recovery in Regime II and $81.9\%$ homochiral yield in Regime III. (c) Trajectories in the (recoverable mass, $\ee$) plane for 1-species ($5.0\%$) and 8-species ($0.63\%$) pure Channel R vs.\ coupled recycling. (d) Open-flow steady $\ee$ across $\alpha_{\mathrm{evap}} = V_{\mathrm{wet}}/V_{\mathrm{dry}}$ and $f_{\mathrm{ex}} = F_{\mathrm{in}}/M_0$, with white boundary $J_\Delta = \ee F_{\mathrm{in}}$ ($\alpha_{\mathrm{crit}} = 1.84$ at $f_{\mathrm{ex}} = 0.045\text{ cycle}^{-1}$).}
\label{fig:coupled_recycling}
\end{figure*}

Under standalone Channel R ($L + D \to P_{LD,\mathrm{solid}}$ with $\Delta q_\rho = 0$), Eq.~\eqref{eq:pathwise_lock} enforces a severe mass-loss ceiling: amplifying $\ee_0 = 5.0\%$ to $60.2\%$ in the 5-state wet--dry ODE at cycle 80 (or $60.0\%$ in the 3-state monomer-sink ODE at cycle 100) retains only $8.3\%$ of soluble mass ($91.7\%$ loss), while the 8-species network retains $1.1\%$ ($98.9\%$ loss; Table~\ref{tab:attribution}). Under open racemic inflow at $f_{\mathrm{ex}} = F_{\mathrm{in}}/M_0 = 0.045\text{ cycle}^{-1}$ ($\ee_{\mathrm{in}} = 0$), Eq.~\eqref{eq:open_washout_law} with $J_\Delta = 0$ drives exponential washout $\ee(N) = \ee_0 \mathrm{e}^{-f_{\mathrm{ex}} N}$ down to $0.22\%$ at $N = 100$ [gray dotted curve in Fig.~\ref{fig:coupled_recycling}(a)].

Coupling phase partitioning with solid- or liquid-phase chiral conversion\cite{Viedma2005,Suwannasang2013,Deck2024,Deck2025,Ribo2025ChemPhysChem,DuKnowles2026Multiplicative,HagelschuerFeringa2026,HeinFeringa2025Supramolecular} overcomes both limits (Fig.~\ref{fig:coupled_recycling}):
\begin{enumerate}
\item \textbf{Regime II (Coupled Channel R + B: Solid Racemization Recycling):} Unlike Regime I racemic-compound precipitation, Viedma ripening\cite{Viedma2005} and the temperature-cycling deracemization of Deck, Hosseinalipour, and Mazzotti\cite{Deck2024,Deck2025} operate on conglomerate-forming crystals $(n_L, n_D)$ ($\Delta q_\rho = \pm 1$) coexisting with a racemic eutectic supernatant ($c_L = c_D = c_{\mathrm{eq}} = 0.0175 M_0$). We integrate the 4-state two-phase mass balance for $(c_L, c_D, n_L, n_D)$ at conserved total inventory $M_{\mathrm{tot}} = n_L + n_D + c_L + c_D = M_0$ (Sec.~S6.B):
\begin{equation}
\frac{\mathrm{d}c_i}{\mathrm{d}t} + \frac{\mathrm{d}n_i}{\mathrm{d}t} = -k_{\mathrm{rac}}\left(c_i - c_j\right), \qquad i \neq j \in \{L, D\}.
\label{eq:deck_recycling}
\end{equation}
Cooling and drying (Channel R) buffer the supernatant at $c_L + c_D = 0.035 M_0$ while crystallizing $96.5\%$ of the solute. Subsequent warming with dissolution--growth Damk\"ohler asymmetry ($a_d = 1.85 > a_g = 0.42$)\cite{Deck2024,Deck2025} and liquid-phase racemization ($k_{\mathrm{rac}} > 0$, Channel B) converts minority $D$-crystals into majority $L$-crystals. Over $N = 100$ cycles from $\ee_0 = 20\%$, Regime II reaches $100\%$ solid-crystal $\ee$ and $96.5\%$ total two-phase $\ee$, yielding $Y_{\mathrm{hom}}(100) = (n_L - n_D)/M_0 = 96.5\%$ recoverable homochiral crystal mass [green curves in Fig.~\ref{fig:coupled_recycling}(a--c); exact shares in Table~\ref{tab:attribution} and Table~S6]. A $2^2$ factorial decomposition across $\{\varnothing, \mathrm{R}, \mathrm{B}, \mathrm{R}+\mathrm{B}\}$ confirms that standalone R or B leaves $\ee_{\mathrm{tot}}(100) = Y_{\mathrm{hom}}(100) = 20.0\%$, so the entire $+76.5\text{ pp}$ gain arises from M\"obius synergy $I_{RB} = +76.5\text{ pp}$ with equal Shapley shares $\phi_R = \phi_B = +38.3\text{ pp}$ (Table~S6).
\item \textbf{Regime III (Coupled Channel R + S + B: Open Evaporative Bifurcation):} Following Rib\'o, Micheau, and Buhse,\cite{Ribo2025ChemPhysChem} we integrate a 6-state open-flow network for monomers, dimers, and racemic precipitate under continuous racemic inflow $f_{\mathrm{ex}} = F_{\mathrm{in}}/M_0 = 0.045\text{ cycle}^{-1}$ ($\ee_{\mathrm{in}} = 0$, reference feed $c_{\mathrm{wet},0} = M_0 = 1.0$; full ODEs in Sec.~S6.B). This network couples reversible heterochiral precipitation ($P_{LD} \rightleftharpoons S_{\mathrm{rac}}$, Channel R), diastereoselective dimerization ($2i \rightleftharpoons P_{ii}$, $L + D \rightleftharpoons P_{LD}$, Channel S), and cooperative autocatalysis at dry-phase concentration $c_{\mathrm{dry}} = \alpha_{\mathrm{evap}} c_{\mathrm{wet},0}$ elevated by evaporation ($\alpha_{\mathrm{evap}} = V_{\mathrm{wet}}/V_{\mathrm{dry}}$). Dividing the chiral-excess rate $\mathrm{d}\Delta_{\mathrm{chiral}}/\mathrm{d}n$ by stationary active chiral inventory $M_{\mathrm{chiral}} = c_L + c_D + 2(P_{LL} + P_{DD})$ (Eq.~(S27) in Sec.~S6.B) gives the intensive Bernoulli ODE for $\ee = \ee_{\mathrm{chiral}} = \Delta_{\mathrm{chiral}}/M_{\mathrm{chiral}}$:
\begin{equation}
\frac{\mathrm{d}\ee}{\mathrm{d}n} = -f_{\mathrm{ex}}\,\ee + \frac{k_B \alpha_{\mathrm{evap}}^2 c_{\mathrm{wet},0}^2}{1 + K_{\mathrm{sat}}\alpha_{\mathrm{evap}}^2 c_{\mathrm{wet},0}^2}\,\ee\left(1 - \ee^2\right).
\label{eq:ribo_open_bifurcation}
\end{equation}
Amplification survives open dilution when $\alpha_{\mathrm{evap}}$ crosses the bifurcation threshold
\begin{equation}
\alpha_{\mathrm{evap}} > \alpha_{\mathrm{crit}} = \frac{1}{c_{\mathrm{wet},0}}\sqrt{\frac{f_{\mathrm{ex}}}{k_B - f_{\mathrm{ex}} K_{\mathrm{sat}}}}.
\label{eq:alpha_crit}
\end{equation}
For $k_B = 0.014\text{ cycle}^{-1}c_{\mathrm{wet},0}^{-2}$, $K_{\mathrm{sat}} = 0.015\,c_{\mathrm{wet},0}^{-2}$, and $f_{\mathrm{ex}} = 0.045\text{ cycle}^{-1}$, Eq.~\eqref{eq:alpha_crit} predicts $\alpha_{\mathrm{crit}} = 1.84$, matching 6-state ODE bisection within $2\times 10^{-10}$ (Sec.~S6.B). Dilute flow at $\alpha_{\mathrm{evap}} = 1.0 < \alpha_{\mathrm{crit}}$ washes out to $\ee_{\mathrm{chiral}}(100) = 1.2\%$ ($Y_{\mathrm{hom}} = 0.87\%$), whereas evaporative concentration at $\alpha_{\mathrm{evap}} = 6.0 > \alpha_{\mathrm{crit}}$ sustains $\ee_{\mathrm{chiral}}(100) = 92.9\%$ and $Y_{\mathrm{hom}} = \Delta_{\mathrm{chiral}}/M_0 = 81.9\%$ with retained soluble fraction $M_{\mathrm{sol}}/M_0 = 0.95$ [Fig.~\ref{fig:coupled_recycling}(d); Table~\ref{tab:attribution}]. Across the $2^3$ factorial decomposition and fixed-evaporation precipitation control at $\alpha_{\mathrm{evap}} = 6.0$ (Sec.~S6.B and Table~S6), both $\text{R}+\text{B}$ and $\text{R}+\text{S}+\text{B}$ obey Eq.~\eqref{eq:ribo_open_bifurcation} and reach $92.9\%$ $\ee_{\mathrm{chiral}}$ through pairwise synergy $I_{RB} = +92.0\text{ pp}$. Without Channel S ($\text{R}+\text{B}$), no achiral $P_{LD}$ or $S_{\mathrm{rac}}$ forms and $Y_{\mathrm{hom}}$ reaches $92.9\%$; adding Channel S ($\text{R}+\text{S}+\text{B}$) contributes $+7.5\text{ pp}$ of within-run solution scavenging ($A_R + A_S$) while sequestering $11.8\%$ of mass into achiral $P_{LD} + S_{\mathrm{rac}}$, lowering $Y_{\mathrm{hom}}$ to $81.9\%$ ($-11.0\text{ pp}$ total, including $-4.2\text{ pp}$ from precipitation alone; Table~S6). Continuous four-channel Aumann--Shapley attribution $(\Phi_R, \Phi_S, \Phi_B, \Phi_{\mathrm{flow}})$ separates chiral-pool $\ee_{\mathrm{chiral}}$ ($0.0000, +0.5242, +4.8958, -4.6913$; sum $+0.7287$) from total-soluble $\ee_{\mathrm{sol}}$ ($+0.2062, 0.0000, +4.4298, -3.9709$; sum $+0.6651$; Sec.~S2.B and Table~S8).
\end{enumerate}

\subsection{Multiphase Observability and Mineral Adsorption Identifiability}
\label{sec:identifiability}

\begin{figure*}[!t]
\centering
\includegraphics[width=0.85\textwidth]{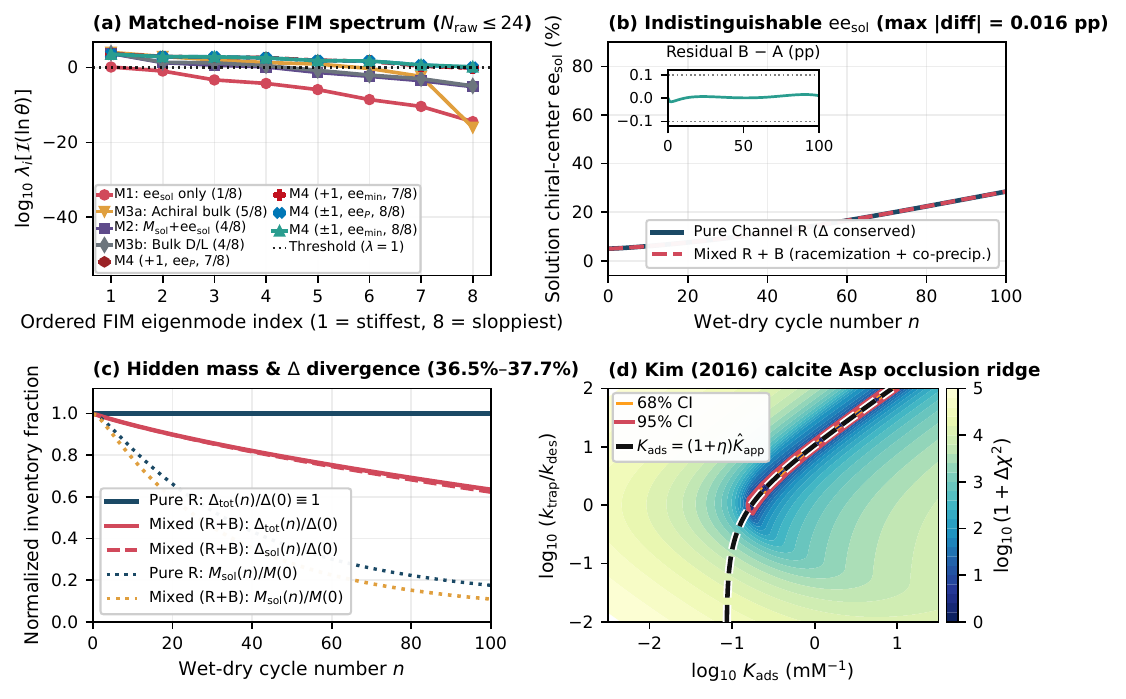}
\caption{Multiphase observability and resolution of the mineral-adsorption identifiability ridge. (a) Ordered FIM eigenvalues $\log_{10}\lambda_i(\mathcal{I}(\ln\bm{\theta}))$ across protocols ($N_{\mathrm{raw}} \le 24$ scalars): ratio-only $\ee_{\mathrm{sol}}$ (M1, practical rank $1/8$), single-phase $(M_{\mathrm{sol}}, \ee_{\mathrm{sol}})$ (M2, rank $4/8$), achiral mineral (M3a, rank $5/8$), bulk acid-hydrolyzed solid (M3b, rank $4/8$), single-enantiomorph crystal (M4 single $+1$, rank $7/8$, $\lambda_{\min} = 0.65$--$0.73$), and antipodal mirror-crystal pair (M4 mirror $+1/-1$, full rank $8/8$, $\lambda_{\min} = 1.01$--$1.55$). (b) Indistinguishable $\ee_{\mathrm{sol}}(n)$ trajectories ($\max|\mathrm{diff}| = 0.016\text{ pp}$) for pure Channel R vs.\ mixed Channel R + B. (c) Hidden $36.5\%$--$37.7\%$ divergence in $\Delta_{\mathrm{tot}}(n)/\Delta(0)$ and $M_{\mathrm{sol}}(n)/M(0)$. (d) Profile-likelihood ridge $K_{\mathrm{ads}} = (1+\eta)\hat{K}_{\mathrm{app}}$ ($\hat{K}_{\mathrm{app}} = 0.0854\text{ mM}^{-1}$, with $68\%$ and $95\%$ confidence contours at $\Delta\chi^2 = 2.30$ and $5.99$) in Kim \emph{et al.}\cite{Kim2016} calcite Asp occlusion data ($\eta = k_{\mathrm{trap}}/k_{\mathrm{des}}$).}
\label{fig:observability}
\end{figure*}

Supernatant-only $\ee_{\mathrm{sol}}(n)$ measurements cannot separate phase removal from coupled racemization and co-precipitation in multiphase reactors [Fig.~\ref{fig:observability}(b,c)]. Comparing a Channel R trajectory ($\Delta_{\mathrm{tot}}(n)/\Delta(0) = 1$) against a mixed Channel R + B trajectory over $N = 100$ cycles yields liquid $\ee_{\mathrm{sol}}(n)$ curves that agree within $0.016\text{ pp}$ [Fig.~\ref{fig:observability}(b)], well below typical chiral HPLC noise ($\sim 0.5\text{ pp}$), even though $\Delta_{\mathrm{tot}}(n)/\Delta(0)$ and $M_{\mathrm{sol}}(n)/M(0)$ diverge by $36.5\%$ and $37.7\%$ [Fig.~\ref{fig:observability}(c)].

\noindent\textbf{Proposition 5 (Mineral Occlusion Non-Identifiability Ridge and Antipodal Conditioning).} An experimental instance of multiphase non-identifiability appears in the mineral occlusion data of Kim \emph{et al.},\cite{Kim2016} who measured D/L-aspartic acid (Asp) incorporation into growing calcite crystals across $5$--$50\text{ mM}$ concentrations and stirring conditions (bleached intracrystalline fraction $P$ vs.\ pooled crystal $B$). Modeling reversible step-edge adsorption ($K_{\mathrm{ads}} = k_{\mathrm{ads}}/k_{\mathrm{des}}$) coupled with irreversible lattice trapping ($\eta = k_{\mathrm{trap}}/k_{\mathrm{des}}$) shows that single-phase and bulk occlusion observables depend on $(K_{\mathrm{ads}}, \eta)$ solely through the effective adsorption-to-desorption/trapping ratio (Sec.~S7.A)
\begin{equation}
\begin{aligned}
K_{\mathrm{app}} &= \frac{K_{\mathrm{ads}}}{1 + \eta} = \hat{K}_{\mathrm{app}} = 0.0854\text{ mM}^{-1} \\
&\Longleftrightarrow K_{\mathrm{ads}} = \left(1 + \eta\right)\hat{K}_{\mathrm{app}},
\end{aligned}
\label{eq:kim_ridge}
\end{equation}
forming a 1D flat valley ($\Delta\chi^2_{\mathrm{profile}} = 0.0$ across $(\log_{10}K_{\mathrm{ads}}, \log_{10}\eta)$ at fixed trapping scale $k_{\mathrm{trap}}$) with invariance $(K_{\mathrm{ads}}, 1+\eta) \mapsto (\alpha K_{\mathrm{ads}}, \alpha(1+\eta))$ across four decades of parameter space [Fig.~\ref{fig:observability}(d)]. Evaluating all $112$ public RP-HPLC rows of Kim \emph{et al.}\cite{Kim2016} ($82$ Asp + $30$ Gly across $57$ batches, $63$ stems, and $29$ conditions) reveals symmetric $7/14$ ($U < B$) versus $7/14$ ($U > B$) unbleached/bleached stem splits driven by carbonate crystal-growth and amino-acid incorporation heterogeneity\cite{Gardella2024,StaniendaPilecki2023,Chomhom2026} ($\sigma_{\mathrm{tech}} = 0.038 \ll \sigma_{\mathrm{split}} = 0.246$). Consequently, hierarchical equal-mean $M_0$ ($U = B$: $\mathrm{BIC}_{112} = -1352.38$, $\mathrm{LOCO\text{-}29}\ \mathrm{ELPD} = 661.87\text{ nats}$, $p_{2\mathrm{s}} = 0.0039$ vs.\ pooled i.i.d.) is favored over surface-excess $M_2$ ($U \ge B$: $\mathrm{BIC}_{112} = -1349.05$, $\Delta\mathrm{BIC} = +3.33$, $\mathrm{ELPD} = 659.49\text{ nats}$, $p_{2\mathrm{s}} = 0.35$; Sec.~S7.A and Table~S12).

Accordingly, to design a forward prospective protocol for enantiomorphic mineral surfaces (rather than bulk achiral calcite) in a coupled 10-state mineral--water--solid network (Sec.~S7.B), we evaluate the matched-noise Fisher Information Matrix $\mathcal{I}(\ln\bm{\theta})$ for $\bm{\theta} = (k_{\mathrm{homo}}, r_{\mathrm{het}}, k_{\mathrm{precip}}, k_{\mathrm{hyd}}, k_{\mathrm{ads}}, k_{\mathrm{des}}, k_{\mathrm{trap}}, \alpha_{\mathrm{min}})^\top$. Here $\alpha_{\mathrm{min}}$ denotes the chiral mineral surface selectivity modulating homochiral dimer adsorption $k_{\mathrm{ads},LL/DD}^{(h)} = k_{\mathrm{ads}}(1 \pm h\,\alpha_{\mathrm{min}})$ on enantiomorphic crystal face $h \in \{+1, -1\}$. Comparing six observation protocols under a fixed budget of $N_{\mathrm{raw}} \le 24$ scalars ($2.5\%$ relative mass noise with $0.5\%\,M_0$ floor and $1.0\text{ pp}$ $\ee$ noise; Table~S7) establishes the identifiability hierarchy in Fig.~\ref{fig:observability}(a):
\begin{enumerate}
\item Supernatant $\ee_{\mathrm{sol}}(t)$ alone (M1) yields structural rank $7/8$ and practical rank $1/8$ with $\lambda_{\min} = 3.6\times 10^{-15}$. Adding liquid mass $M_{\mathrm{sol}}(t)$ (M2, $\lambda_{\min} = 7.8\times 10^{-6}$) or bulk acid-hydrolyzed solid (M3b, $\lambda_{\min} = 1.3\times 10^{-5}$) reaches structural rank $8/8$ but practical rank $4/8$ because acid dissolution mixes surface adsorbate with occluded solid.
\item Resolving desorbed surface adsorbate and bleached intracrystalline solid on a single crystal enantiomorph (M4 single $+1$) reaches structural rank $8/8$ and practical rank $7/8$ at four timepoints ($\lambda_{\min} = 0.61$--$0.73$, rising to $1.31$ on a dense six-timepoint schedule; Table~S7), leaving a weak $k_{\mathrm{des}} \to 0$ zero-rate boundary asymptote\cite{Maiwald2016,PlankSimpson2024} $\Delta\chi^2_{B=1} = 0.89 < 3.84$ ($6/8$ closed at $B = 1$) that requires $B^* = 5$ blocks ($5\times 0.89 = 4.45 > 3.84$, interval $[-2.71, +0.84]$).
\item Splitting the budget across an \textbf{antipodal mirror-crystal pair} (M4 mirror $+1/-1$, such as $(+)$- and $(-)$-quartz or opposing calcite twin faces) flips mineral enantioselectivity $\pm \alpha_{\mathrm{min}}$ while leaving achiral solution rates invariant. This symmetry reversal restores full $8/8$ structural and practical FIM rank with $\lambda_{\min} = 1.01$ at $18$ scalars and $1.10$--$1.55$ at $24$ scalars (Table~S7), closing $7/8$ global $95\%$ profile-likelihood intervals two-sidedly at a single block ($B = 1$, $24$ scalars, with exact zero-rate boundary asymptote $\Delta\chi^2_{B=1}(k_{\mathrm{des}} \to 0) = 1.49 < 3.84$) and closing all $8/8$ two-sided $95\%$ profile-likelihood bounds at $B \ge 3$ replicate mirror blocks ($3\times 1.49 = 4.47 > 3.84$, interval $[-2.62, +0.77]$ at $B = 3$ and $[-1.28, +0.62]$ with $\mathrm{PL}/\mathrm{Wald} \ge 1.0001$ across all $8$ parameters at $B = 5$\cite{Maiwald2016,FischerLewis2021,PlankSimpson2024}; Sec.~S7.B and Fig.~S11).
\end{enumerate}

\section{Step-by-Step Recipe for Experimentalists}
\label{sec:recipe}

Experimentalists can compute $(A_R, A_S, A_B)$ directly from standard analytical measurements (chiral HPLC, quantitative NMR, UV--Vis, or gravimetric inventory) using simple algebra, \textbf{without postulating or fitting a kinetic differential-equation model}.

\subsection{Model-Free Algebraic Protocol}
\label{sec:algebraic_steps}

Given a reactor sampled at $t_0$ and $t_f$:
\begin{enumerate}
\item \textbf{Step 1 (Initial Inventory at $t_0$):} Measure initial liquid volume $V_0$ and soluble concentrations $c_{L,0}, c_{D,0}$ (via hydrolysis or NMR/HPLC) to obtain
\begin{equation}
\begin{aligned}
M_0 &= V_0\left(c_{L,0} + c_{D,0}\right), \quad \Delta_0 = V_0\left(c_{L,0} - c_{D,0}\right), \\
\ee_0 &= \frac{\Delta_0}{M_0}.
\end{aligned}
\label{eq:recipe_step1}
\end{equation}
\item \textbf{Step 2 (Final Inventories at $t_f$):} Measure surviving volume $V_f$, monomer-equivalent soluble chiral-center concentrations $c_{L,f}, c_{D,f}$ (via oligomer hydrolysis or weighted summation $c_{L,f} = [L]_f + 2[P_{LL}]_f + [P_{LD}]_f$, $c_{D,f} = [D]_f + 2[P_{DD}]_f + [P_{LD}]_f$), and subpool concentrations $c_{L,\mathrm{sub},f}, c_{D,\mathrm{sub},f}$ (e.g., $[L]_f, [D]_f$) to obtain
\begin{equation}
\begin{aligned}
M_f &= V_f\left(c_{L,f} + c_{D,f}\right), \quad \Delta_f = V_f\left(c_{L,f} - c_{D,f}\right), \\
\ee_f &= \frac{\Delta_f}{M_f}, \quad \ee_{\mathrm{sub},f} = \frac{\Delta_{\mathrm{sub},f}}{M_{\mathrm{sub},f}}, \\
M_{\mathrm{sub},f} &= V_f\left(c_{L,\mathrm{sub},f} + c_{D,\mathrm{sub},f}\right), \\
\Delta_{\mathrm{sub},f} &= V_f\left(c_{L,\mathrm{sub},f} - c_{D,\mathrm{sub},f}\right).
\end{aligned}
\label{eq:recipe_step2}
\end{equation}
\item \textbf{Step 3 (Algebraic Attribution):} Substitute $(M_0, \Delta_0, \ee_0)$, $(M_f, \Delta_f, \ee_f)$, and $\ee_{\mathrm{sub},f}$ into Eq.~\eqref{eq:ordered_attribution}:
\begin{equation}
\begin{aligned}
A_R &= \ee_0\left(\frac{M_0}{M_f} - 1\right), \quad A_S = \ee_{\mathrm{sub},f} - \ee_f, \\
A_B &= \frac{\Delta_f - \Delta_0}{M_f},
\end{aligned}
\label{eq:recipe_step3}
\end{equation}
and when both $M_f < M_0$ and $\Delta_f \neq \Delta_0$ are substantial (including $\ee_0 = 0$, where ordered $A_R = 0$), report Shapley shares [Eq.~\eqref{eq:shapley_attribution}] or continuous $\Phi_\alpha[\gamma]$ along time-series points.
\end{enumerate}

\subsection{Single Most Stringent Experimental Test}
\label{sec:stringent_test}

The \textbf{single most stringent experimental measurement} to test this framework in any multiphase or wet--dry reactor is a \emph{simultaneous time-resolved two-phase chiral inventory}: at each cycle $n$, separate supernatant from precipitated solid, redissolve the solid, and measure $(M_{\mathrm{sol}}(n), \Delta_{\mathrm{sol}}(n))$ and $(M_{\mathrm{solid}}(n), \Delta_{\mathrm{solid}}(n))$ by chiral HPLC or NMR. This verifies mass closure $M_{\mathrm{sol}}(n) + M_{\mathrm{solid}}(n) = M_0$ and tests physical fractionation ($\Delta_{\mathrm{sol}}(n) + \Delta_{\mathrm{solid}}(n) = \Delta_0$, Channel R) against chiral-center creation or deracemization ($\Delta_{\mathrm{sol}}(n) + \Delta_{\mathrm{solid}}(n) \neq \Delta_0$, Channel B).

\section{Conclusions}
\label{sec:conclusions}

We have established a physical-chemistry framework separating stoichiometric mass-removal and subpool-fractionation effects from net chiral-excess creation in prebiotic amplification:
\begin{enumerate}
\item In any $\mathbb{Z}_2$-symmetric closed network without autocatalytic feedback or epimerization, signed chiral excess is conserved along every trajectory ($\Delta(t) = \Delta(0)$), locking liquid-phase amplification to mass depletion ($M_f/M_0 = \ee_0/\ee_f$) and causing open reactors under racemic inflow to wash out to $\ee \to 0$.
\item The three-channel attribution identity ($\Delta\ee = A_R + A_S + A_B$), continuous path-covariant Aumann--Shapley attribution $\Phi_\alpha[\gamma]$, and Shear--Kramers--Lorentz relative-entropy balance explain why efficiency $\eta_G(\ee)$ decreases across the maturation regimes of mass-depleting Channels R and S (even at $\Omega = 0$: $66.90$ and $98.60\text{ kJ mol}^{-1}$ vs.\ $27.29\text{ kJ mol}^{-1}$ for Channel B at $75\%$ $\ee$, inflated $7.2\times$--$7.8\times$ at $\Omega = 6.0$).
\item Standalone Channel R incurs a $91.7\%$--$98.9\%$ closed-batch mass loss into sediment; however, coupling Channel R eutectic partitioning with Channel B racemization recycling\cite{Deck2024,Deck2025} converts the solid into homochiral product ($96.5\%$ recoverable homochiral crystal mass at $96.5\%$ total and $100\%$ solid $\ee$), and coupling evaporative Channel R concentration with liquid-phase autocatalysis\cite{Ribo2025ChemPhysChem} crosses the open-flow bifurcation threshold ($J_\Delta > \ee F_{\mathrm{in}}$) to sustain $92.9\%$ chiral $\ee$ and $81.9\%$ homochiral yield under continuous racemic re-feeding.
\item The $1 \times K$ non-crossing parity invariant validated across $59$ multicomponent peptide ligation points ($23$ clusters) and $5$-day seeded co-precipitation,\cite{Deng2024} hierarchical Occam analysis across $112$ calcite RP-HPLC rows,\cite{Kim2016} and antipodal mirror-crystal pairing ($B^* = 3$ vs.\ $B^* = 5$ single-crystal blocks) resolve mechanism confounding and parameter non-identifiability while equipping experimentalists with an algebraic, model-free attribution protocol.
\end{enumerate}

\section*{Supplementary Material}

See the supplementary material for proofs of the pathwise conservation theorem, CME parity and Gillespie tables, ordered, discrete Shapley, and continuous Aumann--Shapley derivations, Shear--Kramers--Lorentz free-energy tables, $1 \times K$ non-crossing proofs, $59$-point multicomponent peptide and seeded co-precipitation benchmarks, wet--dry quadrature derivations, and $112$-row calcite Occam, FIM, and profile-likelihood tables.

\begin{acknowledgments}
The physical question that inspired this study originated during the first semester of the first author's freshman year in the multi-lecturer course \emph{Introduction to Frontiers of Medicine} at the Medical College of Tianjin University (Weijin Road Campus), coordinated by Professor Dong Ming, now Vice President of Tianjin University. During a lecture touching upon biomolecular homochirality, the instructor posed the question of why biological molecules are predominantly left-handed. When the first author suggested that a slight random initial majority could be progressively amplified as reactions proceed, the instructor offered warm encouragement and emphasized that any scientific hypothesis should be tested through a rigorous process. Although a considerable span of time passed before the first author was able to translate that early classroom intuition into the mathematical proofs, thermodynamic bounds, and quantitative re-attributions presented here, the intellectual seed planted in that lecture remained the guiding motivation throughout this work.
\end{acknowledgments}

\section*{Author Declarations}

\subsection*{Conflict of Interest}
The authors have no conflicts to disclose.

\subsection*{Author Contributions}
A. Gong and B. Wu contributed equally to this work. \textbf{Zeyu Jia:} Conceptualization (lead); Data curation (lead); Formal analysis (lead); Investigation (lead); Methodology (lead); Software (lead); Validation (lead); Visualization (lead); Writing -- original draft (lead); Writing -- review \& editing (lead). \textbf{Anlin Gong:} Conceptualization (supporting); Investigation (supporting); Validation (supporting); Writing -- review \& editing (supporting). \textbf{Bingmei Wu:} Conceptualization (supporting); Investigation (supporting); Validation (supporting); Writing -- review \& editing (supporting). During the preparation of this work, the authors used Google Gemini and Antigravity to assist with code verification and language editing, and reviewed and edited the content as needed, taking full responsibility for the manuscript.

\section*{Data Availability}
The data that support the findings of this study are available within the article, its supplementary material, and the self-contained reproducibility archive (\texttt{reproducibility\_package\_jcp.zip}), which contains all simulation, data-processing, verification, and figure-generation scripts together with the public experimental benchmark datasets and precomputed outputs.

\bibliographystyle{aipnum4-2}
\bibliography{references}

\clearpage
\onecolumngrid
\begin{center}
{\large\bfseries Supplementary Material for:\\Symmetry-Induced Conservation Laws, Three-Channel Attribution, and Solid-Phase Recycling in Prebiotic Enantiomeric Excess Amplification}\\[6pt]
Zeyu Jia\textsuperscript{1,*}, Anlin Gong\textsuperscript{2,$\dagger$}, and Bingmei Wu\textsuperscript{2,$\dagger$}\\
{\small\itshape \textsuperscript{1}School of Biomedical Engineering and Technology, Tianjin Medical University, Tianjin 300070, China\\
\textsuperscript{2}School of Pharmacy, Tianjin Medical University, Tianjin 300070, China\\
\textsuperscript{$\dagger$}A. Gong and B. Wu contributed equally to this work.}
\end{center}
\setcounter{section}{0}
\setcounter{equation}{0}
\setcounter{figure}{0}
\setcounter{table}{0}
\renewcommand{\thesection}{S\arabic{section}}
\renewcommand{\theequation}{S\arabic{equation}}
\renewcommand{\thefigure}{S\arabic{figure}}
\renewcommand{\thetable}{S\arabic{table}}
\renewcommand{\theHsection}{S\arabic{section}}
\renewcommand{\theHequation}{S\arabic{equation}}
\renewcommand{\theHfigure}{S\arabic{figure}}
\renewcommand{\theHtable}{S\arabic{table}}

\tableofcontents
\newpage

\section{Pathwise \texorpdfstring{$\mathbb{Z}_2$}{Z2} Conservation Proofs, Open-Flow Washout, and Spatially Resolved PDE Verification}
\label{sec:s1_conservation}

\subsection{Stoichiometric Algebra and Proof of the Pathwise Conservation Law}

Consider a reaction network composed of $S$ chemical species indexed by $k \in \{1, \dots, S\}$ partitioned between an active liquid phase $\mathcal{A}$ and one or more solid, adsorbed, or outflow reservoirs $\mathcal{R}$. Each species $k \in \mathcal{A}$ is assigned two integer quantum numbers:
\begin{enumerate}
\item The \emph{chiral-center mass} $m_k = n_{L,k} + n_{D,k} \in \mathbb{Z}_{\ge 0}$, counting the total number of chiral monomer units contained in species $k$ ($m_{L_i} = m_{D_i} = 1$ for monomers; $m_{P_{LL}} = m_{P_{DD}} = m_{P_{LD}} = 2$ for dimers; $m_A = 0$ for achiral precursors or solvent).
\item The \emph{signed chiral charge} $q_k = n_{L,k} - n_{D,k} \in \mathbb{Z}$, counting the net excess of $L$ over $D$ units in species $k$ ($q_{L_i} = +1$, $q_{D_i} = -1$, $q_{P_{LL}} = +2$, $q_{P_{DD}} = -2$, $q_{P_{LD}} = 0$).
\end{enumerate}
For any state vector of molecular copy numbers $\bm{n}(t) = (n_1(t), \dots, n_{|\mathcal{A}|}(t))^\top \in \mathbb{Z}_{\ge 0}^{|\mathcal{A}|}$ within active compartment volume $\Omega(t)$, the total active chiral inventory $M(t)$ and signed chiral excess $\Delta(t)$ are linear functionals of $\bm{n}(t)$:
\begin{equation}
M(t) = \bm{m}^\top \bm{n}(t) = \sum_{k \in \mathcal{A}} m_k n_k(t), \qquad \Delta(t) = \bm{q}^\top \bm{n}(t) = \sum_{k \in \mathcal{A}} q_k n_k(t), \qquad \ee(t) = \frac{\Delta(t)}{M(t)}.
\label{eq:s_inventory_defs}
\end{equation}

Let $\mathcal{E}$ denote the set of elementary reaction channels $\rho$ with active-phase stoichiometric change vectors $\bm{\nu}_\rho \in \mathbb{Z}^{|\mathcal{A}|}$. Under the $\mathbb{Z}_2$ enantiomeric inversion operator $S$ that exchanges every $L$-configured moiety with its mirror-image $D$-configured moiety ($S \bm{m} = \bm{m}$ and $S \bm{q} = -\bm{q}$), elementary channels divide into three stoichiometric classes:
\begin{enumerate}
\item \textbf{Internal Oligomerization and Complexation (Channel S generators):}
\begin{equation}
L_i + L_j \rightleftharpoons P_{L_i L_j}, \qquad D_i + D_j \rightleftharpoons P_{D_i D_j}, \qquad L_i + D_j \rightleftharpoons P_{L_i D_j}.
\label{eq:s_internal_ligation}
\end{equation}
For homochiral dimerization, $\Delta M_\rho = 2 - 1 - 1 = 0$ and $\Delta q_\rho = (+2) - (+1) - (+1) = 0$ (and $0$ for $D_i + D_j$). For heterochiral dimerization, $\Delta M_\rho = 2 - 1 - 1 = 0$ and $\Delta q_\rho = 0 - (+1) - (-1) = 0$. Thus every internal reversible or irreversible oligomerization reaction satisfies $\bm{m}^\top \bm{\nu}_\rho = 0$ and $\bm{q}^\top \bm{\nu}_\rho = 0$ within the pooled active phase $\mathcal{A}$, while redistributing chiral centers between monomer and oligomer subpools.
\item \textbf{Symmetric $1{:}1$ Heterochiral Phase Removal and Achiral Sink Transport (Channel R generators):}
\begin{equation}
L_i + D_i \xrightarrow{k_{\mathrm{precip}}} \mathrm{LD}_{i,(\mathrm{s})} \in \mathcal{R}, \qquad P_{L_i D_j,(\mathrm{aq})} \xrightarrow{k_{\mathrm{cryst}}} P_{L_i D_j,(\mathrm{s})} \in \mathcal{R}.
\label{eq:s_racemic_precip}
\end{equation}
Each precipitation event removes one $L$ and one $D$ unit from $\mathcal{A}$, yielding $\Delta M_\rho = \bm{m}^\top \bm{\nu}_\rho = -2 < 0$ and $\Delta q_\rho = \bm{q}^\top \bm{\nu}_\rho = -(+1) - (-1) = 0$.
\item \textbf{Chiral-Charge-Altering Reactions (Channel B generators):}
Autocatalytic synthesis from an achiral precursor ($A + n L \to (n+1)L$, $\Delta q_\rho = +1$; $A + n D \to (n+1)D$, $\Delta q_\rho = -1$), direct or intermediate-mediated epimerization ($L_i \rightleftharpoons TI_i \rightleftharpoons D_i$, $\Delta q_\rho = \mp 1$ or $\mp 2$), or enantioselective homochiral crystallization/dissolution ($L_{(\mathrm{aq})} \rightleftharpoons L_{(\mathrm{s})}$, $\Delta q_\rho = \mp 1$).
\end{enumerate}

\textbf{Theorem S1 (Pathwise Conservation of Signed Chiral Excess).} \emph{In any closed reaction network restricted to internal oligomerization (Class 1) and $1{:}1$ heterochiral phase removal (Class 2) with zero Channel B flux ($\Delta q_\rho = 0$ for all active channels $\rho \in \mathcal{E}_0$), the signed chiral excess $\Delta(t)$ of the active phase $\mathcal{A}$ is conserved along every individual stochastic jump trajectory $\omega$ at all times $t \ge 0$:}
\begin{equation}
\Delta(t, \omega) = \Delta(0, \omega) + \sum_{\rho \in \mathcal{E}_0} \left(\bm{q}^\top \bm{\nu}_\rho\right) N_\rho(t, \omega) = \Delta(0, \omega),
\label{eq:s_pathwise_proof}
\end{equation}
\emph{where $N_\rho(t, \omega) \in \mathbb{Z}_{\ge 0}$ is the counting process of reaction $\rho$ up to time $t$. Consequently, at any stopping time $t_f$ with surviving active-phase chiral mass $M(t_f, \omega) > 0$,}
\begin{equation}
\ee(t_f, \omega) = \frac{\Delta(t_f, \omega)}{M(t_f, \omega)} = \frac{\Delta(0, \omega)}{M(t_f, \omega)} = \ee(0, \omega)\,\frac{M(0, \omega)}{M(t_f, \omega)}, \qquad \frac{M(t_f, \omega)}{M(0, \omega)} = \frac{\ee(0, \omega)}{\ee(t_f, \omega)}.
\label{eq:s_mass_lock_proof}
\end{equation}
\emph{Proof.} Because $\bm{q}^\top \bm{\nu}_\rho = 0$ for every $\rho \in \mathcal{E}_0$, every term in the stochastic sum of Eq.~\eqref{eq:s_pathwise_proof} vanishes regardless of the propensity functions, spatial heterogeneity, temperature or volume cycling protocols, or waiting-time distributions. Dividing $\Delta(t_f, \omega) = \Delta(0, \omega) = \ee(0, \omega) M(0, \omega)$ by $M(t_f, \omega)$ yields Eq.~\eqref{eq:s_mass_lock_proof}. Note that in multi-species networks ($i = 1, \dots, K$), $\bm{q}^\top\bm{\nu}_\rho = 0$ applies to the pooled total chiral excess $\Delta_{\mathrm{tot}}(t) = \sum_{i=1}^K \Delta_i(t)$ whenever cross-species heterochiral pairings ($L_i + D_j \to P_{L_i D_j,\mathrm{solid}}$ with $i \neq j$) are active, and holds individually for each species' excess $\Delta_i(t)$ when heterochiral pairing is exclusively self-paired ($i = j$). $\square$

\subsection{Derivation of the Open-Flow Quotient Dynamics and Washout Bound}

For a continuous or cyclic open reactor receiving chiral-center mass flux $F_{\mathrm{in}}(t) = F_{L,\mathrm{in}}(t) + F_{D,\mathrm{in}}(t) \ge 0$ at feed enantiomeric excess $\ee_{\mathrm{in}}(t) = [F_{L,\mathrm{in}}(t) - F_{D,\mathrm{in}}(t)] / F_{\mathrm{in}}(t)$ and discharging mass flux $F_{\mathrm{out}}(t) \ge 0$ at outflow excess $\ee_{\mathrm{out}}(t)$, the continuity equations for $M(t)$ and $\Delta(t)$ are
\begin{equation}
\frac{\mathrm{d}M}{\mathrm{d}t} = F_{\mathrm{in}}(t) - F_{\mathrm{out}}(t) + J_M(t), \qquad \frac{\mathrm{d}\Delta}{\mathrm{d}t} = \ee_{\mathrm{in}}(t) F_{\mathrm{in}}(t) - \ee_{\mathrm{out}}(t) F_{\mathrm{out}}(t) + J_\Delta(t),
\label{eq:s_open_continuity}
\end{equation}
where $J_M(t) = \sum_\rho \Delta M_\rho r_\rho(t)$ and $J_\Delta(t) = \sum_\rho \Delta q_\rho r_\rho(t)$ are the internal reaction fluxes. Applying the quotient rule to $\ee(t) = \Delta(t) / M(t)$ yields
\begin{align}
\frac{\mathrm{d}\ee}{\mathrm{d}t} &= \frac{1}{M}\frac{\mathrm{d}\Delta}{\mathrm{d}t} - \frac{\Delta}{M^2}\frac{\mathrm{d}M}{\mathrm{d}t} = \frac{1}{M}\left[\ee_{\mathrm{in}} F_{\mathrm{in}} - \ee_{\mathrm{out}} F_{\mathrm{out}} + J_\Delta - \ee\left(F_{\mathrm{in}} - F_{\mathrm{out}} + J_M\right)\right] \nonumber\\
&= \frac{F_{\mathrm{in}}}{M}\left(\ee_{\mathrm{in}} - \ee\right) - \frac{F_{\mathrm{out}}}{M}\left(\ee_{\mathrm{out}} - \ee\right) - \frac{J_M}{M}\ee + \frac{J_\Delta}{M}.
\label{eq:s_open_quotient_exact}
\end{align}
In a well-mixed liquid phase where outflow samples the bulk active phase ($\ee_{\mathrm{out}}(t) = \ee(t)$) and inflow is racemic ($\ee_{\mathrm{in}}(t) = 0$), the outflow term vanishes:
\begin{equation}
\frac{\mathrm{d}\ee}{\mathrm{d}t} = -\frac{F_{\mathrm{in}}(t) + J_M(t)}{M(t)}\,\ee(t) + \frac{J_\Delta(t)}{M(t)}.
\label{eq:s_open_wellmixed}
\end{equation}
When Channel B is absent ($J_\Delta(t) = 0$) and the reactor operates at stationary active-phase inventory ($\mathrm{d}M/\mathrm{d}t = 0 \implies F_{\mathrm{in}} + J_M = F_{\mathrm{out}} > 0$), Eq.~\eqref{eq:s_open_wellmixed} reduces to $\mathrm{d}\ee/\mathrm{d}t = -(F_{\mathrm{out}}/M_0)\ee$, yielding exponential decay $\ee(t) = \ee_0 \exp(-F_{\mathrm{out}} t / M_0) \to 0$. In the closed-outflow semi-batch limit ($F_{\mathrm{out}} = 0$) where racemic feed compensates internal precipitation loss ($F_{\mathrm{in}} = -J_M > 0$ so $M(t) = M_0$), Eq.~\eqref{eq:s_open_wellmixed} gives $\mathrm{d}\ee/\mathrm{d}t = 0$ ($\Delta(t) = \Delta_0$ and $\ee(t) = \ee_0$ remain constant while precipitated solid accumulates indefinitely); however, in any open reactor with non-zero purge or exchange outflow ($F_{\mathrm{out}} = F_{\mathrm{ex}} > 0$, exchange rate $f_{\mathrm{ex}} = F_{\mathrm{ex}}/M_0 > 0$) required to prevent solid clogging at stationary $M_0$, $F_{\mathrm{in}} + J_M = F_{\mathrm{ex}} > 0$ drives exponential washout $\ee(t) = \ee_0 \exp(-f_{\mathrm{ex}} t) \to 0$.

\subsection{Spatially Resolved 1D and 2D Reaction--Diffusion--Precipitation PDE Verification}

\begin{figure}[!htbp]
\centering
\includegraphics[width=0.92\textwidth]{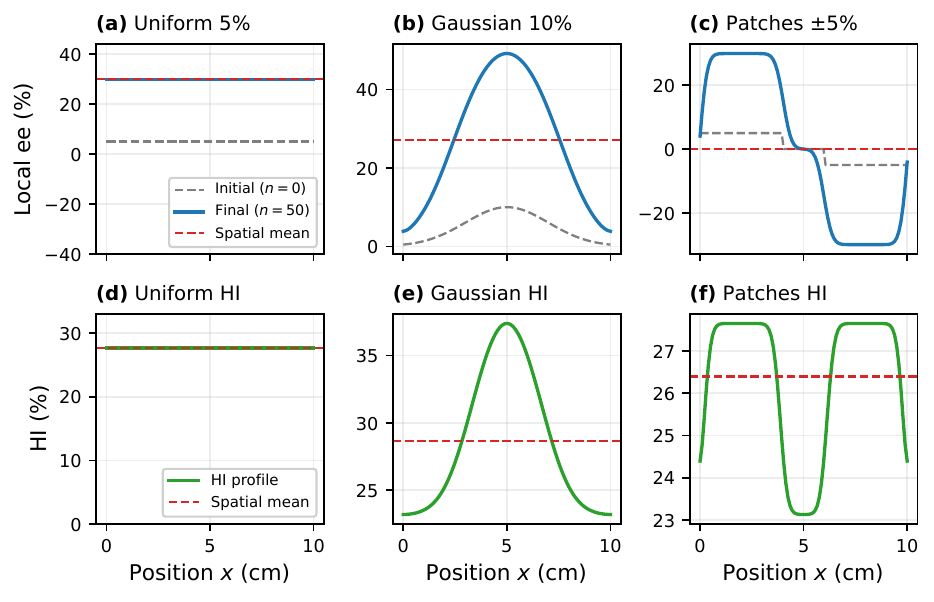}
\caption{Six-panel one-dimensional 5-species ($\{L, D, LL, DD, LD\}$) reaction--diffusion PDE comparison across spatial domain $x \in [0, 10]\text{ cm}$ ($100$ periodic grid points) over $N = 50$ wet--dry cycles. (a--c) Initial ($n=0$, gray dashed) and final ($n=50$, blue solid) local $\ee(x)$ (\%) profiles, with red dashed horizontal lines marking the final ($n=50$) spatial domain means under (a) Uniform ($\ee_0 = 5\% \to \ee_f = 29.9\%$), (b) Gaussian Hotspot ($\ee_{\mathrm{peak}}(0) = 10\% \to \ee_f \in [4.2\%, 49.4\%]$, domain mean $27.3\%$), and (c) Alternating L/D Patches ($\ee_0 = \pm 5\% \to \ee_f \in [-29.9\%, +29.9\%]$, domain mean $0.00\%$). (d--f) Corresponding 50-cycle spatial profiles of the homochiral dimer index $\mathrm{HI}(x)$ (\%) reaching $27.6\%$ (domain mean $27.6\%$), $[23.2\%, 37.9\%]$ (domain mean $28.8\%$), and $[23.1\%, 27.6\%]$ (domain mean $26.4\%$, red dashed horizontal lines).}
\alttext{Six-panel line chart of 50-cycle one-dimensional reaction-diffusion PDE simulations on a 100-point grid across 0 to 10 centimeters under Uniform, Gaussian, and Patches initial conditions showing local enantiomeric excess in panels a to c and homochiral index in panels d to f.}
\label{fig:s_rd1d}
\end{figure}

\begin{figure}[!htbp]
\centering
\includegraphics[width=0.92\textwidth]{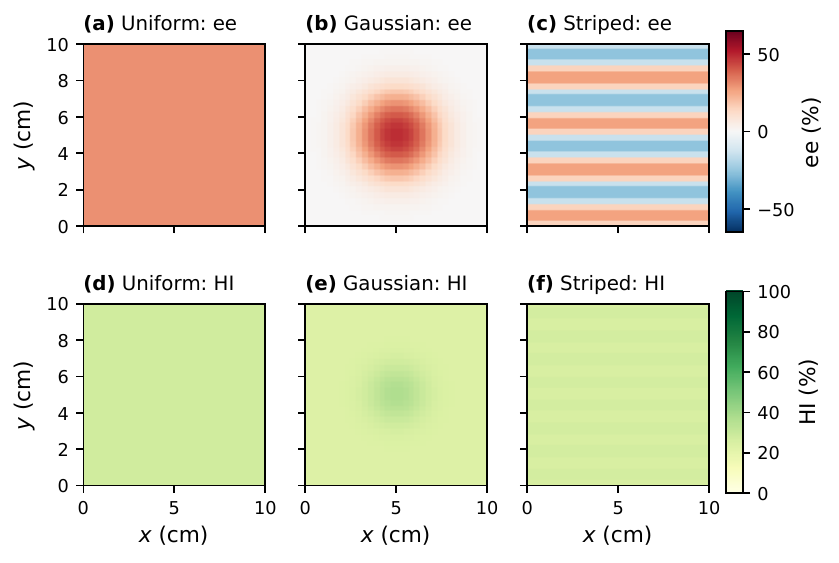}
\caption{Six-panel two-dimensional ($32 \times 32$ periodic grid, $(x,y) \in [0, 10]\times[0, 10]\text{ cm}^2$) 5-species reaction--diffusion PDE comparison after $N = 50$ wet--dry cycles across Uniform ($\ee_0 = 5\%$), Gaussian Hotspot ($\ee_{\mathrm{peak}}(0) = 10\%$), and Striped ($\ee_0 = \pm 5\%$) initial conditions. (a--c) 2D spatial fields of final local $\ee(x,y)$ (\%) ($29.9\%$ uniform in (a), $[7.0\%, 47.4\%]$ with mean $16.5\%$ in (b), and $[-29.8\%, +29.8\%]$ with domain mean $0.00\%$ in (c)). (d--f) Corresponding 2D spatial fields of final homochiral dimer index $\mathrm{HI}(x,y)$ (\%) ($27.6\%$, $[23.4\%, 36.7\%]$, and $[23.1\%, 27.5\%]$).}
\alttext{Six-panel heatmap figure of 50-cycle two-dimensional reaction-diffusion PDE simulations on a 32 by 32 periodic grid across 0 to 10 centimeters under Uniform, Gaussian, and Striped initial conditions showing local enantiomeric excess and homochiral index fields.}
\label{fig:s_rd2d}
\end{figure}

To verify that spatial concentration gradients, wet--dry diffusivity switching, and localized heterodimer precipitation fronts do not violate Theorem~S1 in the absence of Channel B, we solved the spatially resolved 5-species ($\{L, D, LL, DD, LD\}$) reaction--diffusion--precipitation partial differential equations (PDEs) over $N = 50$ wet--dry cycles in 1D ($x \in [0, 10]\text{ cm}$, $n_{\mathrm{sites}} = 100$ grid points, $\Delta x = 0.10\text{ cm}$) and 2D ($(x,y) \in [0, 10]\times[0, 10]\text{ cm}^2$, $n_{\mathrm{sites}} = 32 \times 32 = 1\,024$ grid points, $\Delta x = \Delta y = 0.3125\text{ cm}$) under periodic boundary conditions:
\begin{equation}
\begin{aligned}
\frac{\partial c_L}{\partial t} &= D(t)\nabla^2 c_L - 2 k_{\mathrm{homo}} c_L^2 - k_{\mathrm{het}} c_L c_D + 2 k_{\mathrm{hyd}} c_{LL} + k_{\mathrm{hyd}} c_{LD}, \\
\frac{\partial c_D}{\partial t} &= D(t)\nabla^2 c_D - 2 k_{\mathrm{homo}} c_D^2 - k_{\mathrm{het}} c_L c_D + 2 k_{\mathrm{hyd}} c_{DD} + k_{\mathrm{hyd}} c_{LD}, \\
\frac{\partial c_{LL}}{\partial t} &= D_{\mathrm{dim}}(t)\nabla^2 c_{LL} + k_{\mathrm{homo}} c_L^2 - k_{\mathrm{hyd}} c_{LL}, \quad
\frac{\partial c_{DD}}{\partial t} = D_{\mathrm{dim}}(t)\nabla^2 c_{DD} + k_{\mathrm{homo}} c_D^2 - k_{\mathrm{hyd}} c_{DD}, \\
\frac{\partial c_{LD}}{\partial t} &= D_{\mathrm{dim}}(t)\nabla^2 c_{LD} + k_{\mathrm{het}} c_L c_D - \left(k_{\mathrm{hyd}} + k_{\mathrm{precip}}\right) c_{LD},
\end{aligned}
\label{eq:s_pde_1d2d}
\end{equation}
where each cycle alternates between a dry phase ($t_{\mathrm{dry}} = 1.0$, integrated over $500$ substeps at total dry volume $V_{\mathrm{dry}} = 1.0$, per-grid-cell volume $V_{\mathrm{dry,site}} = V_{\mathrm{dry}}/n_{\mathrm{sites}}$; active bimolecular dimerization $k_{\mathrm{homo}} = 5 \times 10^{-5}$, $k_{\mathrm{het}} = 2 \times 10^{-3}$ [$r = 40$] scaled by $1/V_{\mathrm{dry,site}}$, heterodimer precipitation $k_{\mathrm{precip}} = 5 \times 10^{-2}$ with homochiral co-precipitation factor $\eta_{\mathrm{coprecip}} = 0.1$ on $c_{LL}, c_{DD}$, $k_{\mathrm{hyd}} = 0$ during drying, and dry-phase diffusivity $D(t) = D_{\mathrm{dim}}(t) = D_{\mathrm{dry}} = 10^{-6}\text{ cm}^2\text{ cycle}^{-1}$) and a wet phase ($t_{\mathrm{wet}} = 1.0$, integrated over $50$ substeps at total wet volume $V_{\mathrm{wet}} = 10.0$, $V_{\mathrm{wet,site}} = V_{\mathrm{wet}}/n_{\mathrm{sites}}$; dimer hydrolysis $k_{\mathrm{hyd}} = 10^{-2}$ with $k_{\mathrm{homo}} = k_{\mathrm{het}} = k_{\mathrm{precip}} = 0$ during wetting, and wet-phase diffusivity $D(t) = D_{\mathrm{dim}}(t) = D_{\mathrm{wet}} = 10^{-3}\text{ cm}^2\text{ cycle}^{-1}$), initialized with domain-total monomer inventory $M_0 = 1000$ ($M_{\mathrm{site}}(0) = M_0/n_{\mathrm{sites}}$, $c_{LL}(0) = c_{DD}(0) = c_{LD}(0) = 0$) and matched enantiomeric diffusivities ($D_L = D_D = D(t)$, $D_{LL} = D_{DD} = D_{\mathrm{dim}}(t)$) required by $\mathbb{Z}_2$ symmetry in an achiral medium. Forming the signed chiral-center density $\Delta(\bm{r},t) = (c_L - c_D) + 2(c_{LL} - c_{DD})$ cancels all dimerization, hydrolysis, and $1{:}1$ heterodimer precipitation terms pointwise, and integrating over the periodic domain $\Omega$ yields global chiral-excess conservation:
\begin{equation}
\frac{\mathrm{d}}{\mathrm{d}t}\int_\Omega \Delta(\bm{r},t)\,\mathrm{d}\bm{r} = \int_{\partial\Omega} \bm{n}\cdot\left[D(t)\nabla(c_L - c_D) + 2 D_{\mathrm{dim}}(t)\nabla(c_{LL} - c_{DD})\right]\mathrm{d}S = 0.
\label{eq:s_pde_integral_invariance}
\end{equation}
Figures~\ref{fig:s_rd1d} and \ref{fig:s_rd2d} display the 50-cycle 1D and 2D spatial profiles of local active-phase enantiomeric excess $\ee(\bm{r}) = \Delta(\bm{r})/M(\bm{r})$ (\%) and homochiral dimer index $\mathrm{HI}(\bm{r}) = (c_{LL} + c_{DD})/(c_{LL} + c_{DD} + c_{LD})$ (\%), alongside unweighted spatial grid means $\langle\ee_f\rangle = n_{\mathrm{sites}}^{-1}\sum_{\bm{r}} \ee_f(\bm{r})$, across three initial spatial configurations: (a,d) Uniform initial excess ($\ee_0(\bm{r}) = 5\%$), which amplifies uniformly to $\ee_f = 29.9\%$ and $\mathrm{HI}_f = 27.6\%$ across both 1D and 2D grids via heterodimer phase removal; (b,e) Gaussian Hotspot ($\ee_0(\bm{r}) = \ee_{\mathrm{bg}} + (\ee_{\mathrm{peak}} - \ee_{\mathrm{bg}})\exp[-|\bm{r} - \bm{r}_c|^2/(2\sigma^2)]$ with $\ee_{\mathrm{peak}} = 10\%$ at domain center $\bm{r}_c$, $\sigma = 1.5\text{ cm}$ in 1D [$15$ grid spacings] and $\sigma = 5.0\Delta x = 1.5625\text{ cm}$ in 2D, and background $\ee_{\mathrm{bg}} = 1\%$), which evolves into a diffusive peak-to-background profile with $\ee_f \in [4.2\%, 49.4\%]$ (mean $27.3\%$) and $\mathrm{HI}_f \in [23.2\%, 37.9\%]$ in 1D ($\ee_f \in [7.0\%, 47.4\%]$, mean $16.5\%$, and $\mathrm{HI}_f \in [23.4\%, 36.7\%]$ in 2D); and (c,f) Alternating L-rich/D-rich Patches in 1D ($4$ equal subdomains of width $2.5\text{ cm}$ with $\ee_0 = +5\%, -5\%, +5\%, -5\%$) and Stripes in 2D ($4$ vertical stripes of width $2.5\text{ cm}$ [$8$ grid columns each] with $\ee_0 = \pm 5\%$), which amplify into anti-symmetric local plateaus of $\pm 29.9\%$ ($\pm 29.8\%$ in 2D) separated by diffusive racemic domain walls while preserving global racemic symmetry ($\langle\ee_f\rangle = 0.00\%$, $\Delta_{\mathrm{tot}}(t) = 0$) and elevating local homochiral dimer fraction to $\mathrm{HI}_f \in [23.1\%, 27.6\%]$.

\section{Ordered and Symmetric Shapley Three-Channel Attribution Identities}
\label{sec:s2_attribution}

\subsection{Algebraic Proof of Telescoping and Shapley Decomposition Identities}

Given initial state $(M_0, \Delta_0, \ee_0 = \Delta_0/M_0)$ and final state $(M_f, \Delta_f, \ee_f = \Delta_f/M_f)$ of the active phase $\mathcal{A}$, together with a target measured subpool $\mathcal{A}_{\mathrm{sub}} \subseteq \mathcal{A}$ (such as the free monomer pool $\{L, D\}$) having final inventory $(M_{\mathrm{sub},f}, \Delta_{\mathrm{sub},f}, \ee_{\mathrm{sub},f} = \Delta_{\mathrm{sub},f}/M_{\mathrm{sub},f})$, define the intermediate reference state $\ee_R^* = \Delta_0 / M_f$ corresponding to pure symmetric mass removal from $M_0$ to $M_f$ at fixed $\Delta = \Delta_0$ (generalizing Morowitz's eutectic mass balance\cite{Morowitz1969,Klussmann2006} and Soloshonok's self-disproportionation of enantiomers\cite{Soloshonok2006}), alongside internal subpool partitioning $A_S$ (generalizing the Horeau--Langenbeck and Kagan $\mathrm{ML}_2$ dimer reservoir effect\cite{Guillaneux1994,Girard1998,Blackmond2001}). Adding and subtracting $\ee_R^*$ and $\ee_f$ yields the ordered telescoping identity:
\begin{equation}
\ee_{\mathrm{sub},f} - \ee_0 = \underbrace{\left(\frac{\Delta_0}{M_f} - \frac{\Delta_0}{M_0}\right)}_{A_R} + \underbrace{\left(\frac{\Delta_f}{M_f} - \frac{\Delta_0}{M_f}\right)}_{A_B} + \underbrace{\left(\frac{\Delta_{\mathrm{sub},f}}{M_{\mathrm{sub},f}} - \frac{\Delta_f}{M_f}\right)}_{A_S} = A_R + A_S + A_B.
\label{eq:s_ordered_proof}
\end{equation}
For phase-level attribution ($\ee_{\mathrm{sub},f} = \ee_f$, so $A_S = 0$), $\ee(M, \Delta) = \Delta / M$ is a two-factor function of $M$ and $\Delta$. There are two permutations for transitioning from $(M_0, \Delta_0)$ to $(M_f, \Delta_f)$:
\begin{itemize}
\item Path 1 ($M_0 \to M_f$ first, then $\Delta_0 \to \Delta_f$): marginal contributions are $A_R^{(1)} = \frac{\Delta_0}{M_f} - \frac{\Delta_0}{M_0}$ and $A_B^{(1)} = \frac{\Delta_f - \Delta_0}{M_f}$.
\item Path 2 ($\Delta_0 \to \Delta_f$ first, then $M_0 \to M_f$): marginal contributions are $A_B^{(2)} = \frac{\Delta_f - \Delta_0}{M_0}$ and $A_R^{(2)} = \frac{\Delta_f}{M_f} - \frac{\Delta_f}{M_0}$.
\end{itemize}
Averaging equally over both permutations yields the unique two-player Shapley value decomposition:
\begin{align}
A_R^{\mathrm{Sh}} &= \frac{1}{2}\left(A_R^{(1)} + A_R^{(2)}\right) = \frac{1}{2}\left(\Delta_0 + \Delta_f\right)\left(\frac{1}{M_f} - \frac{1}{M_0}\right), \nonumber\\
A_B^{\mathrm{Sh}} &= \frac{1}{2}\left(A_B^{(1)} + A_B^{(2)}\right) = \frac{1}{2}\left(\Delta_f - \Delta_0\right)\left(\frac{1}{M_0} + \frac{1}{M_f}\right),
\label{eq:s_shapley_proof}
\end{align}
with synergistic interaction difference $\delta_{\mathrm{int}} = A_B^{(1)} - A_B^{\mathrm{Sh}} = \frac{1}{2}(\Delta_f - \Delta_0)\left(\frac{1}{M_f} - \frac{1}{M_0}\right)$.

\begin{figure}[!htbp]
\centering
\includegraphics[width=0.78\textwidth]{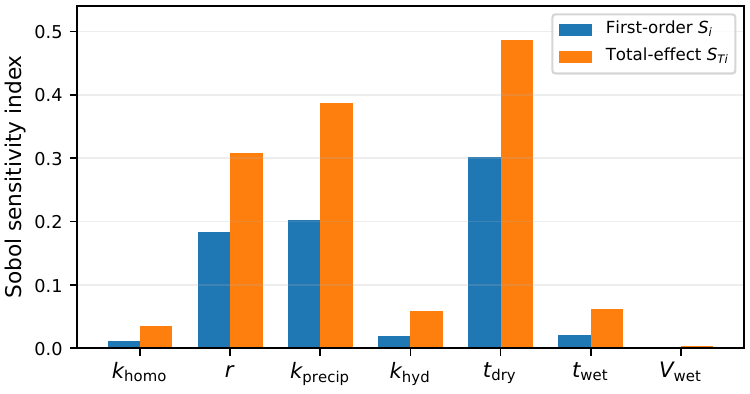}
\caption{First-order ($S_i$, blue) and total-effect ($S_{T_i}$, orange) Saltelli--Sobol variance sensitivity indices ($N_{\mathrm{base}} = 4\,096$, $65\,536$ evaluations) of 100-cycle terminal enantiomeric excess across the 7 wet--dry model parameters $(k_{\mathrm{homo}}, r, k_{\mathrm{precip}}, k_{\mathrm{hyd}}, t_{\mathrm{dry}}, t_{\mathrm{wet}}, V_{\mathrm{wet}})$, led by dry-phase duration $t_{\mathrm{dry}}$ ($S_i = 0.302$, $S_{T_i} = 0.486$), precipitation rate $k_{\mathrm{precip}}$ ($S_i = 0.202$, $S_{T_i} = 0.386$), and heterochiral selectivity $r$ ($S_i = 0.183$, $S_{T_i} = 0.308$).}
\alttext{Grouped bar chart comparing first-order and total-effect Sobol sensitivity indices across seven wet-dry kinetic parameters: homochiral ligation rate, heterochiral selectivity ratio, precipitation rate, hydrolysis rate, dry-phase duration, wet-phase duration, and wet-phase volume for one-hundred-cycle terminal enantiomeric excess.}
\label{fig:s_sobol}
\end{figure}

Table~\ref{tab:s_regime_prevalence} and Fig.~\ref{fig:s_sobol} report the global regime prevalence across the $512\,000$-trajectory parameter survey ($320 \times 320 = 102\,400$ log-spaced grid points per selectivity ratio $r = k_{\mathrm{het}}/k_{\mathrm{homo}} \in \{1.0, 1.5, 2.0, 3.0, 4.0\}$ over $\log_{10}k_{\mathrm{precip}} \in [-4, 0]$, i.e., $k_{\mathrm{precip}} \in [10^{-4}, 10^0]$ with reference $k_{\mathrm{precip},0} = 0.1$, and $\log_{10}\zeta \in [0, 1.505]$, i.e., $\zeta = k_{\mathrm{auto}}/k_{\mathrm{homo}} \in [1, 32]$, matching Main Fig.~1(a--c)) and the variance-based Saltelli--Sobol sensitivity indices ($N_{\mathrm{base}} = 4\,096$, $65\,536$ evaluations across 7 kinetic and cycling parameters). Specifically, in the $512\,000$-trajectory taxonomy survey (Main Fig.~1 and Table~\ref{tab:s_regime_prevalence}), each trajectory integrates the 5-state wet--dry ODE system for extensive copy numbers $(n_L, n_D, n_{LL}, n_{DD}, n_{LD})$ over $N = 100$ cycles from $M_0 = 1000$ and $\ee_0 = 5\%$ ($n_L(0)=525$, $n_D(0)=475$, $n_{LL}(0)=n_{DD}(0)=n_{LD}(0)=0$). During the dry phase ($t_{\mathrm{dry}} = 1.0$, $V_{\mathrm{dry}} = 1.0$, $k_{\mathrm{homo}} = 5\times 10^{-5}$, $\eta_{\mathrm{coprecip}} = 0.1$), the governing equations are
\[
\begin{aligned}
\frac{\mathrm{d}n_L}{\mathrm{d}t} &= -\frac{2 k_{\mathrm{homo}}}{V_{\mathrm{dry}}}n_L^2 - \frac{r k_{\mathrm{homo}}}{V_{\mathrm{dry}}}n_L n_D, &
\frac{\mathrm{d}n_D}{\mathrm{d}t} &= -\frac{2 k_{\mathrm{homo}}}{V_{\mathrm{dry}}}n_D^2 - \frac{r k_{\mathrm{homo}}}{V_{\mathrm{dry}}}n_L n_D, \\
\frac{\mathrm{d}n_{LL}}{\mathrm{d}t} &= \frac{k_{\mathrm{homo}}}{V_{\mathrm{dry}}}n_L^2 - \eta_{\mathrm{coprecip}}k_{\mathrm{precip}}n_{LL}, &
\frac{\mathrm{d}n_{DD}}{\mathrm{d}t} &= \frac{k_{\mathrm{homo}}}{V_{\mathrm{dry}}}n_D^2 - \eta_{\mathrm{coprecip}}k_{\mathrm{precip}}n_{DD}, \\
\frac{\mathrm{d}n_{LD}}{\mathrm{d}t} &= \frac{r k_{\mathrm{homo}}}{V_{\mathrm{dry}}}n_L n_D - k_{\mathrm{precip}}n_{LD}, & &
\end{aligned}
\]
followed by wet-phase dimer hydrolysis ($t_{\mathrm{wet}} = 0.5$, $V_{\mathrm{wet}} = 10.0$, $k_{\mathrm{hyd}} = 0.01$) returning monomers via $\mathrm{d}n_{ij}/\mathrm{d}t = -k_{\mathrm{hyd}}n_{ij}$ ($ij \in \{LL, DD, LD\}$). With Channel B active ($\zeta > 1$), autocatalytic monomer amplification with saturation and precipitation damping drives $A_B = (\Delta_f - \Delta_0)/M_f$ alongside $A_R = \ee_0(M_0/M_f - 1)$ and $A_S = (n_L + n_D)_f\,\ee_{\mathrm{mon},f}/M_f$. Grid points are classified into four mutually exclusive regimes: Washout/Weak if $\Delta\ee_{\mathrm{tot}} = A_R + A_S + A_B < 10^{-3}$; otherwise Channel B if $f_B = A_B/\Delta\ee_{\mathrm{tot}} \ge 0.50$; otherwise Channel R if $A_R \ge A_S$ (and $k_{\mathrm{precip}} \ge 3\times 10^{-3}$); and Channel S otherwise. In the symmetry diagnostic of Main Fig.~2 ($\ee_0 \in [-20\%, +20\%]$), the mirror-covariant Channel B probe tracks an environmental state variable $z(t) \in [-1, 1]$ driven by $\mathrm{d}z/\mathrm{d}t = \kappa_{\mathrm{env}}(\ee(t) - z(t))$ ($\kappa_{\mathrm{env}} = 6.89295$ at $k_{\mathrm{precip}} = 0.05$, $z(0)=0$) that modulates homochiral dimerization as $k_{\mathrm{homo},L/D} = k_{\mathrm{homo}}\max(0, 1 \pm [\lambda z(t) + h_{\mathrm{ext}}])$ ($\lambda \in \{0, 1\}$, $h_{\mathrm{ext}} \in \{0, \pm 0.05\}$). For the Saltelli--Sobol variance decomposition (Fig.~\ref{fig:s_sobol}), the 7 parameters are sampled log-uniformly over $k_{\mathrm{homo}} \in [10^{-5}, 5\times 10^{-4}]$, $r \in [5, 100]$, $k_{\mathrm{precip}} \in [10^{-3}, 0.5]$, $k_{\mathrm{hyd}} \in [10^{-3}, 0.1]$, $t_{\mathrm{dry}} \in [0.2, 5.0]$, $t_{\mathrm{wet}} \in [0.1, 2.0]$, and $V_{\mathrm{wet}} \in [2.0, 50.0]$.

\begin{table}[!htbp]
\caption{Regime prevalence across $512\,000$ ODE integrations ($320 \times 320 = 102\,400$ grid points per selectivity ratio $r = k_{\mathrm{het}}/k_{\mathrm{homo}} \in \{1.0, 1.5, 2.0, 3.0, 4.0\}$) from $\ee_0 = 5\%$ over $N = 100$ cycles.}
\alttext{Data table listing the percentage of parameter grid points dominated by Channel R, Channel S, Channel B, and washout for heterochiral selectivity ratios r equal to 1.0, 1.5, 2.0, 3.0, and 4.0 across 512,000 total ODE integrations.}
\label{tab:s_regime_prevalence}
\begin{ruledtabular}
\begin{tabular}{cccccc}
Selectivity $r$ & Grid Points & Channel R (\%) & Channel S (\%) & Channel B (\%) & Washout / Weak (\%) \\
\colrule
$1.0$ & $102\,400$ & $9.35$ & $21.60$ & $55.72$ & $13.33$ \\
$1.5$ & $102\,400$ & $10.62$ & $21.53$ & $53.14$ & $14.71$ \\
$2.0$ & $102\,400$ & $11.71$ & $21.50$ & $50.89$ & $15.90$ \\
$3.0$ & $102\,400$ & $13.78$ & $21.48$ & $47.08$ & $17.67$ \\
$4.0$ & $102\,400$ & $15.72$ & $21.45$ & $43.88$ & $18.96$ \\
\end{tabular}
\end{ruledtabular}
\end{table}

\subsection{Continuous Path-Covariant Aumann--Shapley Attribution and Chordal Antiderivative}

When a time-resolved state trajectory $\gamma : t \in [0, T] \mapsto \bm{x}(t)$ with additive channel decomposition $\dot{\bm{x}}(t) = \sum_{\alpha \in \mathcal{C}} \bm{f}_\alpha(\bm{x}(t), t)$ is available ($\mathcal{C} = \{R, S, B\}$ for closed networks, or $\mathcal{C} = \{R, S, B, \mathrm{flow}\}$ for open reactors), two-endpoint discrete Shapley replacement [Eq.~\eqref{eq:s_shapley_proof}] averages only over the corners of the hypercube $[0,1]^{|\mathcal{C}|}$ and becomes insensitive to within-interval reaction ordering near racemic starts ($\ee_0 \to 0$). Integrating the directional derivative of any smooth scalar observable $F(\bm{x})$ along $\gamma$ defines the continuous path-covariant Aumann--Shapley attribution:\cite{BelahcenMussard2026}
\[
\begin{aligned}
\Phi_\alpha[\gamma] &= \int_0^T \nabla_{\bm{x}} F\bigl(\bm{x}(t)\bigr) \cdot \bm{f}_\alpha\bigl(\bm{x}(t), t\bigr)\,\mathrm{d}t, \qquad
\sum_{\alpha \in \mathcal{C}} \Phi_\alpha[\gamma] = F\bigl(\bm{x}(T)\bigr) - F\bigl(\bm{x}(0)\bigr).
\end{aligned}
\]
For $F(M, \Delta) = \Delta / M$ with $\nabla_{(M,\Delta)} F = (-\Delta/M^2,\, 1/M)^\top$ sampled at discrete checkpoints $\{(M_k, \Delta_k)\}_{k=0}^m$ ($M_k > 0$), parameterizing each chord $s \in [0, 1]$ by $M(s) = M_k + s\,\delta M_k$ and $\Delta(s) = \Delta_k + s\,\delta\Delta_k$ ($\delta M_k = M_{k+1} - M_k$, $\delta\Delta_k = \Delta_{k+1} - \Delta_k$) yields an exact closed-form rational chordal antiderivative without numerical quadrature error:
\begin{equation}
\begin{aligned}
\Phi_R^{(k)} &= \int_0^1 -\frac{\Delta(s)}{M(s)^2}\,\delta M_k\,\mathrm{d}s = \begin{cases} 0, & \delta M_k = 0, \\ \dfrac{\Delta_k}{M_k} - \dfrac{\Delta_{k+1}}{M_{k+1}} + \Phi_B^{(k)}, & \delta M_k \neq 0, \end{cases} \\
\Phi_B^{(k)} &= \int_0^1 \frac{\delta\Delta_k}{M(s)}\,\mathrm{d}s = \begin{cases} \dfrac{\delta\Delta_k}{M_k}, & \delta M_k = 0, \\ \dfrac{\delta\Delta_k}{\delta M_k}\ln\dfrac{M_{k+1}}{M_k}, & \delta M_k \neq 0, \end{cases}
\end{aligned}
\label{eq:s_aumann_chord}
\end{equation}
with smooth Taylor limit $\Phi_B^{(k)} = \frac{\delta\Delta_k}{\bar{M}_k}\left[1 + \frac{1}{12}(\delta M_k/\bar{M}_k)^2 + \frac{1}{80}(\delta M_k/\bar{M}_k)^4\right]$ ($\bar{M}_k = \tfrac{1}{2}(M_k + M_{k+1})$) whenever $|\delta M_k|/\bar{M}_k < 10^{-6}$. Summing over $k = 0, \dots, m-1$ satisfies exact completeness $\sum_k (\Phi_R^{(k)} + \Phi_B^{(k)}) = \ee_f - \ee_0$ for every chord count $m \ge 1$ (maximum numerical residual $2.69\times 10^{-14}$ across $m \in \{1, 2, 4, 8, 16, 64\}$) and exact odd $\mathbb{Z}_2$ parity $\Phi_\alpha[S\gamma] = -\Phi_\alpha[\gamma]$ (residual $0.0$; Table~\ref{tab:s_aumann_shapley}). On crystallization-induced dynamic resolution\cite{Deck2024,VanDongen2025} from $\ee_0 = 10^{-6} \to 0^+$ to $\ee_f = 1.0000$, two-endpoint Shapley assigns a fixed $0.2500$ share to Channel R regardless of reaction sequence, whereas $\Phi_R[\gamma]$ yields $8.18\times 10^{-5} < 10^{-4}$ ($\Phi_B = 0.9999$) when achiral dissolution precedes deracemization versus $\Phi_R = \Phi_B = 0.5000$ when deracemization precedes dissolution. On open-flow Regime III ($\alpha_{\mathrm{evap}} = 6.0$, $f_{\mathrm{ex}} = 0.045\,\text{cycle}^{-1}$, $N = 100$ cycles), the four-channel attribution $(\Phi_R, \Phi_S, \Phi_B, \Phi_{\mathrm{flow}})$ cleanly distinguishes active chiral-pool $\ee_{\mathrm{chiral}} = \Delta_{\mathrm{chiral}}/M_{\mathrm{chiral}}$ ($0.2000 \to 0.9287$, $\Delta F = +0.7287$: $\Phi_R = 0.0000$, $\Phi_S = +0.5242$, $\Phi_B = +4.8958$, $\Phi_{\mathrm{flow}} = -4.6913$) from total-soluble $\ee_{\mathrm{sol}} = \Delta_{\mathrm{chiral}}/M_{\mathrm{sol}}$ ($0.2000 \to 0.8651$, $\Delta F = +0.6651$: $\Phi_R = +0.2062$, $\Phi_S = 0.0000$, $\Phi_B = +4.4298$, $\Phi_{\mathrm{flow}} = -3.9709$; Table~\ref{tab:s_aumann_shapley}).
\FloatBarrier

\section{2556-State CME Parity Commutator, Gillespie SSA Verification, and Reaction-Resolved Thermodynamics}
\label{sec:s3_cme_thermo}

\subsection{Combinatorial Enumeration of the Full 5-Species and Reduced 2556-State CME Lattices and Parity Commutator Proof}

\begin{figure}[!htbp]
\centering
\includegraphics[width=0.80\textwidth]{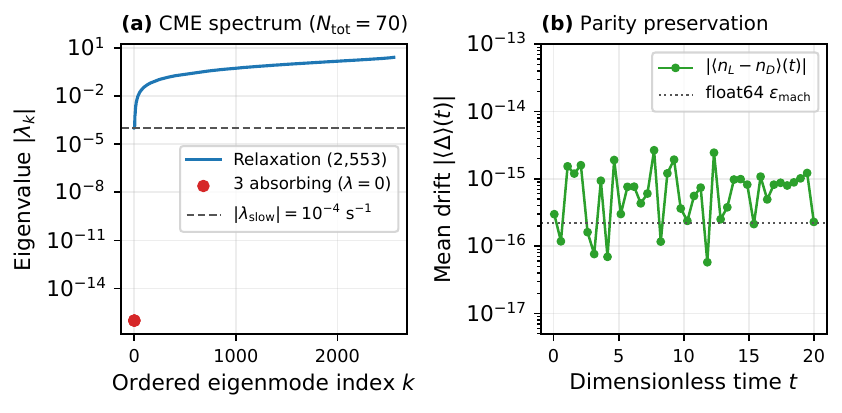}
\caption{$2556 \times 2556$ Chemical Master Equation (CME) parity decomposition at $N_{\mathrm{total}} = 70$ ($\|W - S W S\|_{\max} = 4.44 \times 10^{-16}$). (a) Eigenvalue magnitude $|\lambda_k|$ ($\text{s}^{-1}$) vs.\ ordered eigenmode index $k$, showing $2\,553$ non-absorbing relaxation modes (slowest relaxation rate $|\lambda_{\mathrm{slow}}| = 1.0 \times 10^{-4}\text{ s}^{-1}$) and $3$ absorbing ($\lambda = 0$) modes. (b) Ensemble mean chiral drift $|\langle\Delta\rangle(t)| = |\langle n_L - n_D\rangle(t)|$ remaining below machine precision ($\le 2.66 \times 10^{-15}$, dotted $\texttt{float64}\ \varepsilon_{\mathrm{mach}} = 2.22 \times 10^{-16}$) across dimensionless time $t \in [0.05, 20]$.}
\alttext{Two-panel plot showing the ordered eigenvalue magnitudes of the 2556-state Chemical Master Equation generator with 2553 relaxation modes and 3 absorbing modes on the left and machine-precision zero ensemble mean chiral drift over time on the right.}
\label{fig:s_cme_spectrum}
\end{figure}

For a closed compartment with total chiral-center capacity $N_{\mathrm{total}} = n_L + n_D + 2(n_{LL} + n_{DD} + n_{LD})$ on the 5-species state space $\bm{n} = (n_L, n_D, n_{LL}, n_{DD}, n_{LD}) \in \mathbb{Z}_{\ge 0}^5$,\cite{Kim2020JCP} let $k = n_{LL} + n_{DD} + n_{LD} \in \{0, 1, \dots, \lfloor N_{\mathrm{total}}/2 \rfloor\}$ denote the total number of dimer molecules. For a fixed $k$:
\begin{enumerate}
\item The non-negative integer triple $(n_{LL}, n_{DD}, n_{LD})$ summing to $k$ has $\binom{k + 3 - 1}{3 - 1} = \binom{k+2}{2} = \frac{(k+1)(k+2)}{2}$ distinct compositions.
\item The remaining $N_{\mathrm{total}} - 2k$ chiral centers are distributed between monomers $(n_L, n_D)$ summing to $N_{\mathrm{total}} - 2k$, which admits $N_{\mathrm{total}} - 2k + 1$ distinct compositions ($n_L \in \{0, 1, \dots, N_{\mathrm{total}} - 2k\}$).
\end{enumerate}
Summing over $k \in \{0, \dots, \lfloor N_{\mathrm{total}}/2 \rfloor\}$ yields the 5-species state-space cardinality $|\mathcal{S}_5(N_{\mathrm{total}})|$ (valid for all even and odd $N_{\mathrm{total}} \in \mathbb{Z}_{\ge 0}$), while coarse-graining dimer states into the total sink pair count $k = (N_{\mathrm{total}} - n_L - n_D)/2$ projects the active monomer dynamics onto the 2-DOF monomer-plus-sink simplex $\mathcal{S}_{\mathrm{red}}(N_{\mathrm{total}}) = \{(n_L, n_D) \in \mathbb{Z}_{\ge 0}^2 : n_L + n_D \le N_{\mathrm{total}}\}$:
\begin{align}
|\mathcal{S}_5(N_{\mathrm{total}})| &= \sum_{k=0}^{\lfloor N_{\mathrm{total}}/2 \rfloor} \frac{(k+1)(k+2)}{2}\left(N_{\mathrm{total}} - 2k + 1\right) \nonumber\\
&= \binom{\lfloor N_{\mathrm{total}}/2\rfloor + 4}{4} + \binom{\lceil N_{\mathrm{total}}/2\rceil + 3}{4}, \nonumber\\
|\mathcal{S}_{\mathrm{red}}(N_{\mathrm{total}})| &= \binom{N_{\mathrm{total}} + 2}{2} = \frac{(N_{\mathrm{total}} + 1)(N_{\mathrm{total}} + 2)}{2}.
\label{eq:s_cme_cardinality_proof}
\end{align}
Evaluating Eq.~\eqref{eq:s_cme_cardinality_proof} at $N_{\mathrm{total}} = 70$ ($\lfloor N_{\mathrm{total}}/2 \rfloor = \lceil N_{\mathrm{total}}/2 \rceil = 35$) gives $|\mathcal{S}_5(70)| = \binom{39}{4} + \binom{38}{4} = 82\,251 + 73\,815 = 156\,066$ full 5-species microstates and $|\mathcal{S}_{\mathrm{red}}(70)| = \binom{72}{2} = 2556$ reduced monomer-simplex states. In the irreversible pair-removal limit ($k_{\mathrm{hyd}} = 0$, with $2L \to LL$ at propensity $k_{\mathrm{homo}}n_L(n_L-1)/2$, $2D \to DD$ at $k_{\mathrm{homo}}n_D(n_D-1)/2$, and $L+D \to LD$ at $k_{\mathrm{het}}n_L n_D$, $k_{\mathrm{homo}} = 10^{-4}\text{ s}^{-1}$, $k_{\mathrm{het}} = 2\times 10^{-3}\text{ s}^{-1}$), forward monomer pair removal depends only on $(n_L, n_D)$, so $W_{\mathrm{red}}$ is the autonomous monomer-projected Markovian generator ($2553$ non-absorbing relaxation modes with slowest rate $-2k_{\mathrm{homo}} = -1.0\times 10^{-4}\text{ s}^{-1}$ and $3$ absorbing monomer states $(0,0), (1,0), (0,1)$). On the full 5-species space $\mathcal{S}_5(70)$ at $k_{\mathrm{hyd}} = 0$ and $k_{\mathrm{precip}} = 0$, any microstate with $n_L + n_D \in \{0, 1\}$ is absorbing across the $36$ dimer partitions ($n_{LL}+n_{DD}+n_{LD}=35$, giving $\binom{37}{2}=666$ absorbing microstates for $n_L=n_D=0$ and $2\binom{36}{2}=1260$ absorbing microstates for $n_L+n_D=1$, or $1926$ absorbing microstates in total that project onto the $3$ terminal monomer states $(0,0),(1,0),(0,1)$); when reversible hydrolysis is active ($k_{\mathrm{hyd}} = 10^{-4}\text{ s}^{-1} > 0$, as included in the full $156\,066 \times 156\,066$ generator $W_5$ commutator check), monomer regeneration depends on $(n_{LL}, n_{DD}, n_{LD})$, so Markovian closure operates on $W_5$. Because each pair-removal event changes $n_L + n_D$ by $-2$, $\mathcal{S}_{\mathrm{red}}(70)$ decomposes into two uncoupled parity sublattices: the even-sum sector ($n_L + n_D \equiv 0 \pmod 2$, containing $36^2 = 1\,296$ states that absorb into $(0,0)$) and the odd-sum sector ($n_L + n_D \equiv 1 \pmod 2$, containing $36\times 35 = 1\,260$ states that absorb into $(1,0)$ and $(0,1)$). Thus all $3$ absorbing modes appear in the full spectrum of $W_{\mathrm{red}}$ [Fig.~\ref{fig:s_cme_spectrum}(a)], whereas trajectories initialized at even $N_0 = 70$ evolve exclusively within the $1\,296$-state even sector and absorb at $(0,0)$.

Let $S$ be the $\mathbb{Z}_2$ parity permutation operator defined by $S(n_L, n_D, n_{LL}, n_{DD}, n_{LD}) = (n_D, n_L, n_{DD}, n_{LL}, n_{LD})$ on $\mathcal{S}_5$ and $S(n_L, n_D) = (n_D, n_L)$ on $\mathcal{S}_{\mathrm{red}}$, satisfying $S^2 = I$ and $S^\top = S$. In the absence of chiral bias ($\lambda = 0$, $h_{\mathrm{ext}} = 0$), every transition propensity satisfies $a_{S(\rho)}(S\bm{n}) = a_\rho(\bm{n})$, which implies that the CME transition rate matrix $W$ commutes with $S$:
\begin{equation}
[W, S] = W S - S W = 0 \quad \Longleftrightarrow \quad S W S = W.
\label{eq:s_cme_commutator}
\end{equation}
Numerical sparse assembly at $N_{\mathrm{total}} = 70$ confirms $\|W - S W S\|_{\max} = 4.44 \times 10^{-16}$ on \emph{both} the $2556 \times 2556$ reduced monomer-simplex generator $W_{\mathrm{red}}$ and the full $156\,066 \times 156\,066$ 5-species generator $W_5$ (Fig.~\ref{fig:s_cme_spectrum}). Setting $\ee = 0$ by convention on the empty monomer boundary $n_L + n_D = 0$, both $\ee(\bm{n})$ and $\Delta(\bm{n})$ are odd functions under $S$ ($\Delta(S\bm{n}) = -\Delta(\bm{n})$, $\ee(S\bm{n}) = -\ee(\bm{n})$); any parity-even initial distribution $P(\bm{n}, 0) = P(S\bm{n}, 0)$ evolves into $P(\bm{n}, t) = (\mathrm{e}^{Wt} P(0))(\bm{n}) = (S \mathrm{e}^{Wt} S P(0))(\bm{n}) = P(S\bm{n}, t)$, ensuring $|\langle\Delta\rangle(t)| \le 2.66 \times 10^{-15}$ and $\mathbb{E}[\ee(t)] = 0$ at all $t \ge 0$.

\subsection{Finite-Count Gillespie SSA Wet--Dry Verification and Endpoint Dispersion}

\begin{figure}[!htbp]
\centering
\includegraphics[width=0.92\textwidth]{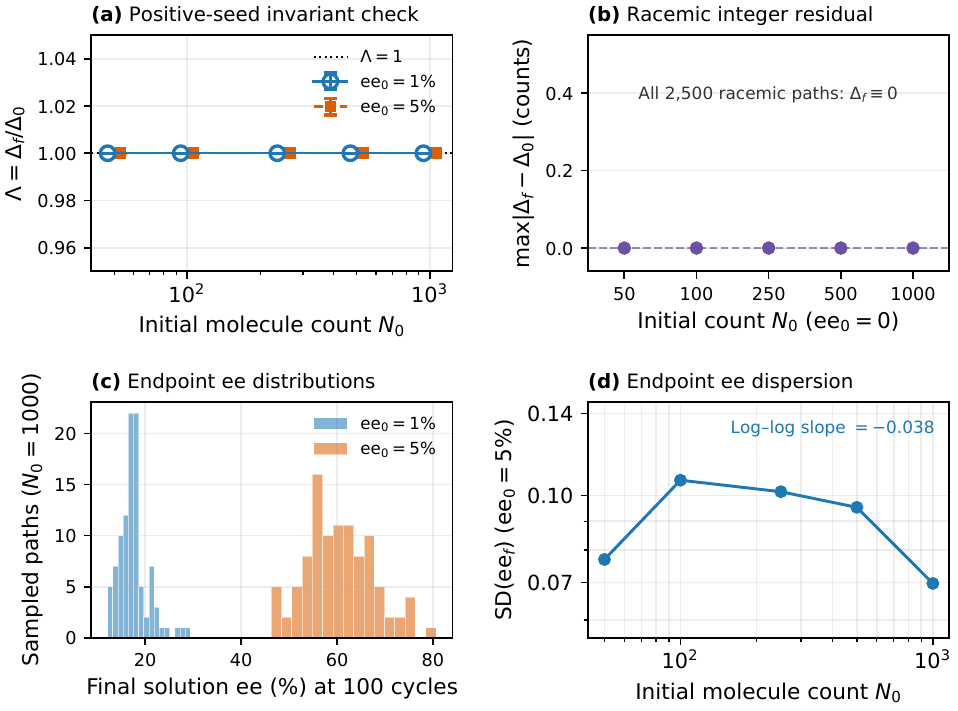}
\caption{Four-panel Direct Gillespie SSA verification across $7\,500$ trajectories over $N = 100$ wet--dry cycles. (a) Positive-seed invariant check confirming $\Lambda = \Delta_f/\Delta_0 = 1.0000$ across all $N_0 \in \{50, 100, 250, 500, 1000\}$ for $\ee_0 = 1\%$ (open blue circles) and $5\%$ (filled orange squares). (b) Racemic integer residual confirming $\max|\Delta_f - \Delta_0| = 0$ across all $2\,500$ racemic ($\ee_0 = 0$) trajectories. (c) Endpoint $\ee_f$ distributions at $N_0 = 1000$ for $\ee_0 = 1\%$ and $\ee_0 = 5\%$. (d) Log--log scaling of endpoint dispersion $\mathrm{SD}(\ee_f)$ vs.\ $N_0$ at $\ee_0 = 5\%$ (overall slope $-0.038$ across $[50, 1000]$; the step from $7.69\%$ at $N_0 = 50$ to $10.65\%$ at $N_0 = 100$ reflects integer parity rounding $\ee_{0,\mathrm{real}} = 4.00\% \to 6.00\%$, followed by monotone decrease across $N_0 \in [100, 1000]$ with slope $-0.185$).}
\alttext{Four-panel figure showing positive-seed conservation ratio Lambda equal to 1.000 in panel a, zero racemic integer residuals across 2,500 paths in panel b, endpoint enantiomeric excess histograms at initial count 1000 in panel c, and endpoint dispersion versus initial count with log-log slope minus 0.038 in panel d.}
\label{fig:s_ssa_wetdry}
\end{figure}

Table~\ref{tab:s_ssa_scaling} and Fig.~\ref{fig:s_ssa_wetdry} summarize the $7\,500$ Direct Gillespie Stochastic Simulation Algorithm (SSA) trajectories over $N = 100$ wet--dry cycles across 15 cells of initial molecule counts $N_0 \in \{50, 100, 250, 500, 1000\}$ and requested initial excesses $\ee_0 \in \{0.0\%, 1.0\%, 5.0\%\}$ ($500$ trajectories per cell, initialized with monomers $n_L(0)+n_D(0)=N_0$ and $n_{LL}(0)=n_{DD}(0)=n_{LD}(0)=0$). Each cycle alternates between a dry phase ($t_{\mathrm{dry}} = 1.0$, volume $\Omega_{\mathrm{dry}} = V_{\mathrm{dry}} = 1.0$) with discrete propensities $a_{LL} = (k_{\mathrm{homo}}/V_{\mathrm{dry}})n_L(n_L-1)$, $a_{DD} = (k_{\mathrm{homo}}/V_{\mathrm{dry}})n_D(n_D-1)$, $a_{LD} = (k_{\mathrm{het}}/V_{\mathrm{dry}})n_L n_D$, and $a_{\mathrm{precip},LD} = k_{\mathrm{precip}}n_{LD}$ ($k_{\mathrm{homo}} = 5\times 10^{-5}$, $k_{\mathrm{het}} = 2\times 10^{-3}$ [$r = 40$], $k_{\mathrm{precip}} = 5\times 10^{-2}$, and zero homochiral co-precipitation $\eta_{\mathrm{coprecip}} = 0$ so $\Delta(t) = n_L - n_D + 2(n_{LL} - n_{DD})$ is a pathwise invariant), and a wet phase ($t_{\mathrm{wet}} = 0.5$, volume $\Omega_{\mathrm{wet}} = V_{\mathrm{wet}} = 10.0$) with dimer hydrolysis propensities $a_{\mathrm{hyd},ij} = k_{\mathrm{hyd}}n_{ij}$ ($k_{\mathrm{hyd}} = 10^{-2}$, $ij \in \{LL, DD, LD\}$). Because molecular copy numbers $n_L(0), n_D(0)$ must be non-negative integers satisfying $n_L(0) + n_D(0) = N_0$, their difference $\Delta_0 = n_L(0) - n_D(0)$ must share the parity of $N_0$ ($\Delta_0 \equiv N_0 \pmod 2$). For $\ee_0 = 0.0\%$, $\Delta_0 = 0$; for $\ee_0 > 0$, the target excess $\Delta^* = N_0 \ee_0$ is rounded (half-up) to the nearest integer sharing the parity of $N_0$ with a minimum positive seed $\Delta_0 = 2$, setting $n_L(0) = (N_0 + \Delta_0)/2$, $n_D(0) = (N_0 - \Delta_0)/2$, and realized initial excess $\ee_{0,\mathrm{real}} = \Delta_0 / N_0$. Along every single trajectory under Channels R and S, $\max_t |\Delta(t) - \Delta_0| = 0$ (zero violations), confirming $\Lambda = \Delta_f/\Delta_0 = 1.0000$ on all $5\,000$ positive-seed trajectories and $\Delta_f = 0$ on all $2\,500$ racemic trajectories. Because $\Delta_f = \Delta_0$ is fixed while the surviving active molecule count $M_f$ fluctuates across trajectories, the endpoint standard deviation $\mathrm{SD}(\ee_f) = |\Delta_0|\,\mathrm{SD}(1/M_f)$ reflects finite-count mass-depletion dispersion, decreasing from $6.05\%$ ($N_0 = 50$) to $3.30\%$ ($N_0 = 1000$, log--log slope $-0.171$) for $\ee_0 = 1.0\%$ (with a discrete parity-rounding step between $N_0 = 250$, $\Delta_0 = 2$, $\ee_{0,\mathrm{real}} = 0.80\%$, $\langle\ee_f\rangle = 9.32\%$, $\mathrm{SD}(\ee_f) = 2.48\%$ and $N_0 = 500$, $\Delta_0 = 6$, $\ee_{0,\mathrm{real}} = 1.20\%$, $\langle\ee_f\rangle = 17.94\%$, $\mathrm{SD}(\ee_f) = 4.02\%$). For requested $\ee_0 = 5.0\%$, parity rounding gives $\Delta^* = 2.5 \to \Delta_0 = 2$ ($\ee_{0,\mathrm{real}} = 4.00\%$, $\mathrm{SD}(\ee_f) = 7.69\%$) at $N_0 = 50$ and $\Delta^* = 5.0 \to \Delta_0 = 6$ ($\ee_{0,\mathrm{real}} = 6.00\%$, $\mathrm{SD}(\ee_f) = 10.65\%$) at $N_0 = 100$, after which $\mathrm{SD}(\ee_f)$ decreases monotonically across $N_0 \in \{100, 250, 500, 1000\}$ ($10.65\% \to 10.16\% \to 9.53\% \to 6.97\%$; log--log slope $-0.038$ across $N_0 \in [50, 1000]$ and $-0.185$ across $N_0 \in [100, 1000]$).

\begin{table}[!htbp]
\caption{Finite-count Direct Gillespie SSA ensemble statistics across $7\,500$ trajectories ($15$ cells of $N_0 \in \{50, 100, 250, 500, 1000\} \times \ee_0 \in \{0.0\%, 1.0\%, 5.0\%\}$, $500$ trajectories per cell, $N = 100$ wet--dry cycles) under Channels R + S ($\max|\Delta_f - \Delta_0| = 0$).}
\alttext{Table of Direct Gillespie stochastic simulation statistics across five initial molecule counts from 50 to 1000 and three initial enantiomeric excesses showing exact conservation of integer chiral excess, mean surviving mass, mean final ee, and endpoint standard deviation.}
\label{tab:s_ssa_scaling}
\footnotesize
\begin{ruledtabular}
\begin{tabular}{cccccccc}
Initial $N_0$ & $\ee_0$ (req.\ / real.\ \%) & $\Delta_0$ & Runs & $\langle M_f\rangle$ & $\langle\Delta_f\rangle$ & $\langle\ee_f\rangle$ (\%) & $\mathrm{SD}(\ee_f)$ (\%) \\
\colrule
$50$ & $0.0 / 0.00$ & $0$ & $500$ & $11.34$ & $0.00$ & $0.00$ & $0.00$ \\
$100$ & $0.0 / 0.00$ & $0$ & $500$ & $14.78$ & $0.00$ & $0.00$ & $0.00$ \\
$250$ & $0.0 / 0.00$ & $0$ & $500$ & $22.58$ & $0.00$ & $0.00$ & $0.00$ \\
$500$ & $0.0 / 0.00$ & $0$ & $500$ & $33.85$ & $0.00$ & $0.00$ & $0.00$ \\
$1000$ & $0.0 / 0.00$ & $0$ & $500$ & $56.89$ & $0.00$ & $0.00$ & $0.00$ \\
\colrule
$50$ & $1.0 / 4.00$ & $2$ & $500$ & $11.60$ & $2.00$ & $18.77$ & $6.05$ \\
$100$ & $1.0 / 2.00$ & $2$ & $500$ & $14.90$ & $2.00$ & $14.45$ & $4.19$ \\
$250$ & $1.0 / 0.80$ & $2$ & $500$ & $22.76$ & $2.00$ & $9.32$ & $2.48$ \\
$500$ & $1.0 / 1.20$ & $6$ & $500$ & $34.98$ & $6.00$ & $17.94$ & $4.02$ \\
$1000$ & $1.0 / 1.00$ & $10$ & $500$ & $58.21$ & $10.00$ & $17.71$ & $3.30$ \\
\colrule
$50$ & $5.0 / 4.00$ & $2$ & $500$ & $11.91$ & $2.00$ & $18.43$ & $7.69$ \\
$100$ & $5.0 / 6.00$ & $6$ & $500$ & $15.66$ & $6.00$ & $40.72$ & $10.65$ \\
$250$ & $5.0 / 4.80$ & $12$ & $500$ & $25.60$ & $12.00$ & $48.86$ & $10.16$ \\
$500$ & $5.0 / 5.20$ & $26$ & $500$ & $44.83$ & $26.00$ & $59.48$ & $9.53$ \\
$1000$ & $5.0 / 5.00$ & $50$ & $500$ & $82.89$ & $50.00$ & $61.10$ & $6.97$ \\
\end{tabular}
\end{ruledtabular}
\end{table}

\subsection{Reaction-Resolved Entropy Production and Marginal Dissipation Slope Proof}

For a network of reversible elementary reactions $\rho$ satisfying local detailed balance at temperature $T = 298.15\text{ K}$ ($RT = 2.479\text{ kJ mol}^{-1}$), the instantaneous entropy production rate is $\sigma(t) = R \sum_\rho (J_\rho^+(t) - J_\rho^-(t)) \ln(J_\rho^+(t)/J_\rho^-(t)) \ge 0$. Along any monotonic amplification trajectory ($\mathrm{d}\ee/\mathrm{d}t > 0$), specific free-energy dissipation normalized per mole of surviving chiral product $M_{\mathrm{prod}}(t)$ is tracked via the dimensionless quadrature state $g_\alpha(t) = \Delta G_{\mathrm{diss},\alpha}(t)/(RT)$ ($\alpha \in \{R, S, B\}$):
\begin{equation}
\Delta G_{\mathrm{diss}}(\ee) = \int_0^{t(\ee)} \frac{T\sigma(s)}{M_{\mathrm{prod}}(s)}\,\mathrm{d}s = RT \int_{\ee_0}^{\ee} c(u)\,\mathrm{d}u, \qquad c(u) = \frac{\sigma(u)/R}{M_{\mathrm{prod}}(u)\,(\mathrm{d}u/\mathrm{d}t)} > 0,
\label{eq:s_c_u_def}
\end{equation}
and the dimensionless thermodynamic amplification efficiency is $\eta_G(\ee) = \frac{\ee - \ee_0}{\Delta G_{\mathrm{diss}}(\ee)/(RT)} = \left[\frac{1}{\ee - \ee_0}\int_{\ee_0}^{\ee} c(u)\,\mathrm{d}u\right]^{-1}$. Differentiating $\eta_G(\ee)$ with respect to $\ee$ yields
\begin{equation}
\frac{\mathrm{d}\eta_G(\ee)}{\mathrm{d}\ee} = \frac{\int_{\ee_0}^{\ee} c(u)\,\mathrm{d}u - (\ee - \ee_0) c(\ee)}{\left[\int_{\ee_0}^{\ee} c(u)\,\mathrm{d}u\right]^2} = -\frac{\int_{\ee_0}^{\ee} (u - \ee_0) c'(u)\,\mathrm{d}u}{\left[\int_{\ee_0}^{\ee} c(u)\,\mathrm{d}u\right]^2}.
\label{eq:s_eta_derivative_proof}
\end{equation}
Because the kernel $(u - \ee_0)$ is non-negative on $[\ee_0, \ee]$, Eq.~\eqref{eq:s_eta_derivative_proof} proves that:
\begin{enumerate}
\item Whenever $c'(u) > 0$ across $[\ee_0, \ee]$ (or whenever the $(u - \ee_0)$-weighted integral $\int_{\ee_0}^{\ee}(u-\ee_0)c'(u)\,\mathrm{d}u > 0$, as in buffered mass-depleting Channel R with $B_0/M_0 \ge 2.5$ where $M_{\mathrm{prod}}(u) = M_0 \ee_0 / u \to 0$ and buffered Channel S with $B_0/M_0 = 5.0$ where homochiral side-flux grows as $(1+u^2)/2$), $\mathrm{d}\eta_G/\mathrm{d}\ee < 0$, so efficiency \emph{decreases monotonically} with $\ee$.
\item Whenever $c'(u) < 0$ across an amplification interval (or $\int_{\ee_0}^{\ee}(u-\ee_0)c'(u)\,\mathrm{d}u < 0$, as in driven autocatalytic Channel B where $M_{\mathrm{prod}}(t) = c_L(t) + c_D(t)$ is integrated autonomously under chemostatted precursor bath $A_{\mathrm{chem}} = 1.0$ and heterodimer sink bath $\mathrm{LD}_{\mathrm{bath}} = 0.05$ with cumulative turnover counters $n_{A,\mathrm{cons}}, n_{LD,\mathrm{form}}$ satisfying $(c_L + c_D) - n_{A,\mathrm{cons}} + 2 n_{LD,\mathrm{form}} = M_0$, while $\mathrm{d}u/\mathrm{d}t \propto u(1-u^2)$ accelerates away from the unstable racemic fixed point and antagonistic heterodimer dissipation $\propto (1-u^2)$ subsides), $\mathrm{d}\eta_G/\mathrm{d}\ee > 0$, so efficiency \emph{increases} with $\ee$ (Table~\ref{tab:s_thermo_table}).
\end{enumerate}
Under two-step phase removal or sequestration ($L + D \rightleftharpoons S_1 \rightleftharpoons S_2$ for Channel R; $L + D \rightleftharpoons P_{LD} \rightleftharpoons Q_{LD}$ and $2i \rightleftharpoons P_{ii} \rightleftharpoons Q_{ii}$ for Channel S),\cite{HagelschuerFeringa2026,HeinFeringa2025Supramolecular,BilancioniEsposito2025Gears,PengHong2026Entropy,GhosalBisker2026PCCP} the marginal specific dissipation $c(u)$ decomposes into primary phase/dimer formation $c_1(u)$ plus secondary lattice maturation or conformational sequestration $c_2(u)$ divided by the shrinking surviving pool $M_{\mathrm{prod}}(u) = M_0\ee_0/u$:
\begin{equation}
c_R(u) = c_{R,1}(u) + c_{R,2}(u) = \frac{1}{2 u} \ln\frac{k_{1f} c_L c_D}{k_{1r} S_1} + \frac{J_2^+ - J_2^-}{2 u\,J_1^{\mathrm{net}}} \ln\frac{J_2^+}{J_2^-},
\label{eq:s_c_r_two_term}
\end{equation}
where $J_1^{\mathrm{net}}(u) = k_{1f} M_0^2 \ee_0^2 (1 - u^2)/(4 u^2) - k_{1r} S_1$ and $M_{\mathrm{prod}}(u)\,\mathrm{d}u/\mathrm{d}t = 2 u J_1^{\mathrm{net}}(u)$. Specifically, the three autonomous reversible mass-action ODE systems integrated from $M_0 = 1.0$, $\ee_0 = 3\%$ ($c_L(0) = 0.515$, $c_D(0) = 0.485$) are:
\begin{itemize}
\item \textbf{Channel R (5-state ODE for $[c_L, c_D, S_1, S_2, g_R]$):} Two-step reversible heterochiral crystallization ($L + D \rightleftharpoons S_1 \rightleftharpoons S_2$) proceeds with elementary fluxes and rate equations
\[
\begin{aligned}
J_1^+ &= k_{1f}c_L c_D, \quad J_1^- = k_{1r}S_1, \quad J_2^+ = k_{2f}S_1, \quad J_2^- = k_{2r}S_2, \\
\frac{\mathrm{d}c_L}{\mathrm{d}t} &= \frac{\mathrm{d}c_D}{\mathrm{d}t} = -\left(J_1^+ - J_1^-\right), \quad
\frac{\mathrm{d}S_1}{\mathrm{d}t} = \left(J_1^+ - J_1^-\right) - \left(J_2^+ - J_2^-\right), \\
\frac{\mathrm{d}S_2}{\mathrm{d}t} &= J_2^+ - J_2^-, \quad
\frac{\mathrm{d}g_R}{\mathrm{d}t} = \frac{\sigma_R / R}{c_L + c_D},
\end{aligned}
\]
with rate constants $k_{1f} = 0.80$, $k_{1r} = 10^{-5}$, $k_{2f} = 0.012$, $k_{2r} = 5.0\times 10^{-4}$ (matching the crystallization equilibrium ratio $k_{1f}M_{\mathrm{eq}}^2 k_{2f}/(k_{1r}k_{2r}) = 24$ at $M_{\mathrm{eq}} = 0.005$), and baseline initial buffer $S_1(0) = B_0 = 6.0 M_0$, $S_2(0) = 0.02 M_0$.
\item \textbf{Channel S (9-state ODE for $[c_L, \allowbreak c_D, \allowbreak P_{LL}, \allowbreak P_{DD}, \allowbreak P_{LD}, \allowbreak Q_{LL}, \allowbreak Q_{DD}, \allowbreak Q_{LD}, \allowbreak g_S]$):} Reversible homochiral and heterochiral dimerization coupled with secondary dimer sequestration ($ij \in \{LL, DD, LD\}$) obeys
\[
\begin{aligned}
J_{LL}^+ &= k_{\mathrm{homo}}c_L^2, \quad J_{LL}^- = k_{\mathrm{diss,homo}}P_{LL}, \quad
J_{DD}^+ = k_{\mathrm{homo}}c_D^2, \quad J_{DD}^- = k_{\mathrm{diss,homo}}P_{DD}, \\
J_{LD}^+ &= 2 r_{\mathrm{sel}}k_{\mathrm{homo}}c_L c_D, \quad J_{LD}^- = k_{\mathrm{diss,het}}P_{LD}, \quad
J_{Q,ij}^+ = k_{\mathrm{seq},f}P_{ij}, \quad J_{Q,ij}^- = k_{\mathrm{seq},r}Q_{ij}, \\
\frac{\mathrm{d}c_L}{\mathrm{d}t} &= -2\left(J_{LL}^+ - J_{LL}^-\right) - \left(J_{LD}^+ - J_{LD}^-\right), \quad
\frac{\mathrm{d}c_D}{\mathrm{d}t} = -2\left(J_{DD}^+ - J_{DD}^-\right) - \left(J_{LD}^+ - J_{LD}^-\right), \\
\frac{\mathrm{d}P_{ij}}{\mathrm{d}t} &= \left(J_{ij}^+ - J_{ij}^-\right) - \left(J_{Q,ij}^+ - J_{Q,ij}^-\right), \quad
\frac{\mathrm{d}Q_{ij}}{\mathrm{d}t} = J_{Q,ij}^+ - J_{Q,ij}^-, \quad
\frac{\mathrm{d}g_S}{\mathrm{d}t} = \frac{\sigma_S / R}{c_L + c_D},
\end{aligned}
\]
with rate constants $k_{\mathrm{homo}} = 0.25$, $r_{\mathrm{sel}} = 10.0$, $k_{\mathrm{diss,homo}} = 2.0\times 10^{-6}$, $k_{\mathrm{diss,het}} = 4.0\times 10^{-6}$, $k_{\mathrm{seq},f} = 0.012$, $k_{\mathrm{seq},r} = 4.0\times 10^{-4}$, and baseline initial buffer $P_{LD}(0) = B_0 = 5.0 M_0$, $Q_{LD}(0) = 0.02 M_0$, $P_{LL}(0) = P_{DD}(0) = Q_{LL}(0) = Q_{DD}(0) = 10^{-4} M_0$.
\item \textbf{Channel B (5-state ODE for $[c_L, c_D, n_{A,\mathrm{cons}}, n_{LD,\mathrm{form}}, g_B]$):} Driven cooperative autocatalysis ($i \in \{L, D\}$), epimerization, and heterochiral antagonism proceed via
\[
\begin{aligned}
J_{0,i}^+ &= k_{0f}A_{\mathrm{chem}}, \quad J_{0,i}^- = k_{0r}c_i, \quad
J_{a,i}^+ = k_{af}A_{\mathrm{chem}}c_i^2, \quad J_{a,i}^- = k_{ar}c_i^3, \\
J_{\mathrm{rac}}^+ &= k_{\mathrm{rac}}c_L, \quad J_{\mathrm{rac}}^- = k_{\mathrm{rac}}c_D, \quad
J_{\mathrm{inh}}^+ = k_{\mathrm{inh},f}c_L c_D, \quad J_{\mathrm{inh}}^- = k_{\mathrm{inh},r}\mathrm{LD}_{\mathrm{bath}}, \\
\frac{\mathrm{d}c_L}{\mathrm{d}t} &= \left(J_{0,L}^+ - J_{0,L}^-\right) + \left(J_{a,L}^+ - J_{a,L}^-\right) - \left(J_{\mathrm{rac}}^+ - J_{\mathrm{rac}}^-\right) - \left(J_{\mathrm{inh}}^+ - J_{\mathrm{inh}}^-\right), \\
\frac{\mathrm{d}c_D}{\mathrm{d}t} &= \left(J_{0,D}^+ - J_{0,D}^-\right) + \left(J_{a,D}^+ - J_{a,D}^-\right) + \left(J_{\mathrm{rac}}^+ - J_{\mathrm{rac}}^-\right) - \left(J_{\mathrm{inh}}^+ - J_{\mathrm{inh}}^-\right), \\
\frac{\mathrm{d}n_{A,\mathrm{cons}}}{\mathrm{d}t} &= \sum_{i \in \{L,D\}}\left[\left(J_{0,i}^+ - J_{0,i}^-\right) + \left(J_{a,i}^+ - J_{a,i}^-\right)\right], \\
\frac{\mathrm{d}n_{LD,\mathrm{form}}}{\mathrm{d}t} &= J_{\mathrm{inh}}^+ - J_{\mathrm{inh}}^-, \qquad
\frac{\mathrm{d}g_B}{\mathrm{d}t} = \frac{\sigma_B / R}{c_L + c_D},
\end{aligned}
\]
with parameters $A_{\mathrm{chem}} = 1.0$, $\mathrm{LD}_{\mathrm{bath}} = 0.05$, $k_{0f} = 0.08$, $k_{0r} = 0.22$, $k_{af} = 2.75$, $k_{ar} = 1.60$, $k_{\mathrm{rac}} = 0.005$, $k_{\mathrm{inh},f} = 2.30$, and $k_{\mathrm{inh},r} = 0.01$ (chemostatted driving affinity $\Delta\mu_{\mathrm{drive}} = RT\ln[(k_{af}/k_{ar})/(k_{0f}/k_{0r})] \approx RT\ln 4.73$, coupled with mutual antagonism matching the effective driving scale $\Delta\mu_{\mathrm{eff}} = RT\ln 24 \approx 3.18\,RT$).
\end{itemize}
The lower panel of Table~\ref{tab:s_thermo_table} sweeps $B_0/M_0 = S_1(0)/M_0 \in \{0.0, 1.0, 2.5, 5.0, 6.0\}$ for Channel R ($S_2(0) = 0.02\,(B_0/6M_0)$) and $B_0/M_0 = P_{LD}(0)/M_0 \in \{0.0, 1.0, 2.5, 5.0\}$ for Channel S ($Q_{LD}(0) = 0.02\,(B_0/5M_0)$) while keeping all elementary rate constants invariant. At $B_0/M_0 = 0.0$ (unbuffered monomer-only initial states), an initial nucleation-affinity boundary-layer burst ($\ln(J_1^+/J_1^-) \gg 1$ as $S_1(0^+), P_{LD}(0^+) \to 0$) depresses $\eta_G(10\%)$ ($10^3\eta_{G,R} = 10.62$, $10^3\eta_{G,S} = 7.24$) before reaching $10^3\eta_{G,R} = 24.84 \to 26.68$ at $50\% \to 75\%$ ($\Delta G_{\mathrm{diss},R} = 66.90\text{ kJ mol}^{-1}$) and $10^3\eta_{G,S} = 19.01 \to 18.10$ ($\Delta G_{\mathrm{diss},S} = 98.60\text{ kJ mol}^{-1}$, already decreasing over $[50\%, 75\%]$ with slope $-3.62\times 10^{-3}$). At $B_0/M_0 = 1.0$, both channels exhibit negative maturation slopes over $\ee \in [50\%, 75\%]$ ($10^3\eta_{G,R}: 11.04 \to 14.91 \to 13.66$; $10^3\eta_{G,S}: 8.07 \to 13.17 \to 9.30$). For buffered states ($B_0/M_0 \ge 2.5$ for R; $B_0/M_0 = 5.0$ for S), secondary maturation/sequestration over shrinking $M_{\mathrm{prod}}(u) = M_0\ee_0/u$ dominates the entire interval, enforcing monotone decrease across the benchmarks $\eta_G(10\%) > \eta_G(50\%) > \eta_G(75\%)$ (with $\eta_{G,R}$ at $B_0/M_0 = 6.0$ rising slightly from $6.33\times 10^{-3}$ at $5\%$ $\ee$ to $6.96\times 10^{-3}$ at $10\%$ $\ee$ before declining monotonically across $\ee \in [10\%, 90\%]$, and $\eta_{G,S}$ at $B_0/M_0 = 5.0$ exhibiting a brief initial rise to $9.00\times 10^{-3}$ near $\ee \approx 18\%$ before declining monotonically across $\ee \in [18\%, 90\%]$), with $B_0/M_0 = 6.0$ (Channel R, $487.15\text{ kJ mol}^{-1}$) and $B_0/M_0 = 5.0$ (Channel S, $602.80\text{ kJ mol}^{-1}$) matching the baseline upper panel of Table~\ref{tab:s_thermo_table} and Main Fig.~3 within $< 10^{-12}\text{ kJ mol}^{-1}$. Across all $B_0/M_0 \ge 0$, driven autocatalytic Channel B ($\eta_{G,B}(75\%) = 65.41\times 10^{-3}$, $\overline{\mathrm{d}\eta_{G,B}/\mathrm{d}\ee} = +73.86\times 10^{-3} > 0$) exceeds passive efficiency at $75\%$ $\ee$ by $2.45\times$--$17.85\times$ (Channel R) and $3.61\times$--$22.09\times$ (Channel S).

\begin{table}[!htbp]
\caption{Reaction-resolved specific free-energy dissipation $\Delta G_{\mathrm{diss}}(\ee)$ ($\text{kJ mol}^{-1}$ of surviving product), dimensionless thermodynamic efficiency $\eta_G(\ee) = (\ee - \ee_0) / [\Delta G_{\mathrm{diss}}(\ee)/(RT)]$, and baseline-invariant initial-buffer sensitivity across $B_0/M_0 \in \{0.0, 1.0, 2.5, 5.0, 6.0\}$ from $\ee_0 = 3\%$ at $T = 298.15\text{ K}$ ($RT = 2.479\text{ kJ mol}^{-1}$).}
\alttext{Table comparing specific dissipated free energy in kilojoules per mole and dimensionless thermodynamic efficiency at 10, 50, and 75 percent enantiomeric excess for Channel R, Channel S, and Channel B across initial buffer ratios.}
\label{tab:s_thermo_table}
\scriptsize
\begin{ruledtabular}
\begin{tabular}{ccccccc}
Target $\ee$ & $\Delta G_{\mathrm{diss},R}$ & $\eta_{G,R}$ & $\Delta G_{\mathrm{diss},S}$ & $\eta_{G,S}$ & $\Delta G_{\mathrm{diss},B}$ & $\eta_{G,B}$ \\
\colrule
$10.0\%$ (Main Fig.~3) & $24.92$ & $0.0070$ & $22.11$ & $0.0078$ & $9.97$ & $0.0174$ \\
$50.0\%$ (Main Fig.~3) & $259.87$ & $0.0045$ & $217.13$ & $0.0054$ & $23.41$ & $0.0498$ \\
$75.0\%$ (Main Fig.~3) & $487.15$ & $0.0037$ & $602.80$ & $0.0030$ & $27.29$ & $0.0654$ \\
\colrule
Buffer $B_0/M_0$ & $\Delta G_{\mathrm{diss},R}$\textsuperscript{a} & $10^3\eta_{G,R}$\textsuperscript{b} & $\Delta G_{\mathrm{diss},S}$\textsuperscript{a} & $10^3\eta_{G,S}$\textsuperscript{b} & Slope R\textsuperscript{c} & Slope S\textsuperscript{d} \\
\colrule
$0.0$ (monomer-only) & $66.90$\textsuperscript{a} & $10.62 \to 24.84 \to 26.68$\textsuperscript{b} & $98.60$\textsuperscript{a} & $7.24 \to 19.01 \to 18.10$\textsuperscript{b} & $+7.35$\textsuperscript{c} & $-3.62$\textsuperscript{d} \\
$1.0$ (intermediate) & $130.66$\textsuperscript{a} & $11.04 \to 14.91 \to 13.66$\textsuperscript{b} & $191.90$\textsuperscript{a} & $8.07 \to 13.17 \to 9.30$\textsuperscript{b} & $-5.01$\textsuperscript{c} & $-15.46$\textsuperscript{d} \\
$2.5$ (buffered) & $233.99$\textsuperscript{a} & $9.64 \to 8.90 \to 7.63$\textsuperscript{b} & $341.60$\textsuperscript{a} & $8.15 \to 8.60 \to 5.22$\textsuperscript{b} & $-5.09$\textsuperscript{c} & $-13.48$\textsuperscript{d} \\
$5.0$ (baseline S) & $413.40$\textsuperscript{a} & $7.60 \to 5.24 \to 4.32$\textsuperscript{b} & $602.80$\textsuperscript{a} & $7.85 \to 5.37 \to 2.96$\textsuperscript{b} & $-3.68$\textsuperscript{c} & $-9.62$\textsuperscript{d} \\
$6.0$ (baseline R) & $487.15$\textsuperscript{a} & $6.96 \to 4.48 \to 3.66$\textsuperscript{b} & --- & --- & $-3.28$\textsuperscript{c} & --- \\
\end{tabular}
\end{ruledtabular}
\raggedright\scriptsize
\textsuperscript{a}$\Delta G_{\mathrm{diss}}(75\%)$ in $\text{kJ mol}^{-1}$ under unmodified baseline rate constants. \textsuperscript{b}Values of $10^3\,\eta_G(\ee)$ at $\ee = 10\% \to 50\% \to 75\%$ ($B_0/M_0 = 6.0$ for R and $B_0/M_0 = 5.0$ for S reproduce the upper panel; upper-panel $\eta_G$ entries are rounded to $4$ decimal places, i.e., $10^{-4}$). \textsuperscript{c}Maturation-regime slope $10^3\,\mathrm{d}\eta_{G,R}/\mathrm{d}\ee$ across $50\%$--$75\%$ $\ee$ (mean $10\%$--$75\%$ slopes are $+24.71, +4.03, -3.10, -5.04, -5.08\times 10^{-3}$ for $B_0/M_0 = 0.0, 1.0, 2.5, 5.0, 6.0$). \textsuperscript{d}Maturation-regime slope $10^3\,\mathrm{d}\eta_{G,S}/\mathrm{d}\ee$ across $50\%$--$75\%$ $\ee$ (mean $10\%$--$75\%$ slopes are $+16.71, +1.89, -4.49, -7.52\times 10^{-3}$ for $B_0/M_0 = 0.0, 1.0, 2.5, 5.0$; vs.\ $+73.86\times 10^{-3}$ for Channel B).
\end{table}

\subsection{Shear--Kramers--Lorentz Relative-Entropy Identity and Unbuffered \texorpdfstring{$\Omega = 0$}{Omega = 0} Baseline}

For any closed isothermal ideal-dilute mass-action network with concentration vector $\bm{Z}(t) \in \mathbb{R}_{>0}^S$ relaxing toward a positive detailed-balanced equilibrium $\bm{Z}_{\mathrm{eq}}$ ($J_\rho^+(\bm{Z}_{\mathrm{eq}}) = \allowbreak J_\rho^-(\bm{Z}_{\mathrm{eq}})$ for every reversible reaction $\rho$), the non-equilibrium Helmholtz or Gibbs free-energy excess above $\bm{Z}_{\mathrm{eq}}$ equals $RT$ times the Shear--Kramers--Lorentz relative entropy (f-divergence):\cite{Rao2016,RaoEsposito2018,Remlein2026}
\[
\begin{aligned}
\mathcal{F}_{\mathrm{tot}}\bigl(\bm{Z}\bigr) - \mathcal{F}_{\mathrm{tot}}\bigl(\bm{Z}_{\mathrm{eq}}\bigr) &= RT\,\mathcal{L}\bigl(\bm{Z}\big\|\bm{Z}_{\mathrm{eq}}\bigr) = RT\sum_{i=1}^S \left[Z_i \ln\frac{Z_i}{Z_{i,\mathrm{eq}}} - Z_i + Z_{i,\mathrm{eq}}\right] \ge 0, \\
\frac{\mathrm{d}\mathcal{F}_{\mathrm{tot}}}{\mathrm{d}t} &= RT\sum_{i=1}^S \dot{Z}_i \ln\frac{Z_i}{Z_{i,\mathrm{eq}}} = -RT\sum_\rho \left(J_\rho^+ - J_\rho^-\right)\ln\frac{J_\rho^+}{J_\rho^-} = -T\sigma(t) \le 0.
\end{aligned}
\]
Symbolic differentiation verifies $\mathrm{d}\mathcal{F}_{\mathrm{tot}}/\mathrm{d}t + T\sigma(t) = 0$ identically, and numerical quadrature from $\ee_0 = 3\%$ to $\ee_f = 75\%$ across $\Omega = B_0/M_0 \in \{0.0, 0.5, 2.0, 6.0\}$ confirms $|\Delta\mathcal{F}_{\mathrm{tot}} + \int_0^t T\sigma(s)\,\mathrm{d}s| \le 5.47\times 10^{-13}\text{ kJ mol}^{-1}$ across both Channel R and Channel S (Table~\ref{tab:s_shear_kl_omega}). Table~\ref{tab:s_shear_kl_omega} separates intrinsic monomer-depletion cost ($\Omega = 0$, seeded at thermodynamic nucleation floor $10^{-12} M_0$) from pre-loaded intermediate-buffer maturation inflation ($\Omega > 0$) under both the physical parameter set of Main Fig.~3 and an isometric chemical-affinity matched control where all three channels operate at identical $\Delta\mu_{\mathrm{tot}} = RT\ln 24 \approx 3.178\,RT = 7.88\text{ kJ mol}^{-1}$. Even at $\Omega = 0$, reaching $75\%$ $\ee$ requires $66.90\text{ kJ mol}^{-1}$ ($\eta_G = 0.0267$) for Channel R and $98.60\text{ kJ mol}^{-1}$ ($\eta_G = 0.0181$) for Channel S versus $27.29\text{ kJ mol}^{-1}$ ($\eta_G = 0.0654$) for open-chemostat Channel B ($61.30$ and $95.03\text{ kJ mol}^{-1}$ vs.\ $23.24\text{ kJ mol}^{-1}$, $\eta_G = 0.0768$, under isometric $\Delta\mu = RT\ln 24$), while pre-loading $\Omega = 6.0$ inflates $\Delta G_{\mathrm{diss}}(75\%)$ by $7.28\times$ ($487.15\text{ kJ mol}^{-1}$) and $7.21\times$ ($710.99\text{ kJ mol}^{-1}$; or $7.79\times$ and $7.76\times$ under isometric $\ln 24$).

\section{Unified Re-Attribution Data Tables Across Seven Experimental Systems}
\label{sec:s4_experiments}

Table~\ref{tab:s_seven_systems_full} provides the complete inventory inputs $(M_0, \Delta_0, \ee_0, M_f, \Delta_f, \ee_f, \ee_{\mathrm{sub},f})$ and both ordered $(A_R, A_S, A_B)$ and symmetric Shapley $(A_R^{\mathrm{Sh}}, A_B^{\mathrm{Sh}})$ attributions for the seven experimental benchmarks and multiphase recycling regimes analyzed in Section~IV of the main text. In ``Regime I: 8-Species Pure R'', cross-heterodimer precipitation $L_i + D_j \to P_{L_i D_j,\mathrm{solid}}$ across 8 distinct chiral monomer species at $k_x/k_{\mathrm{het}} = 1.0$ conserves the global 8-species chiral excess $\Delta_{\mathrm{tot},f} = \sum_{i=1}^8 \Delta_{i,f} = \sum_{i=1}^8 \Delta_{i,0} = \Delta_{\mathrm{tot},0}$ ($\Delta_{\mathrm{tot},f}/\Delta_{\mathrm{tot},0} = 1.000$ and $M_{\mathrm{tot},f}/M_{\mathrm{tot},0} = \ee_0/\ee_f = 0.05/0.5714 = 0.0875$), on which the pooled attribution shares $(A_R = +52.14\%, A_S = 0.00\%, A_B = 0.00\%)$ are evaluated; for single species $i=1$, asymmetric cross-pairings ($j \neq 1$) remove $\Delta_1$ ($\Delta_{1,f}/\Delta_{1,0} = 0.126$) and reduce surviving species-1 monomer fraction to $M_{1,f}/M_{1,0} = (0.05/0.5714)\times 0.126 = 0.0110 \approx (0.05/0.5714)/8$ (for which single-species Eq.~\eqref{eq:s_ordered_proof} yields $A_{R,1} = \ee_0(M_{1,0}/M_{1,f} - 1) = +449.55\%$ and cross-species chirality transfer $A_{B,1}^{\mathrm{cross}} = (\Delta_{1,f} - \Delta_{1,0})/M_{1,f} = -397.41\%$, telescoping to $A_{R,1} + A_{B,1}^{\mathrm{cross}} = +52.14\%$).

\begin{table}[!htbp]
\caption{Complete inventory and ordered vs.\ symmetric Shapley three-channel re-attribution across seven experimental benchmarks and multiphase regimes. Here the $\ee_f$ column reports $\ee_{\mathrm{sub},f}$ when $A_S \neq 0$ (while $A_R^{\mathrm{Sh}} + A_B^{\mathrm{Sh}} = \ee_{\mathrm{tot},f} - \ee_0 = A_R + A_B$ decomposes the active-phase enrichment prior to $A_S$), and em-dashes (---) denote unmeasured phase-split inventories (with $A_B$ reporting $\ee_f - \ee_0$ conditional on unmeasured phase split) and do not imply zero channel activity.}
\alttext{Comprehensive table listing initial and final enantiomeric excess, retained mass fraction, ordered attribution shares for Channels R, S, and B, and Shapley shares for seven experimental literature benchmarks and multiphase regimes.}
\label{tab:s_seven_systems_full}
\scriptsize\setlength{\tabcolsep}{2.5pt}
\begin{ruledtabular}
\begin{tabular}{lcccccccc}
System / Reference & $\ee_0$ (\%) & $\ee_f$ (\%) & $M_f/M_0$ & $\Delta_f/\Delta_0$ & $A_R$ (\%) & $A_S$ (\%) & $A_B$ (\%) & $A_R^{\mathrm{Sh}} / A_B^{\mathrm{Sh}}$ \\
\colrule
Soai (1995)\cite{Soai1995}\textsuperscript{e} & $0.50$ & $99.50$ & $1.0000$ & $186.4$ & $0.00$ & $+6.30$ & $+92.70$ & $0.00 / +92.70$ \\
Blackmond (2001)\cite{Blackmond2001}\textsuperscript{e} & $10.00$ & $68.00$ & $1.0000$ & $1.000$ & $0.00$ & $+58.00$ & $0.00$ & $0.00 / 0.00$ \\
Klussmann (2006)\cite{Klussmann2006} & $2.00$ & $99.00$ & $0.0202$ & $1.000$ & $+97.00$ & $0.00$ & $0.00$ & $+97.00 / 0.00$ \\
Deng (2024)\cite{Deng2024}\textsuperscript{e} & $3.20$ & $42.20$ & --- & $5.50$ & --- & $+24.60$ & $+14.40$ & --- \\
van Dongen (2025)\cite{VanDongen2025} & $0.00$ & $91.00$ & $0.4600$ & $\infty$ & $0.00$\textsuperscript{a} & $0.00$ & $+91.00$\textsuperscript{a} & $+24.57 / +66.43$ \\
Viedma (2005)\cite{Viedma2005} & $5.00$ & $99.50$ & $1.0000$\textsuperscript{b} & $19.90$ & --- & --- & $+94.50$\textsuperscript{b} & $0.00 / +94.50$ \\
Noorduin (2008)\cite{Noorduin2008} & $0.00$ & $99.00$ & $1.0000$\textsuperscript{b} & $\infty$ & --- & --- & $+99.00$\textsuperscript{b} & $0.00 / +99.00$ \\
\colrule
Regime I: 5-state Pure R (cycle 80) & $5.00$ & $60.20$ & $0.0831$ & $1.000$ & $+55.20$ & $0.00$ & $0.00$ & $+55.20 / 0.00$ \\
Regime I: 8-species Pure R (cycle 100)\textsuperscript{d} & $5.00$ & $57.14$ & $0.0110$\textsuperscript{d} & $0.126$\textsuperscript{d} & $+52.14$\textsuperscript{d} & $0.00$ & $0.00$\textsuperscript{d} & $+52.14 / 0.00$\textsuperscript{d} \\
Regime II: R + B (Total)\cite{Deck2024,Deck2025} & $20.00$ & $96.50$\textsuperscript{b} & $1.0000$\textsuperscript{b} & $4.825$ & $0.00$\textsuperscript{b} & $0.00$ & $+76.50$\textsuperscript{b} & $0.00 / +76.50$ \\
Regime II: R + B (Solid)\cite{Deck2024,Deck2025} & $20.00$ & $100.00$ & $0.9650$ & $4.825$ & $+0.73$ & $0.00$ & $+79.27$ & $+2.11 / +77.89$ \\
Regime III: R + S + B\cite{Ribo2025ChemPhysChem} & $20.00$ & $92.87$ & $0.9467$\textsuperscript{c} & $4.095$ & $+1.13$ & $+6.36$ & $+65.38$ & $+2.87 / +63.64$ \\
\end{tabular}
\end{ruledtabular}
\raggedright\scriptsize
\textsuperscript{a}Ordered attribution from $\ee_0 = 0$; symmetric Shapley Eq.~(7) [Eq.~\eqref{eq:s_shapley_proof}] gives $A_R^{\mathrm{Sh}} = +24.57\%$ and $A_B^{\mathrm{Sh}} = +66.43\%$. \textsuperscript{b}Total two-phase inventory ($M_{\mathrm{tot}}/M_0 = 1.0000$); in Regime II, the solid crystal phase ($M_{\mathrm{solid}}/M_0 = 0.9650$) reaches $\ee_{\mathrm{solid}} = 100.00\%$ ($A_{R,\mathrm{solid}} = +0.73\%$, $A_{B,\mathrm{solid}} = +79.27\%$; Shapley $+2.11\% / +77.89\%$). \textsuperscript{c}Soluble inventory $M_{\mathrm{sol}}/M_0 = 0.9467$ ($M_{\mathrm{chiral}}/M_0 = 0.8818$ at $\ee_{\mathrm{chiral}} = 92.87\%$, $\ee_{\mathrm{sol}} = 86.51\%$, $Y_{\mathrm{hom}} = 81.90\%$) under continuous racemic inflow ($f_{\mathrm{ex}} = 0.045\text{ cycle}^{-1}$, $\alpha_{\mathrm{evap}} = 6.0$). \textsuperscript{d}Columns $M_f/M_0$ and $\Delta_f/\Delta_0$ list species-1 monomer fractions ($M_{1,f}/M_{1,0} = (0.05/0.5714)\times 0.126 = 0.0110$, $\Delta_{1,f}/\Delta_{1,0} = 0.126$) at $k_x/k_{\mathrm{het}} = 1.0$, whereas $(A_R = +52.14\%, A_S = 0, A_B = 0)$ are evaluated on the pooled 8-species inventory ($\Delta_{\mathrm{tot},f}/\Delta_{\mathrm{tot},0} = 1.000$, $M_{\mathrm{tot},f}/M_{\mathrm{tot},0} = 0.0875$; single-species Eq.~\eqref{eq:s_ordered_proof} gives $A_{R,1} = +449.55\%$ and $A_{B,1}^{\mathrm{cross}} = -397.41\%$, summing to $+52.14\%$). \textsuperscript{e}When $A_S \neq 0$, intermediate active-phase excesses are $\ee_{\mathrm{tot},f} = 93.20\%$ for Soai\cite{Soai1995} ($\Delta_{\mathrm{tot},f}/\Delta_0 = 93.20/0.50 = 186.4$ at $M_f/M_0 = 1.000$, $A_B = +92.70\%$, active catalyst subpool $\ee_{\mathrm{sub},f} = 99.50\%$, $A_S = +6.30\%$), $\ee_{\mathrm{tot},f} = \ee_0 = 10.00\%$ for Blackmond\cite{Blackmond2001} ($\Delta_{\mathrm{tot},f}/\Delta_0 = 1.000$, $A_B = 0.00\%$, monomer catalyst subpool $\ee_{\mathrm{sub},f} = 68.00\%$, $A_S = +58.00\%$), and $\ee_{\mathrm{tot},f} = 17.60\%$ for Deng \emph{et al.}\cite{Deng2024} (Table~S81 at $90\%$ conversion: $\ee_0 = 3.20\%$, $\Delta_{\mathrm{tot},f}/\Delta_0 = 17.60/3.20 = 5.50$, $A_B = +14.40\%$, unreacted monomer subpool $\ee_{\mathrm{sub},f} = 42.20\%$, $A_S = +24.60\%$).
\end{table}

\section{Dual-Monomer Non-Crossing Parity Invariant, Thioimidate Mechanistic Decoupling, and RPKA Identifiability}
\label{sec:s5_deng}

\subsection{Proof of the Dual-Monomer Non-Crossing Invariant}

Consider two distinct chiral amino acid or aminonitrile monomer species $i \in \{1, 2\}$ (such as Leu and Val in Table~S79 of Deng \emph{et al.}\cite{Deng2024}) with enantiomeric concentrations $(L_1, D_1)$ and $(L_2, D_2)$, monomer inventories $M_i(t) = L_i(t) + D_i(t)$, and signed monomer excesses $\Delta_i(t) = L_i(t) - D_i(t)$. Under general parity-symmetric homochiral ($k_{\mathrm{homo}}(t)$) and heterochiral ($k_{\mathrm{het}}(t) = r k_{\mathrm{homo}}(t)$) couplings without epimerization ($J_{\Delta_i} = 0$), we analyze both symmetric self- plus cross-ligation and selective electrophile--nucleophile cross-ligation:
\begin{enumerate}
\item \textbf{Symmetric Self- plus Cross-Ligation:} When monomer species $i$ reacts with both monomer pools ($L_1 + L_2$ and $D_1 + D_2$), the kinetic equations are
\begin{align}
\frac{\mathrm{d}L_i}{\mathrm{d}t} &= -k_{\mathrm{homo}}(t)\,L_i\left(L_1 + L_2\right) - k_{\mathrm{het}}(t)\,L_i\left(D_1 + D_2\right), \nonumber\\
\frac{\mathrm{d}D_i}{\mathrm{d}t} &= -k_{\mathrm{homo}}(t)\,D_i\left(D_1 + D_2\right) - k_{\mathrm{het}}(t)\,D_i\left(L_1 + L_2\right).
\label{eq:s_dual_monomer_odes}
\end{align}
Using $L_i(L_1 + L_2) - D_i(D_1 + D_2) = \frac{1}{2}[\Delta_i M + M_i \Delta]$ and $L_i(D_1 + D_2) - D_i(L_1 + L_2) = \frac{1}{2}[\Delta_i M - M_i \Delta]$ ($M = M_1 + M_2$, $\Delta = \Delta_1 + \Delta_2$), we obtain
\begin{align}
\frac{\mathrm{d}\Delta_i}{\mathrm{d}t} &= -\frac{k_{\mathrm{homo}}(t) + k_{\mathrm{het}}(t)}{2}\,M(t)\,\Delta_i(t) - \frac{k_{\mathrm{homo}}(t) - k_{\mathrm{het}}(t)}{2}\,M_i(t)\,\Delta(t), \nonumber\\
\frac{\mathrm{d}M_i}{\mathrm{d}t} &= -\frac{k_{\mathrm{homo}}(t) + k_{\mathrm{het}}(t)}{2}\,M(t)\,M_i(t) - \frac{k_{\mathrm{homo}}(t) - k_{\mathrm{het}}(t)}{2}\,\Delta(t)\,\Delta_i(t).
\label{eq:s_delta_i_ode}
\end{align}
Applying the quotient rule to $\ee_i = \Delta_i / M_i$ gives $\mathrm{d}\ee_i/\mathrm{d}t = \frac{1}{2}(k_{\mathrm{het}} - k_{\mathrm{homo}})\Delta(t)(1 - \ee_i^2)$, so for any initial ratio $M_1(0)/M_2(0)$, the intensive difference satisfies $\frac{\mathrm{d}(\ee_1 - \ee_2)}{\mathrm{d}t} = -\frac{1}{2}(k_{\mathrm{het}} - k_{\mathrm{homo}})\Delta(t)(\ee_1 + \ee_2)(\ee_1 - \ee_2)$, which is proportional to $(\ee_1 - \ee_2)$ and cannot cross zero.
\item \textbf{Selective Electrophile--Nucleophile Cross-Ligation:} In the experimental protocol of Deng \emph{et al.},\cite{Deng2024} Substrate 1 is the activated electrophile ($\text{Ac-Leu-CN}$) and Substrate 2 is the amino acid nucleophile ($\text{Val-OH}$), coupling via $v_{LL} = k_{\mathrm{homo}} L_1 L_2$, $v_{DD} = k_{\mathrm{homo}} D_1 D_2$, $v_{LD} = k_{\mathrm{het}} L_1 D_2$, and $v_{DL} = k_{\mathrm{het}} D_1 L_2$. Subtracting species 2 from species 1 cancels homochiral fluxes $(v_{LL} - v_{DD})$ and yields the general extensive law
\begin{equation}
\frac{\mathrm{d}\left(\Delta_1 - \Delta_2\right)}{\mathrm{d}t} = -k_{\mathrm{het}}(t)\left[M_2(t)\,\Delta_1(t) - M_1(t)\,\Delta_2(t)\right].
\label{eq:s_noncrossing_ode_proof}
\end{equation}
For equimolar initial monomer pools ($\delta_M = (M_{1,0} - M_{2,0})/M_{2,0} = 0 \implies M_1(t) = M_2(t) = M_c(t) = M(t)/2$), Eq.~\eqref{eq:s_noncrossing_ode_proof} and the quotient rule yield
\begin{align}
\frac{\mathrm{d}\left(\Delta_1 - \Delta_2\right)}{\mathrm{d}t} &= -k_{\mathrm{het}}(t)\,M_c(t)\left(\Delta_1(t) - \Delta_2(t)\right), \nonumber\\
\frac{\mathrm{d}\left(\ee_1 - \ee_2\right)}{\mathrm{d}t} &= -\frac{k_{\mathrm{het}}(t) - k_{\mathrm{homo}}(t)}{2}\,M_c(t)\left(1 + \ee_1\ee_2\right)\left(\ee_1 - \ee_2\right), \nonumber\\
\frac{\ee_1(t) - \ee_2(t)}{\ee_1(0) - \ee_2(0)} &= \frac{M_c(0)}{M_c(t)}\exp\left(-\int_0^t k_{\mathrm{het}}(s)\,M_c(s)\,\mathrm{d}s\right) > 0.
\label{eq:s_noncrossing_exact_solution}
\end{align}
\end{enumerate}
Because the exponential factor and $M_c(0)/M_c(t)$ are positive at all finite conversions ($M_c(t) > 0$), $\Delta_1(t) - \Delta_2(t)$ and $\ee_1(t) - \ee_2(t)$ \textbf{cannot change sign or cross zero} under any parity-symmetric equimolar Channel S network. When an initial stoichiometric offset $\delta_M = (M_{1,0} - M_{2,0})/M_{2,0} = -16.95\%$ ($H_{\mathrm{stoich}}$, $M_{1,0} = 0.8305 M_{2,0}$) is introduced into Eq.~\eqref{eq:s_noncrossing_ode_proof}, $M_1(t) = M_2(t) + \delta_M M_{2,0} < M_2(t)$ approaches zero as conversion nears $83\%$, causing $M_2(t)\Delta_1(t) - M_1(t)\Delta_2(t)$ to invert sign on S79 ($\mathrm{RMSE} = 1.41\text{ pp}$) while failing on held-out S81 HPLC where $\ee_1(0) = 3.05\% \ll \ee_2(0) = 49.66\%$ ($\mathrm{RMSE} = 5.35\text{ pp}$).

\subsection{Conversion-Domain tQSSA Cancellation Proof and Clock-Time RPKA Closure}

When reversible thioimidate intermediate formation ($L_1 + \mathrm{RSH} \underset{k_{-1}}{\overset{k_{\mathrm{on}}}{\rightleftharpoons}} TI_L$, $D_1 + \mathrm{RSH} \underset{k_{-1}}{\overset{k_{\mathrm{on}}}{\rightleftharpoons}} TI_D$, with forward pseudo-first-order rate $k_1 = k_{\mathrm{on}}[\mathrm{RSH}]$, reverse cleavage rate $k_{-1} = k_{\mathrm{off,TI}}$, and equilibrium association constant $K_{\mathrm{TI}} = k_{\mathrm{on}}/k_{\mathrm{off,TI}}$) and thioimidate epimerization ($TI_L \underset{k_{\mathrm{epim,TI}}}{\overset{k_{\mathrm{epim,TI}}}{\rightleftharpoons}} TI_D$) are coupled with selective electrophile--nucleophile cross-ligation between Substrate 1 ($\text{Ac-Leu-CN}$) and Substrate 2 ($\text{Val-OH}$), let $M_1^{\mathrm{tot}}(t) = L_1(t) + D_1(t) + TI_L(t) + TI_D(t)$ and $\Delta_1^{\mathrm{tot}}(t) = L_1(t) - D_1(t) + TI_L(t) - TI_D(t)$ denote the total unreacted Substrate-1 inventory and signed chiral excess recovered upon hydrolytic quenching of $TI_{L,D}$ back to monomer prior to chiral HPLC/NMR analysis, and let $M_2(t) = L_2(t) + D_2(t)$, $\Delta_2(t) = L_2(t) - D_2(t)$ denote Substrate 2. Specifically, the 11-state clock-time ODE system for $H_{\mathrm{joint}}$ tracks the state vector $\bm{y}_{11} = (L_1, \allowbreak D_1, \allowbreak L_2, \allowbreak D_2, \allowbreak TI_L, \allowbreak TI_D, \allowbreak \mathrm{RSH}, \allowbreak p_{LL}, \allowbreak p_{DD}, \allowbreak p_{LD}, \allowbreak p_{DL})^\top$ with 4 parameters $(k_{\mathrm{cat,homo}}, \allowbreak r, \allowbreak k_{\mathrm{epim,TI}}, \allowbreak K_{\mathrm{TI}})$:
\[
\begin{aligned}
\frac{\mathrm{d}L_1}{\mathrm{d}t} &= -k_{\mathrm{on}} L_1[\mathrm{RSH}] + k_{\mathrm{off,TI}} TI_L, \qquad
\frac{\mathrm{d}D_1}{\mathrm{d}t} = -k_{\mathrm{on}} D_1[\mathrm{RSH}] + k_{\mathrm{off,TI}} TI_D, \\
\frac{\mathrm{d}TI_L}{\mathrm{d}t} &= k_{\mathrm{on}} L_1[\mathrm{RSH}] - k_{\mathrm{off,TI}} TI_L - v_{LL} - v_{LD} - k_{\mathrm{epim,TI}}\left(TI_L - TI_D\right), \\
\frac{\mathrm{d}TI_D}{\mathrm{d}t} &= k_{\mathrm{on}} D_1[\mathrm{RSH}] - k_{\mathrm{off,TI}} TI_D - v_{DD} - v_{DL} + k_{\mathrm{epim,TI}}\left(TI_L - TI_D\right), \\
\frac{\mathrm{d}L_2}{\mathrm{d}t} &= -v_{LL} - v_{DL}, \qquad
\frac{\mathrm{d}D_2}{\mathrm{d}t} = -v_{DD} - v_{LD}, \qquad
\frac{\mathrm{d}p_{ab}}{\mathrm{d}t} = v_{ab}, \\
\frac{\mathrm{d}[\mathrm{RSH}]}{\mathrm{d}t} &= -k_{\mathrm{on}}\left(L_1 + D_1\right)[\mathrm{RSH}] + k_{\mathrm{off,TI}}\left(TI_L + TI_D\right) + \sum_{a,b\in\{L,D\}} v_{ab},
\end{aligned}
\]
where the dipeptide ligation fluxes are $v_{LL} = k_{\mathrm{cat,homo}} TI_L L_2$, $v_{DD} = k_{\mathrm{cat,homo}} TI_D D_2$, $v_{LD} = r k_{\mathrm{cat,homo}} TI_L D_2$, and $v_{DL} = r k_{\mathrm{cat,homo}} TI_D L_2$. Under the total quasi-steady-state approximation (tQSSA),\cite{Byun2024} $[TI_{L,D}](t) = f_{\mathrm{TI}}([\mathrm{RSH}])\,(1 \pm \ee_1(t))\,M_1^{\mathrm{tot}}(t)/2$ with $f_{\mathrm{TI}}([\mathrm{RSH}]) = \frac{K_{\mathrm{TI}}[\mathrm{RSH}]}{1 + K_{\mathrm{TI}}[\mathrm{RSH}]}$. At equimolar initial stoichiometry $C_0 \equiv M_{1,0} = M_{2,0} = M_0/2$ (where $M_0 = M_{1,0} + M_{2,0}$ is the pooled two-monomer concentration, so $M_1^{\mathrm{tot}}(t) = M_2(t) = M(t)/2 = C_0(1 - c(t))$ and $P_{\mathrm{tot}}(t) = p_{LL} + p_{DD} + p_{LD} + p_{DL} = C_0 c(t)$), the selective cross-ligation rates are $\mathrm{d}M_i/\mathrm{d}t = -\bar{k}_{\mathrm{lig}}(\ee) M_1^{\mathrm{tot}} M_2 = -\frac{1}{4}\bar{k}_{\mathrm{lig}}(\ee) M(t)^2$, $\mathrm{d}c/\mathrm{d}t = -(1/M_0)\mathrm{d}M/\mathrm{d}t = \frac{1}{2}\bar{k}_{\mathrm{lig}}(\ee) M_0(1-c)^2 = \bar{k}_{\mathrm{lig}}(\ee) C_0(1-c)^2$, and $\mathrm{d}\ee_i/\mathrm{d}t = \frac{1}{4}k_{\mathrm{cat,homo}}f_{\mathrm{TI}}(r-1)M_j\ee_j(1-\ee_i^2) - 2k_{\mathrm{epim,TI}}f_{\mathrm{TI}}\delta_{i,1}\ee_1$ ($j \neq i$), where $\bar{k}_{\mathrm{lig}}(\ee) = \frac{1}{4}k_{\mathrm{cat,homo}} f_{\mathrm{TI}}[(r+1) - (r-1)\ee_1\ee_2]$. Dividing $\mathrm{d}\ee_i/\mathrm{d}t$ by $\mathrm{d}c/\mathrm{d}t$ cancels $f_{\mathrm{TI}}([\mathrm{RSH}])$ between numerator and denominator:
\begin{equation}
\frac{\mathrm{d}\ee_i}{\mathrm{d}c} = \frac{1}{1 - c}\,\frac{(r - 1)\,\ee_j\left(1 - \ee_i^2\right)}{(r + 1) - (r - 1)\ee_1\ee_2} - \frac{\kappa_{\mathrm{epim},i}^{\mathrm{eff}}(\ee)}{(1 - c)^2}\,\ee_i \qquad (j \neq i \in \{1, 2\}),
\label{eq:s_tqssa_cancellation}
\end{equation}
where $\kappa_{\mathrm{epim},2}^{\mathrm{eff}}(\ee) = 0$ (since nucleophile Substrate 2 does not form a thioimidate intermediate) and $\kappa_{\mathrm{epim},1}^{\mathrm{eff}}(\ee) = \frac{2 k_{\mathrm{epim,TI}} f_{\mathrm{TI}}([\mathrm{RSH}])}{\bar{k}_{\mathrm{lig}}(\ee) C_0} = \frac{4 k_{\mathrm{epim,TI}} f_{\mathrm{TI}}([\mathrm{RSH}])}{\bar{k}_{\mathrm{lig}}(\ee) M_0} = \frac{\kappa_{\mathrm{epim,TI}}(r+1)}{(r+1)-(r-1)\ee_1\ee_2}$ with constant $\kappa_{\mathrm{epim,TI}} = \frac{8 k_{\mathrm{epim,TI}}}{(r+1)k_{\mathrm{cat,homo}} M_0} = \frac{4 k_{\mathrm{epim,TI}}}{(r+1)k_{\mathrm{cat,homo}} C_0}$ (independent of $f_{\mathrm{TI}}$ and $K_{\mathrm{TI}}$, reducing to constant $\kappa_{\mathrm{epim},1}^{\mathrm{eff}} \approx \kappa_{\mathrm{epim,TI}}$ when $|\ee_1\ee_2| \ll 1$), thereby reducing the 3 conversion-domain parameters $(r, \kappa_{\mathrm{epim,TI}}, K_{\mathrm{TI}})$ of $H_{\mathrm{ti,tQSSA}}$ to practical rank $2/3$ on conversion-only S79 data ($\max|\Delta\ee_{\mathrm{tQSSA}} - \Delta\ee_{\mathrm{full}}| = 3.17 \times 10^{-4}\text{ pp}$). Main Fig.~4(a,b,d) and Table~\ref{tab:s_deng_comparison} compare the dual-monomer S79/S81 candidate model set (whereas Main Fig.~4(c) displays the 7-model 59-point multicomponent benchmark of Sec.~S5.C and Table~\ref{tab:s_deng_multicomp}): closed $H_0$ ($r = k_{\mathrm{het}}/k_{\mathrm{homo}} = 2.00$, corresponding to statistical non-selective coupling when the combinatorial factor of $2$ is absorbed into $k_{\mathrm{het}}$), fitted ligation $H_1$ ($\hat{r}=3.47$), Val-only epimerization control $H_{\mathrm{Val}}$ ($\rho_{\mathrm{epim},2}$), pre-equilibrated thioimidate $H_{\mathrm{ti,pre}}$ ($\eta_{\mathrm{pre},1}$), substrate-2 thioimidate $H_{\mathrm{m2}}$, initial stoichiometric offset $H_{\mathrm{stoich}}$ ($\delta_M = -16.95\%$), direct Leu epimerization $H_{\mathrm{Leu}}$ ($\rho_{\mathrm{epim},1}$), Leu hydrolytic shunt $H_{\mathrm{shunt}}$ ($\rho_{\mathrm{shunt},1}$), 3-parameter conversion-domain tQSSA $H_{\mathrm{ti,tQSSA}}$, and 4-parameter clock-time ODE $H_{\mathrm{joint}}$. Table~\ref{tab:s_deng_comparison} summarizes the quantitative model comparison across S79 ($n=9$), held-out S81 HPLC ($n=6$), S81 $^{13}\text{C}$-NMR ($n=3$), and unique pooled validation points ($n=17$, comprising $6$ S81 HPLC points, $3$ S81 NMR points, $4$ unique S80 NMR points, and $4$ unique S82 NMR points after deduplicating the $90\%$-conversion NMR point shared with S79/S81), as well as the quenched clock-time Reaction Progress Kinetic Analysis (RPKA) protocol for $(k_{\mathrm{cat,homo}}, \allowbreak r, \allowbreak k_{\mathrm{epim,TI}}, \allowbreak K_{\mathrm{TI}})$ of $H_{\mathrm{joint}}$ across $3$ initial concentrations $\{50, 150, 400\}\text{ mM} \times 4\text{ observables}$ ($M_1^{\mathrm{tot}}, M_2, \ee_1 - \ee_2, [TI_1]+[TI_2]$), which breaks conversion scaling invariance and achieves full $4/4$ parameter identifiability ($\lambda_{\min}(\mathrm{FIM}) = 9.62 > 1.0$ at $4\text{ timepoints} = 48$ scalars; $\lambda_{\min} = 15.11$ at $6\text{ timepoints} = 72$ scalars).

\begin{table}[!htbp]
\caption{Quantitative comparison of competing hypotheses on Deng \emph{et al.}\cite{Deng2024} dual-monomer peptide ligation datasets S79 ($n=9$), held-out S81 HPLC ($n=6$), S81 $^{13}\text{C}$-NMR ($n=3$), and unique pooled points ($n=17$).}
\alttext{Table comparing root-mean-square error across S79, held-out S81 HPLC, S81 carbon-13 NMR, and unique pooled datasets alongside Delta AICc and practical Fisher Information Matrix rank across six kinetic hypotheses.}
\label{tab:s_deng_comparison}
\scriptsize\setlength{\tabcolsep}{3.5pt}
\begin{ruledtabular}
\begin{tabular}{lccccccc}
Model Hypothesis & Channels & S79 & S81 HPLC & S81 $^{13}\text{C}$\textsuperscript{b} & Pooled\textsuperscript{b} & $\Delta\mathrm{AIC}_c$ & Pract.\ Rank \\
\colrule
$H_0$: Ligation ($r=2.00$) & S only & $3.02$ & $1.85$ & $0.71$ & $2.12$ & $+11.95$ & $1/1$ \\
$H_1$: Fitted Ligation ($\hat{r}=3.47$) & S only & $2.30$ & $1.38$\textsuperscript{a} & --- & --- & --- & $1/1$ \\
$H_{\mathrm{stoich}}$: Offset ($\delta_M = -16.95\%$) & S + offset & $1.41$ & $5.35$ & $1.88$ & $3.80$ & $+1.68$ & $2/2$ \\
$H_{\mathrm{Leu}}$: Direct Leu Epim. & S + B & $1.29$ & $2.15$ & $0.83$ & $2.30$ & $+0.13$ & $2/2$ \\
$H_{\mathrm{ti,tQSSA}}$: Lumped TI ($r=2.119$) & S + B & $0.98$ & $1.26$ & $0.27$ & $2.01$ & $0.00$ & $2/3$ (conv.) \\
$H_{\mathrm{joint}}$: Full TI ($r=2.118$) & S + B & $0.99$ & $1.23$ & $0.30$ & $2.00$ & $+0.04$ & $4/4$ (RPKA) \\
\end{tabular}
\end{ruledtabular}
\raggedright\footnotesize
\textsuperscript{a}Combined S81 ($n=9$) RMSE for $H_1$ ($\hat{r}=3.47$); single-amino-acid control RMSEs are $0.62\text{ pp}$ on Val (S75) and $2.38\text{ pp}$ on Leu (S76) vs.\ $0.72\text{ pp}$ and $1.84\text{ pp}$ for $H_0$ ($r=2$).
\textsuperscript{b}Column ``S81 $^{13}\text{C}$'' ($n=3$) lists split-pot $^{13}\text{C}$-NMR nominal-initialization RMSEs ($\ee_1(0)=0\%, \ee_2(0)=50\%$:\ $0.71, 1.88, 0.83, 0.27, 0.30\text{ pp}$; under unlabeled LC-SFC initializations $\ee_1(0)=3.05\%, \ee_2(0)=49.66\%$, the $n=3$ RMSEs are $1.45, 1.06, 1.37, 1.83, 1.78\text{ pp}$). Column ``Pooled'' ($n=17$) lists unlabeled-initialization unique held-out RMSEs ($2.12, 3.80, 2.30, 2.01, 2.00\text{ pp}$; under split-pot $^{13}\text{C}$-NMR nominal initializations, the $n=17$ RMSEs are $2.74, 4.42, 2.93, 2.53, 2.53\text{ pp}$ at fixed S79 parameters and $2.74, 4.40, 2.93, 2.51, 2.51\text{ pp}$ under split-pot S79 calibration). Across all $11$ held-out clusters ($2^{11}=2048$ sign flips), $H_{\mathrm{stoich}}$ degrades generalization relative to $H_0$ (two-sided $p = 0.0449$ unlabeled-init; $p = 0.0078$ split-pot), whereas $H_{\mathrm{ti,tQSSA}}$ and $H_{\mathrm{joint}}$ outperform $H_{\mathrm{stoich}}$ on $3/3$ multi-timepoint S81 clusters and $7/11$ total clusters (one-sided $p = 0.031\text{--}0.033$, two-sided $p = 0.062\text{--}0.066$) and outperform $H_0$ on $8/11$ clusters (two-sided split-pot $p = 0.032\text{--}0.033$; unlabeled-init $p = 0.197\text{--}0.238$).
\end{table}

\subsection{\texorpdfstring{$1 \times K$}{1 x K} Multicomponent Non-Crossing Theorem, 59-Point S80--S86 Benchmark, and Adduct NMR Probes}

Extending Proposition~4 to competitive coupling between one $\alpha$-aminonitrile electrophile $(L_1, D_1)$ and $K$ amino-acid nucleophiles $\{(L_{2,k}, D_{2,k})\}_{k=1}^K$ with arbitrary parity-even rate functions $k_{\mathrm{homo},k}(t,\bm{x}), k_{\mathrm{het},k}(t,\bm{x}) > 0$ and zero epimerization yields exact symbolic closure:\cite{Deng2024,Barat2024}
\[
\begin{aligned}
\frac{\mathrm{d}}{\mathrm{d}t}\!\left(\Delta_1 - \sum_{k=1}^K \Delta_{2,k}\right) &= -\sum_{k=1}^K k_{\mathrm{het},k}(t,\bm{x})\Bigl(M_{2,k}\Delta_1 - M_1\Delta_{2,k}\Bigr), \\
\frac{\mathrm{d}}{\mathrm{d}t}\!\left(\Delta_1 + \sum_{k=1}^K \Delta_{2,k} + 2\sum_{k=1}^K \Delta_{P,\mathrm{homo},k}\right) &= 0,
\end{aligned}
\]
where $\Delta_{P,\mathrm{homo},k} = P_{LL,k} - P_{DD,k}$ ($0/240$ sign flips and maximum conservation residual $6.66\times 10^{-16}$ across $240$ randomized $1\times 2$ and $1\times 3$ non-autonomous laws). When all $K$ nucleophiles share homogeneous kinetics ($k_{\mathrm{homo},k} = k_{\mathrm{homo}}$, $k_{\mathrm{het},k} = k_{\mathrm{het}}$) at balanced stoichiometry $M_1(0) = \sum_{k=1}^K M_{2,k}(0)$, pooling $M_{2,\mathrm{tot}} = \sum_k M_{2,k}$ and $\Delta_{2,\mathrm{tot}} = \sum_k \Delta_{2,k}$ integrates to $\Delta_1(t) - \Delta_{2,\mathrm{tot}}(t) = [\Delta_1(0) - \Delta_{2,\mathrm{tot}}(0)]\exp(-\int_0^t k_{\mathrm{het}}(s) M_1(s)\,\mathrm{d}s)$, forbidding zero crossings of $\ee_1(t) - \ee_{2,\mathrm{tot}}(t)$.

To evaluate zero-shot generalization across multicomponent mixtures without refitting nucleophile parameters, we extract the single-amino-acid constants reported independently by Deng \emph{et al.}\cite{Deng2024} under identical aqueous conditions ($\mathrm{pD}\ 9.5$, $5\,^\circ\mathrm{C}$, $\mathrm{D_2O}$, $400\text{ mM}$ $\mathrm{Ac\text{-}Cys\text{-}OH}$): Figs.~S238, S240, and S242 give $k_{LL,\mathrm{Val}} = 0.032$, $k_{DL,\mathrm{Val}} = 0.064$ ($r_{\mathrm{Val}} = 2.0000$), $k_{LL,\mathrm{Leu}} = 0.033$, $k_{DL,\mathrm{Leu}} = 0.050$ ($r_{\mathrm{Leu}} = 1.5152$, $\rho_{\mathrm{Leu}} = k_{LL,\mathrm{Leu}}/k_{LL,\mathrm{Val}} = 1.0000$ in mixed competition), Table~S49 gives Ac-Ala-CN + Ac-Cys-OH $K_{\mathrm{eq}} = 0.8279\text{ M}^{-1}$ ($K_{\mathrm{TI}}(200\text{ mM}) = 0.16558$), and p.~S256 chiral SFC gives $^{13}\text{C}$-labeled Ac-Leu-CN stock $\ee_1(0) = 3.05\%$. Calibrating only on Table~S79 ($n = 9$) and evaluating zero-shot across all $59$ independent validation points ($23$ conversion clusters) spanning binary Tables~S80--S82 ($n = 17$, $11$ clusters) and three-component OOD Tables~S83--S86 ($n = 42$, $12$ clusters, including all $27$ unaveraged triplicate HPLC points in Table~S86; Main Fig.~4(c) and Table~\ref{tab:s_deng_multicomp}) shows that stoichiometric offset $H_{\mathrm{stoich}}$ ($\delta_M = +11.90\%$) degrades total OOS RMSE from $2.18\text{ pp}$ ($H_0$, $0$ fitted parameters) to $3.58\text{ pp}$ ($3.40\text{ pp}$ on 3-component OOD; exact $2^{23}$ cluster sign-flip $p = 1.09\times 10^{-4}$ vs.\ $H_{\mathrm{ti,tQSSA}}$). By contrast, ongoing Ac-Leu-CN thioimidate epimerization\cite{Deng2024,Barat2024} improves OOS RMSE to $2.12$/$2.14\text{ pp}$ for 1-DOF $H_{\mathrm{Leu}}$/$H_{\mathrm{ti,tQSSA}}$ ($1.83$/$1.85\text{ pp}$ on 3-component OOD) and to $1.92$/$1.94\text{ pp}$ for 2-DOF $H_{\mathrm{Leu},2\mathrm{p}}$/$H_{\mathrm{ti},2\mathrm{p}}$ ($r_{\mathrm{Val}} = 2.125$, $1.65$/$1.66\text{ pp}$ on 3-component OOD; exact $2^{23}$ sign-flip $p = 0.0127$ vs.\ $H_0$, with $H_{\mathrm{Leu},2\mathrm{p}}$ vs.\ $H_{\mathrm{ti},2\mathrm{p}}$ indistinguishable on fixed-thiol conversion trajectories at two-sided $p = 0.0772$).

Crucially, exact symbolic reduction of the 12-state conversion-rescaled ODE proves why $H_{\mathrm{Leu}}$ and $H_{\mathrm{ti,tQSSA}}$ coincide at fixed $[\mathrm{RSH}]_0 = 400\text{ mM}$: rapid thioimidate equilibration rescales clock time uniformly while mapping simultaneous free-nitrile ($k_{\mathrm{epim,free}}$) and thioimidate ($k_{\mathrm{epim,TI}}$) epimerization onto $k_{\mathrm{epim,eff}}(\xi) = k_{\mathrm{epim,TI}} + k_{\mathrm{epim,free}}/(K_{\mathrm{TI}}[\mathrm{RSH}](\xi))$ (symbolic residual $0$ across all $12$ components). Varying initial thiol $[\mathrm{RSH}]_0$ from $400\text{ mM}$ down to $50\text{ mM}$ breaks this tQSSA invariance,\cite{Byun2024} yielding a $3.13\text{ pp}$ Briggs--Haldane shift ($k_{\mathrm{cat}}/k_{\mathrm{off}}$ breakdown) and a $21.77\text{ pp}$ divergence between free-nitrile and thioimidate epimerization at $\xi = 0.90$. Direct $^{1}\text{H}$-NMR fitting on Deng \emph{et al.} Table~S50 (Ac-Ala-CN + Val + Ac-Cys-OH: $k_{\mathrm{off}} = 2.79\text{ h}^{-1}$, $k_{\mathrm{cat}} = 3.76\text{ M}^{-1}\text{ h}^{-1}$, Briggs--Haldane breakdown ratio $k_{\mathrm{cat}}[N]_0/k_{\mathrm{off}} = 0.257$) reduces $[\mathrm{Adduct}](t)$ RMSE $5.1\times$ from $5.82\text{ mM}$ (instantaneous pre-equilibrium) to $1.14\text{ mM}$, and on Table~S52 (Cys capture: $K_{\mathrm{eq}} = 0.4469\text{ M}^{-1}$, $k_{\mathrm{off}} = 5.65\text{ h}^{-1}$, ratio $0.111$) reduces RMSE $4.1\times$ from $2.55\text{ mM}$ to $0.62\text{ mM}$.

\subsection{Seeded Multicomponent Dipeptide Co-Precipitation Across Tables S90 and S93--S97}

Coupling heterochiral peptide or azapeptide ligation with diastereoselective phase separation (as in Deng \emph{et al.} Figs.~S297--S299 and related heterochiral coupling cascades)\cite{Deng2024,Yan2024} links solution diastereoselectivity to solid-state crystallization. In Deng \emph{et al.} Tables~S90 and S93--S97 ($5$-day crystallization of scalemic dipeptide mixtures Ac-Val-Val-OH, Ac-Ala-Val-OH, and Ac-Leu-Val-OH in $\mathrm{D_2O}$),\cite{Deng2024,Ahn2022,Nahas2026} unseeded binary mixtures S93 ($42.56\text{ mM}$ total heterochiral $LD+DL$) and S97 ($41.48\text{ mM}$) produce $0.0\text{ mM}$ precipitate over $5$ days despite exceeding seeded eutectic solubility, establishing a primary nucleation metastability threshold $c_{\mathrm{nuc,LD,tot}} = 53.75\text{ mM}$ bypassed only by crystal seeding (S90, S94--S96). Above nucleation or under seeding, pure-compound independent racemic precipitation ($k_p = 0$, $s_{\mathrm{Val}} = 10.31\text{ mM}$) fails on mixed supernatants because trace guest heterochiral dipeptides (Ac-Ala-Val-OH and Ac-Leu-Val-OH) co-precipitate into the host Ac-Val-Val-OH racemic crystal lattice via multicomponent solid-solution incorporation.\cite{Ma2024,Sun2023} Modeling guest incorporation either by Hill--Sheng linear lattice partition $(c_{\mathrm{solid},k}/c_{\mathrm{solid,Val}}) = k_{p,k}(c_{\mathrm{aq},k}/s_{\mathrm{Val}})$ ($s_{\mathrm{Val}} = 10.31\text{ mM}$, $k_{p,\mathrm{Ala}} = 0.1842$, $k_{p,\mathrm{Leu}} = 0.7583$) or by symmetrical Guggenheim excess-free-energy solid solutions\cite{Ma2024,Sun2023} $G^{\mathrm{E}} = RT\sum_{i<j} a_{0,ij} x_i x_j \le 0$ ($s_{\mathrm{Val}} = 12.24\text{ mM}$, $a_{0,\mathrm{Ala}} = 0.00$, $a_{0,\mathrm{Leu}} = -1.3777$) cuts held-out Day-5 binary supernatant $\ee$ RMSE on S94+S96 ($n = 4$ endpoints, calibrated on Days~0--4) from $12.58\text{ pp}$ ($k_p = 0$; $9.19\text{ pp}$ across all $6$ tables [$8$ Day-5 observations]) to $2.20\text{ pp}$ ($2.77\text{ pp}$ across all $6$ tables [$8$ Day-5 observations]) for Hill--Sheng and $3.85\text{ pp}$ ($3.15\text{ pp}$ across all $6$ tables [$8$ Day-5 observations]) for Guggenheim, and under zero-refit cross-peptide transfer to Table~S96 ($n = 10$ time-series points using Tanford hydrophobic $\mathrm{CH}_2$ scaling from non-S96 pots) cuts $\ee$ RMSE from $11.06\text{ pp}$ ($15.03\text{ pp}$ on guest Leu) to $3.24\text{ pp}$ ($1.49\text{ pp}$) and $2.40\text{ pp}$ ($0.95\text{ pp}$; Table~\ref{tab:s_deng_coprecip}).

\section{Analytical Wet--Dry Quadrature, Three-Regime Coupled Recycling, and Cluster Control}
\label{sec:s6_wetdry_recycling}

\subsection{Derivation of the Analytical Wet--Dry Differential Law and Quadrature}

\begin{figure}[!htbp]
\centering
\includegraphics[width=0.78\textwidth]{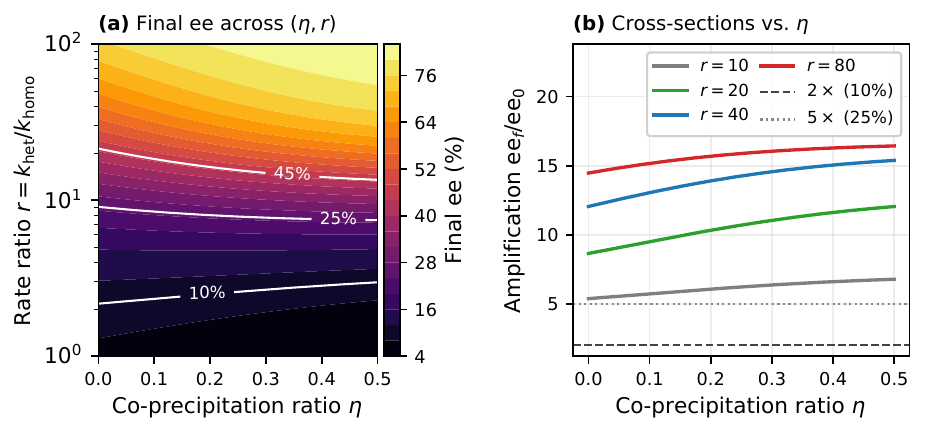}
\caption{Two-panel co-precipitation robustness analysis across homochiral-to-heterochiral dimer precipitation ratio $\eta_{\mathrm{coprecip}} = k_{\mathrm{precip,homo}} / \allowbreak k_{\mathrm{precip,het}} \in [0, 0.5]$ (labeled $\eta$ on axis) and diastereoselectivity $r = k_{\mathrm{het}}/k_{\mathrm{homo}} \in [1, 100]$ over $N = 100$ cycles from $\ee_0 = 5\%$. (a) Parameter-plane heatmap of final $\ee_f$ (\%) with white iso-contours at $10\%, 25\%,$ and $45\%$ ($\ee_f = 60.26\%$ at $\eta_{\mathrm{coprecip}} = 0, r = 40$; $66.19\%$ at $\eta_{\mathrm{coprecip}} = 0.12$; $76.93\%$ at $\eta_{\mathrm{coprecip}} = 0.50$). (b) Cross-sections of amplification factor $\ee_f/\ee_0$ vs.\ $\eta_{\mathrm{coprecip}}$ at fixed selectivity ratios $r \in \{10, 20, 40, 80\}$.}
\alttext{Two-panel figure showing a heatmap of final enantiomeric excess across co-precipitation ratio eta from 0 to 0.5 and selectivity r from 1 to 100 on the left and amplification factor curves versus eta at four selectivity values on the right.}
\label{fig:s_coprecip}
\end{figure}

\begin{figure}[!htbp]
\centering
\includegraphics[width=0.78\textwidth]{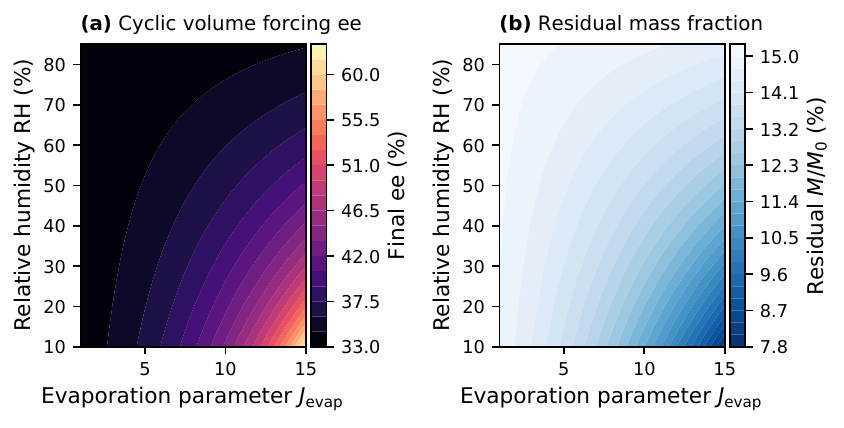}
\caption{Two-panel parameter-space analysis of continuous evaporation across evaporation flux parameter $J_{\mathrm{evap}} \in [1.0, 15.0]$ and relative humidity $\mathrm{RH} \in [10\%, 85\%]$ ($N = 100$ cycle-equivalent duration, $\ee_0 = 5\%$). (a) Final enantiomeric excess $\ee$ (\%) reaching up to $62.57\%$ at high $J_{\mathrm{evap}}$ and low $\mathrm{RH}$. (b) Corresponding residual soluble mass fraction $M/M_0$ (\%) dropping to $7.99\%$ in the high-amplification corner.}
\alttext{Two-panel heatmap across evaporation flux parameter from 1 to 15 and relative humidity from 10 to 85 percent showing final enantiomeric excess up to 62.57 percent in panel a and residual soluble mass fraction down to 7.99 percent in panel b.}
\label{fig:s_cont_evap}
\end{figure}

\begin{figure}[!htbp]
\centering
\includegraphics[width=0.88\textwidth]{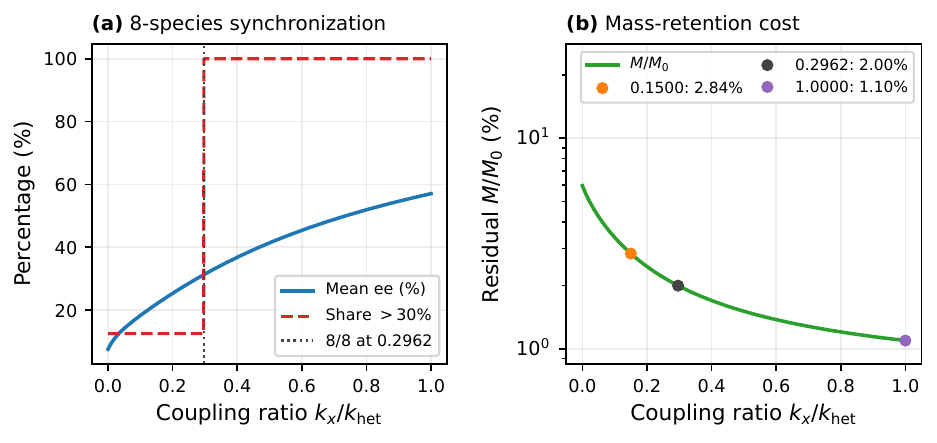}
\caption{Two-panel 8-species cross-dimerization coupling sweep ($100\,000$ ODE evaluations) across dimensionless cross-coupling strength $k_x/k_{\mathrm{het}} \in [0, 1]$ from diverse initial excesses $\ee_{i,0} \in [1.5\%, 7.0\%]$. (a) Ensemble mean final $\ee$ (\%, solid blue curve, reaching $57.06\%$ at $k_x/k_{\mathrm{het}} = 1.0$) and percentage share of the 8 species exceeding $30\%$ $\ee$ (dashed red curve on the same vertical percentage scale, reaching $8/8 = 100\%$ consensus at threshold $k_x/k_{\mathrm{het}} = 0.2962$, marked by the vertical dotted line). (b) Log-scale residual soluble mass fraction $M/M_0$ (\%) ($2.84\%$ at $k_x/k_{\mathrm{het}} = 0.1500$, $2.00\%$ at $0.2962$, and $1.10\%$ at $1.0000$).}
\alttext{Two-panel line plot across cross-coupling ratio k-x over k-het from 0 to 1 showing mean final enantiomeric excess and fraction of eight species above 30 percent ee in panel a and log-scale residual soluble mass percentage in panel b.}
\label{fig:s_eight_species}
\end{figure}

\begin{figure}[!htbp]
\centering
\includegraphics[width=0.78\textwidth]{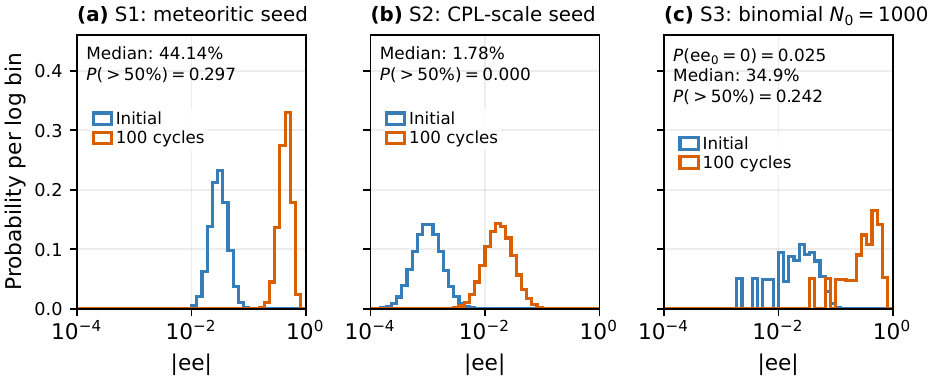}
\caption{Three-panel distribution of final enantiomeric excess across initial seed scenarios over $N = 100$ cycles. (a) S1: meteoritic seed distribution ($\ee_0 \sim \mathcal{N}(2\%, 1.5\%)$). (b) S2: circularly polarized light (CPL)-scale seed ($\ee_0 = 10^{-4}$). (c) S3: racemic binomial finite-number seed at $N_0 = 1000$ ($\ee_0 = 0$, mean $|\ee|$ amplified from $2.52\%$ to $10.67\%$).}
\alttext{Three-panel histogram comparing final enantiomeric excess distributions under meteoritic seed S1 in panel a, CPL-scale seed S2 in panel b, and binomial finite-molecule seed S3 at initial count 1000 in panel c.}
\label{fig:s_seed_scenarios}
\end{figure}

During cycle $n \to n+1$, let the dry phase precipitate fraction $\mu_{\mathrm{het}}(M)$ of $1{:}1$ $L{:}D$ pairs ($\Delta q = 0$) and fraction $\mu_{\mathrm{homo}}(M)$ of homochiral solute ($\Delta_{n+1} = [1 - \mu_{\mathrm{homo}}(M)]\Delta_n$), while total soluble mass transitions as $M_{n+1} = [1 - \mu_{\mathrm{het}}(M) - \mu_{\mathrm{homo}}(M)] M_n + \mu_{\mathrm{het}}(M) \ee_n^2 M_n$ due to heterochiral pairing propensity $\propto 4 c_L c_D / M^2 = 1 - \ee^2$. Dividing $\Delta_{n+1}$ by $M_{n+1}$ yields the single-step difference $\ee_{n+1} - \ee_n = \frac{\mu_{\mathrm{het}}(M)}{1 - \mu_{\mathrm{homo}}(M) - \mu_{\mathrm{het}}(M)(1-\ee_n^2)}\ee_n(1-\ee_n^2)$; in the small-step continuum limit ($\mu_{\mathrm{het}}, \mu_{\mathrm{homo}} \ll 1$, replacing $\mu_{\mathrm{het}}(M)(1-\ee_n^2)$ by $\mu_{\mathrm{het}}(M)$ in the denominator), this reduces to
\begin{equation}
\frac{\mathrm{d}\ee}{\mathrm{d}n} = C(M)\,\ee\left(1 - \ee^2\right), \qquad C(M) = \frac{\mu_{\mathrm{het}}(M)}{1 - \mu_{\mathrm{het}}(M) - \mu_{\mathrm{homo}}(M)}.
\label{eq:s_wetdry_ode_derivation}
\end{equation}
Separating variables for a constant $C(M) = C_{\mathrm{const}} = 0.01696$ gives $\int_{\ee_0}^{\ee} [u(1-u^2)]^{-1}\mathrm{d}u = \frac{1}{2}\ln\frac{\ee^2(1-\ee_0^2)}{\ee_0^2(1-\ee^2)} = C_{\mathrm{const}} N$, which inverts to Eq.~(16) of the main text. In the pre-saturation Channel R regime ($\mu_{\mathrm{homo}} = 0 \implies \Delta = \Delta_0$ and $M = \Delta_0/\ee = M_0\ee_0/\ee$), quasi-steady dimer partitioning during the dry phase forms homochiral and heterochiral dimers in proportion to $n_{LL} + n_{DD} \propto k_{\mathrm{homo}}(n_L^2 + n_D^2) = \frac{1}{2}k_{\mathrm{homo}}M^2(1+\ee^2)$ and $n_{LD} \propto k_{\mathrm{het}}n_L n_D = \frac{1}{4}r k_{\mathrm{homo}}M^2(1-\ee^2)$ ($r = k_{\mathrm{het}}/k_{\mathrm{homo}} = 40$). Thus the fraction of dimerized mass residing in precipitable $LD$ heterodimers is $g(\ee) = \frac{2n_{LD}}{2(n_{LL}+n_{DD}+n_{LD})} = \frac{r(1-\ee^2)}{(r+2) - (r-2)\ee^2}$, while bimolecular dimer formation across dry phase $t_{\mathrm{dry}}$ saturates with total inventory $M$ as $\frac{C_0 M}{M + K} = \frac{C_0}{1 + K/M}$ (with half-saturation constant $K$ and dimensionless ratio $\kappa \equiv K/M_0$). Combining these two factors gives the per-cycle fractional removal rate $-\frac{1}{M}\frac{\mathrm{d}M}{\mathrm{d}n} = \frac{1}{\ee}\frac{\mathrm{d}\ee}{\mathrm{d}n} = C(M)(1-\ee^2)$ with
\begin{equation}
C(M) = \frac{r C_0}{\left[(r + 2) - (r - 2)\ee^2\right]\left(1 + \frac{K}{M}\right)} = \frac{r C_0}{\left[(r + 2) - (r - 2)\ee^2\right]\left(1 + \frac{\kappa\,\ee}{\ee_0}\right)},
\label{eq:s_C_M_explicit}
\end{equation}
where we substituted $M/M_0 = \ee_0/\ee$. Separating variables in $\mathrm{d}n = \frac{\mathrm{d}\ee}{\ee(1-\ee^2)C(\Delta_0/\ee)} = \frac{(r+2)-(r-2)\ee^2}{r C_0\,\ee(1-\ee^2)}\left(1 + \frac{\kappa}{\ee_0}\ee\right)\mathrm{d}\ee$ cancels $\ee$ in the $\kappa/\ee_0$ term and integrates in closed form [Eq.~(17) of the main text] (plotted directly as the solid curves in Main Fig.~5 at calibrated $C_0 = 0.0431$ and $\kappa = K/M_0 = 0.1417$, i.e., $K = 141.7$ at $M_0 = 1000$, median out-of-sample error $7.40\%$):
\begin{equation}
\ln\frac{\ee_f}{\ee_0} + \frac{2}{r}\ln\!\left[\frac{\ee_f\left(1-\ee_0^2\right)}{\ee_0\left(1-\ee_f^2\right)}\right] + \frac{\kappa}{r\,\ee_0}\Bigl[\left(r-2\right)\left(\ee_f - \ee_0\right) + 4\left(\operatorname{atanh}\ee_f - \operatorname{atanh}\ee_0\right)\Bigr] = N C_0.
\label{eq:s_quadrature_closed}
\end{equation}
Differentiating the left-hand side of Eq.~\eqref{eq:s_quadrature_closed} with respect to $\ee_f$ via $\frac{\mathrm{d}}{\mathrm{d}u}\left[\ln u + \frac{2}{r}\ln\frac{u}{1-u^2}\right] = \frac{(r+2)-(r-2)u^2}{r\,u(1-u^2)}$ and $\frac{\mathrm{d}}{\mathrm{d}u}\left[(r-2)u + 4\operatorname{atanh}u\right] = \frac{(r+2)-(r-2)u^2}{1-u^2}$ reproduces $\frac{C_0}{\ee_f(1-\ee_f^2)C(\Delta_0/\ee_f)} = C_0\,\frac{\mathrm{d}N}{\mathrm{d}\ee_f}$. Figures~\ref{fig:s_coprecip}--\ref{fig:s_seed_scenarios} evaluate sensitivity across the homochiral-to-heterochiral dimer co-precipitation ratio $\eta_{\mathrm{coprecip}} = k_{\mathrm{precip,homo}} / \allowbreak k_{\mathrm{precip,het}} \in [0, 0.5]$ and diastereoselectivity $r = k_{\mathrm{het}}/k_{\mathrm{homo}} \in [1, 100]$, continuous evaporation across evaporation flux $J_{\mathrm{evap}} \in [1.0, 15.0]$ and relative humidity $\mathrm{RH} \in [10\%, 85\%]$, 8-species cross-dimerization coupling $k_x/k_{\mathrm{het}} \in [0, 1]$, and prebiotic initial seed scenarios (S1--S3).

\subsection{Three-Regime Coupled Solid-Phase Racemization Recycling and Open-Flow Bifurcation}

Table~\ref{tab:s_three_regimes} summarizes the quantitative parameters, complete $2^2$ (Regime II) and $2^3$ (Regime III) factorial channel decompositions, and M\"obius/Shapley interaction decompositions of the three multiphase regimes (Regimes I--III) analyzed in Section~V.A and Fig.~6 of the main text, uniting the temperature-cycling solid racemization mechanism of Deck, Hosseinalipour, and Mazzotti\cite{Deck2024,Deck2025} (Regime II, Channel R + B) and the open-flow evaporative autocatalytic bifurcation mechanism of Rib\'o, Micheau, and Buhse\cite{Ribo2025ChemPhysChem} alongside cooperative asymmetric catalysis and deracemization\cite{DuKnowles2026Multiplicative,HagelschuerFeringa2026} (Regime III, Channel R + S + B).

In Regime II (4-state two-phase ODE for liquid $[c_L, c_D]$ and conglomerate crystals $[n_L, n_D]$ with $c_{\mathrm{eq}} = 0.0175 M_0$, $k_{\mathrm{ex}} = 0.45\text{ cycle}^{-1}$, $k_{\mathrm{rac}} = 2.50\text{ cycle}^{-1}$, $a_d = 1.85$, $a_g = 0.42$, and cycle-averaged Damk\"ohler ripening coefficient $\beta_{\mathrm{asym}} = 0.32(a_d - a_g)$),\cite{Saito2013,Deck2024,Deck2025} the governing equations for $i \neq j \in \{L, D\}$ are
\[
\begin{aligned}
\frac{\mathrm{d}c_i}{\mathrm{d}t} &= -J_{\mathrm{cryst},i} - k_{\mathrm{rac}}\left(c_i - c_j\right), \qquad
\frac{\mathrm{d}n_i}{\mathrm{d}t} = J_{\mathrm{cryst},i} \qquad (i \neq j \in \{L, D\}), \\
J_{\mathrm{cryst},i} &= k_{\mathrm{ex}}\left[f_i\frac{c_i}{c_{\mathrm{eq}}} - f_i\left(1 - \beta_{\mathrm{asym}} f_j \left(f_i - f_j\right)\right)\right], \qquad
f_i = \frac{n_i}{n_L + n_D + \varepsilon},
\end{aligned}
\]
with dry-out regularizer $\varepsilon = 10^{-15}\,M_0$ (satisfying $f_L + f_D = 1 - \mathcal{O}(10^{-15})$ for $n_L + n_D = \mathcal{O}(M_0)$), signed solid-phase excess $\phi_s = f_L - f_D = -(f_D - f_L)$, and eutectic initial vector $\bm{y}(0) = (c_L(0), c_D(0), n_L(0), n_D(0)) = (c_{\mathrm{eq}}, c_{\mathrm{eq}}, \tfrac{1}{2}(1+\ee_0)M_0 - c_{\mathrm{eq}}, \tfrac{1}{2}(1-\ee_0)M_0 - c_{\mathrm{eq}})$ ($\beta_{\mathrm{asym}}$ is gated by $\mathbb{I}(k_{\mathrm{ex}} > 0 \wedge k_{\mathrm{rac}} > 0)$ since ripening requires both solid--liquid cycling and solution racemization). Evaluating all $2^2 = 4$ factorial combinations ($\varnothing$, R only, B only, $\text{R}+\text{B}$) shows that disabling either Channel R ($k_{\mathrm{ex}} = 0$) or Channel B ($k_{\mathrm{rac}} = 0$) leaves terminal $\ee_{\mathrm{tot}}(100) = Y_{\mathrm{hom}}(100) = 20.00\%$ ($m_R = m_B = 0.00\text{ pp}$). Enabling both channels ($\text{R}+\text{B}$) yields $\ee_{\mathrm{tot}}(100) = 96.50\%$ ($\ee_{\mathrm{solid}}(100) = 99.9981\%$, reported as $100.00\%$ at two decimal places in Table~I and Tables~\ref{tab:s_seven_systems_full} and~\ref{tab:s_three_regimes}) and $Y_{\mathrm{hom}}(100) = 96.50\%$, so the entire $+76.50\text{ pp}$ gain over the baseline arises from two-channel M\"obius synergy $I_{RB} = +76.50\text{ pp}$ with equal factorial Shapley shares $\phi_R = \phi_B = +38.25\text{ pp}$ (while Eq.~(7) [Eq.~\eqref{eq:s_shapley_proof}] inventory-resolved Shapley attributions on the harvested solid phase $M_{\mathrm{solid}}/M_0 = 0.9650$ are $A_R^{\mathrm{Sh}} = +2.11\%$ and $A_B^{\mathrm{Sh}} = +77.89\%$).

In Regime III (6-state open-flow ODE for $c_L, \allowbreak c_D, \allowbreak P_{LL}, \allowbreak P_{DD}, \allowbreak P_{LD}, \allowbreak S_{\mathrm{rac}}$ with continuous time $t$ measured in wet--dry cycles, $\Delta t = 1\text{ cycle}$ so $\mathrm{d}/\mathrm{d}t \equiv \mathrm{d}/\mathrm{d}n$, at exchange rate $f_{\mathrm{ex}} = 0.045\,\text{cycle}^{-1}$, reference feed concentration $c_{\mathrm{wet},0} = M_0 = 1.0$, and evaporative concentration factor $\alpha_{\mathrm{evap}} = \alpha_{\mathrm{eff}} = 6.0$ when factor R is active or $1.0$ when R is off),\cite{Ribo2025ChemPhysChem,DuKnowles2026Multiplicative,HagelschuerFeringa2026,HeinFeringa2025Supramolecular} the kinetic parameters are $k_{\mathrm{homo},f} = 0.22[1 + 0.10(\alpha_{\mathrm{eff}} - 1)]$ ($= 0.33$ at $\alpha_{\mathrm{eff}} = 6.0$ and $0.22$ at $\alpha_{\mathrm{eff}} = 1.0$), $k_{\mathrm{homo},r} = k_{\mathrm{diss}} = 1.10$, $r_{\mathrm{sel}} = 3.0$, $k_{\mathrm{precip},f} = 0.90\max(0, (\alpha_{\mathrm{eff}} - 1)/5)$ ($= 0.90$ at $\alpha_{\mathrm{eff}} = 6.0$ and $0.0$ at $\alpha_{\mathrm{eff}} = 1.0$), $k_{\mathrm{precip},r} = 1.05$, $k_B = 0.014\,\text{cycle}^{-1}c_{\mathrm{wet},0}^{-2}$, and $K_{\mathrm{sat}} = 0.015\,c_{\mathrm{wet},0}^{-2}$. Let $M_{\mathrm{chiral}} = c_L + c_D + 2(P_{LL} + P_{DD})$, $x_L = (c_L + 2P_{LL})/M_{\mathrm{chiral}}$, $x_D = (c_D + 2P_{DD})/M_{\mathrm{chiral}}$, and $r_B(\alpha_{\mathrm{evap}}) = \frac{2 k_B \alpha_{\mathrm{evap}}^2 c_{\mathrm{wet},0}^2}{1 + K_{\mathrm{sat}}\alpha_{\mathrm{evap}}^2 c_{\mathrm{wet},0}^2}$. The reaction fluxes and governing 6-state ODEs are
\[
\begin{aligned}
J_{ii} &= k_{\mathrm{homo},f} c_i^2 - k_{\mathrm{homo},r} P_{ii} \quad (i \in \{L, D\}), \qquad
J_{LD} = 2 r_{\mathrm{sel}} k_{\mathrm{homo},f} c_L c_D - k_{\mathrm{homo},r} P_{LD}, \\
J_{\mathrm{precip}} &= k_{\mathrm{precip},f} P_{LD} - k_{\mathrm{precip},r} S_{\mathrm{rac}}, \qquad
J_{\mathrm{auto,mon}} = r_B(\alpha_{\mathrm{evap}})\left(c_D x_L^2 - c_L x_D^2\right), \\
J_{\mathrm{auto,dim}} &= r_B(\alpha_{\mathrm{evap}})\left(P_{DD} x_L^2 - P_{LL} x_D^2\right), \qquad
\frac{\mathrm{d}c_i}{\mathrm{d}t} = \frac{1}{2}f_{\mathrm{ex}} M_0 - f_{\mathrm{ex}} c_i - 2J_{ii} - J_{LD} \pm J_{\mathrm{auto,mon}}, \\
\frac{\mathrm{d}P_{ii}}{\mathrm{d}t} &= J_{ii} - f_{\mathrm{ex}} P_{ii} \pm J_{\mathrm{auto,dim}}, \qquad
\frac{\mathrm{d}P_{LD}}{\mathrm{d}t} = J_{LD} - J_{\mathrm{precip}} - f_{\mathrm{ex}} P_{LD}, \\
\frac{\mathrm{d}S_{\mathrm{rac}}}{\mathrm{d}t} &= J_{\mathrm{precip}} - f_{\mathrm{ex}} S_{\mathrm{rac}},
\end{aligned}
\]
where the upper ($+$) sign applies to $L / LL$ and the lower ($-$) sign applies to $D / DD$, integrated from the monomer-seeded initial vector $\bm{y}(0) = (c_L(0), \allowbreak c_D(0), \allowbreak P_{LL}(0), \allowbreak P_{DD}(0), \allowbreak P_{LD}(0), \allowbreak S_{\mathrm{rac}}(0)) = \allowbreak (\tfrac{1}{2}(1+\ee_0)M_0, \allowbreak \tfrac{1}{2}(1-\ee_0)M_0, \allowbreak 0, 0, 0, 0)$. Because $(c_D x_L^2 - c_L x_D^2) + 2(P_{DD} x_L^2 - P_{LL} x_D^2) = M_{\mathrm{chiral}} x_L x_D(x_L - x_D) = \frac{1}{4}M_{\mathrm{chiral}}\ee_{\mathrm{chiral}}(1 - \ee_{\mathrm{chiral}}^2)$, the extensive chiral subpool excess $\Delta_{\mathrm{chiral}} = (c_L - c_D) + 2(P_{LL} - P_{DD}) = M_{\mathrm{chiral}}\,\ee_{\mathrm{chiral}}$ obeys the differential equation
\begin{equation}
\frac{\mathrm{d}\Delta_{\mathrm{chiral}}}{\mathrm{d}t} = -f_{\mathrm{ex}}\Delta_{\mathrm{chiral}} + \frac{k_B \alpha_{\mathrm{evap}}^2 c_{\mathrm{wet},0}^2}{1 + K_{\mathrm{sat}}\alpha_{\mathrm{evap}}^2 c_{\mathrm{wet},0}^2}\,M_{\mathrm{chiral}}\,\ee_{\mathrm{chiral}}\left(1 - \ee_{\mathrm{chiral}}^2\right),
\label{eq:s_quotient_closure}
\end{equation}
so dividing Eq.~\eqref{eq:s_quotient_closure} by $M_{\mathrm{chiral}}(t)$ at stationary $M_{\mathrm{chiral}}(t)$ ($\mathrm{d}M_{\mathrm{chiral}}/\mathrm{d}t = 0$) yields the identical intensive Bernoulli ODE [Eq.~(19) of the main text] for both $\text{R}+\text{B}$ ($M_{\mathrm{chiral}}/M_0 = 1.0000$) and $\text{R}+\text{S}+\text{B}$ ($M_{\mathrm{chiral}}/M_0 = 0.8818$). Pure 6-state ODE trace-seed bisection ($\ee_{\mathrm{init}} = 10^{-6}$, testing $\Delta_{\mathrm{chiral}}(100) > \Delta_{\mathrm{chiral}}(0)$ over 32 bisection steps) yields $\alpha_{\mathrm{crit}}^{\mathrm{ODE}} = 1.83769168$, matching the analytical Bernoulli threshold $\alpha_{\mathrm{crit}}^{\mathrm{ana}} = 1.83769168$ [Eq.~(20) of the main text] within $|\Delta\alpha_{\mathrm{crit}}| = 1.71\times 10^{-10}$, and matching terminal $\ee_{\mathrm{chiral}}(100) = 92.87\%$ at $\alpha_{\mathrm{evap}} = 6.0$ within $1.58\times 10^{-7}\text{ pp}$. Evaluating all $2^3 = 8$ factorial channel combinations ($\varnothing$, R, S, B, R+S, R+B, S+B, and R+S+B) alongside a fixed-$\alpha_{\mathrm{evap}} = 6.0$ heterochiral precipitation control ($k_{\mathrm{precip},f} \in \{0.0, 0.90\}$ across S-off and S-on; Table~\ref{tab:s_three_regimes}) reveals both the mass-partitioning trade-off between active chiral purity $\ee_{\mathrm{chiral}}$ and recoverable homochiral yield $Y_{\mathrm{hom}} = \Delta_{\mathrm{chiral}}/M_{\mathrm{tot}}$ and the separation between achiral evaporative concentration ($\alpha_{\mathrm{evap}}$) and $1{:}1$ heterochiral solid precipitation ($P_{LD} \rightleftharpoons S_{\mathrm{rac}}$):
\begin{enumerate}
\item Both $\text{R}+\text{B}$ ($S$-off) and $\text{R}+\text{S}+\text{B}$ ($S$-on) obey Eq.~\eqref{eq:s_quotient_closure} at $\alpha_{\mathrm{eff}} = 6.0$ and converge to $\ee_{\mathrm{chiral}}(100) = 92.87\%$ (pairwise M\"obius synergy $I_{RB}(\ee_{\mathrm{chiral}}) = +92.00\text{ pp}$, $I_{RSB}(\ee_{\mathrm{chiral}}) = -0.37\text{ pp}$, across-run 3-factor Shapley shares $\phi_R = +45.91\text{ pp}$, $\phi_S = +0.10\text{ pp}$, $\phi_B = +46.64\text{ pp}$). Under $\text{R}+\text{B}$ ($S$-off), no heterodimers $P_{LD}$ or heterochiral precipitate $S_{\mathrm{rac}}$ form ($M_{\mathrm{chiral}}/M_0 = M_{\mathrm{sol}}/M_0 = 1.0000$), giving $Y_{\mathrm{hom}}(100) = 92.87\%$ ($A_R = A_S = 0.00\%$, $A_B = +72.87\%$).
\item Adding Channel S ($\text{R}+\text{S}+\text{B}$) activates diastereoselective heterodimer formation and heterochiral precipitation, providing $+7.49\text{ pp}$ of within-run subpool scavenging ($A_R = +1.13\%$, $A_S = \ee_{\mathrm{chiral}} - \ee_{\mathrm{sol}} = +6.36\%$, $A_B = +65.38\%$) while partitioning $11.82\%$ of total mass into achiral heterodimers ($2P_{LD} = 6.49\%$) and heterochiral precipitate ($2S_{\mathrm{rac}} = 5.33\%$, $M_{\mathrm{sol}}/M_0 = 0.9467$, $M_{\mathrm{chiral}}/M_0 = 0.8818$). Consequently, recoverable homochiral mass is $Y_{\mathrm{hom}}(100) = 81.90\%$ ($-10.97\text{ pp}$ vs.\ $\text{R}+\text{B}$, $I_{RB}(Y_{\mathrm{hom}}) = +92.00\text{ pp}$, $I_{RSB}(Y_{\mathrm{hom}}) = -10.97\text{ pp}$, $\phi_R = +42.34\text{ pp}$, $\phi_S = -3.66\text{ pp}$, $\phi_B = +42.99\text{ pp}$).
\item Because factor ``R'' in the $2^3$ factorial table bundles achiral evaporative concentration ($\alpha_{\mathrm{eff}}: 1.0 \to 6.0$) with heterochiral solid precipitation ($k_{\mathrm{precip},f}: 0 \to 0.90$), holding $\alpha_{\mathrm{evap}} = 6.0$ fixed while varying $k_{\mathrm{precip},f} \in \{0.0, 0.90\}$ isolates the two physical mechanisms: when S is off, $P_{LD} = 0$ so $k_{\mathrm{precip},f}$ has zero effect ($\ee_{\mathrm{chiral}} = Y_{\mathrm{hom}} = 92.87\%$ for both $k_{\mathrm{precip},f} = 0$ and $0.90$, showing that the $+92.00\text{ pp}$ $I_{RB}$ bifurcation gain is driven by achiral evaporative concentration $\alpha_{\mathrm{evap}} = 6.0 > \alpha_{\mathrm{crit}} = 1.84$ accelerating nonlinear Channel B); when S is on at $\alpha_{\mathrm{evap}} = 6.0$, turning on heterochiral precipitation ($k_{\mathrm{precip},f}: 0.0 \to 0.90$) leaves $\ee_{\mathrm{chiral}}(100) = 92.87\%$ invariant, increases soluble $\ee_{\mathrm{sol}}$ from $86.08\%$ to $86.51\%$ (shifting ordered attribution from $A_R = 0.00\%, A_S = +6.79\%, A_B = +66.08\%$ at $k_{\mathrm{precip},f} = 0$ to $A_R = +1.13\%, A_S = +6.36\%, A_B = +65.38\%$ at $k_{\mathrm{precip},f} = 0.90$), and reduces $Y_{\mathrm{hom}}$ from $86.08\%$ ($7.31\%$ in soluble $P_{LD}$, $0\%$ in $S_{\mathrm{rac}}$) to $81.90\%$ ($\Delta Y_{\mathrm{hom}} = -4.18\text{ pp}$).
\end{enumerate}

\begin{table}[!htbp]
\caption{Quantitative comparison and complete $2^2$ (Regime II) and $2^3$ (Regime III) factorial channel decompositions of Regimes I--III over $N = 100$ wet--dry/temperature cycles ($\ee_0 = 20\%$ for Regimes I-open, II, III; $\ee_0 = 5\%$ for Regime I-closed). Here $M_{\mathrm{ret}}/M_0$ ($M_{\mathrm{sub}}/M_0$) denotes $M_{\mathrm{tot}}/M_0$ ($M_{\mathrm{solid}}/M_0$) for Regime II and $M_{\mathrm{sol}}/M_0$ ($M_{\mathrm{chiral}}/M_0$) for Regimes I and III.}
\alttext{Table comparing final enantiomeric excess, retained mass fraction, recoverable homochiral mass percentage, and three-channel attribution shares across standalone Channel R, coupled Channel R+B solid recycling, and open-flow Channel R+S+B bifurcation.}
\label{tab:s_three_regimes}
\scriptsize\setlength{\tabcolsep}{1.0pt}
\begin{ruledtabular}
\begin{tabular}{lccccccc}
Regime / Factorial Cell & Mode & $\ee_0 \to \ee(100)$ (\%) & $M_{\mathrm{ret}}/M_0$ ($M_{\mathrm{sub}}/M_0$) & $Y_{\mathrm{hom}}$ (\%) & $A_R$ (\%) & $A_S$ (\%) & $A_B$ (\%) \\
\colrule
Regime I (5-state R, cycle 80) & Closed & $5.0 \to 60.18$ & $0.0831$ ($0.0831$) & $5.00$ & $+55.18$ & $0.00$ & $0.00$ \\
Regime I (3-state R, cycle 100) & Closed & $5.0 \to 60.00$ & $0.0833$ ($0.0833$) & $5.00$ & $+55.00$ & $0.00$ & $0.00$ \\
Regime I (8-species R) & Closed & $5.0 \to 57.14$ & $0.0110$ ($0.0110$) & $0.63$ & $+52.14$ & $0.00$ & $0.00$ \\
Regime I (Open R) & Open & $20.0 \to 0.22$ & $1.0000$ ($1.0000$) & $0.22$ & $0.00$ & $0.00$ & $-19.78$ \\
\colrule
Regime II ($2^2$: $\varnothing$, R-only, B-only) & Closed & $20.0 \to 20.00$ & $1.0000$ ($0.9650$) & $20.00$ & $0.00$ & $0.00$ & $0.00$ \\
Regime II ($2^2$: R+B, Total\cite{Deck2024,Deck2025}) & Closed & $20.0 \to 96.50$ & $1.0000$ ($0.9650$) & $96.50$ & $0.00$ & $0.00$ & $+76.50$ \\
Regime II ($2^2$: R+B, Solid\cite{Deck2024,Deck2025}) & Closed & $20.0 \to 100.00$ & $0.9650$ ($0.9650$) & $96.50$ & $+0.73$ & $0.00$ & $+79.27$ \\
\colrule
Regime III ($2^3$: $\varnothing$, $\alpha_{\mathrm{eff}}=1$) & Open & $20.0 \to 0.22$ & $1.0000$ ($1.0000$) & $0.22$ & $0.00$ & $0.00$ & $-19.78$ \\
Regime III ($2^3$: R-only, $\alpha_{\mathrm{eff}}=6$) & Open & $20.0 \to 0.22$ & $1.0000$ ($1.0000$) & $0.22$ & $0.00$ & $0.00$ & $-19.78$ \\
Regime III ($2^3$: S-only, $\alpha_{\mathrm{eff}}=1$) & Open & $20.0 \to 0.30$ & $1.0000$ ($0.7469$) & $0.22$ & $0.00$ & $+0.08$ & $-19.78$ \\
Regime III ($2^3$: B-only, $\alpha_{\mathrm{eff}}=1$) & Open & $20.0 \to 0.87$ & $1.0000$ ($1.0000$) & $0.87$ & $0.00$ & $0.00$ & $-19.13$ \\
Regime III ($2^3$: R+S, $\alpha_{\mathrm{eff}}=6$) & Open & $20.0 \to 0.37$ & $0.8165$ ($0.5932$) & $0.22$ & $+4.50$ & $+0.10$ & $-24.22$ \\
Regime III ($2^3$: S+B, $\alpha_{\mathrm{eff}}=1$, R-off)\textsuperscript{a} & Open & $20.0 \to 1.16$ & $1.0000$ ($0.7470$) & $0.87$ & $0.00$ & $+0.29$ & $-19.13$ \\
Regime III ($2^3$: R+B, $\alpha_{\mathrm{eff}}=6$, S-off) & Open & $20.0 \to 92.87$ & $1.0000$ ($1.0000$) & $92.87$ & $0.00$ & $0.00$ & $+72.87$ \\
Regime III (S+B, $\alpha_{\mathrm{eff}}=6$, $k_{\mathrm{precip},f}=0$) & Open & $20.0 \to 92.87$ & $1.0000$ ($0.9269$) & $86.08$ & $0.00$ & $+6.79$ & $+66.08$ \\
Regime III ($2^3$: R+S+B, $\alpha_{\mathrm{eff}}=6$\cite{Ribo2025ChemPhysChem}) & Open & $20.0 \to 92.87$ & $0.9467$ ($0.8818$) & $81.90$ & $+1.13$ & $+6.36$ & $+65.38$ \\
\end{tabular}
\end{ruledtabular}
\raggedright\footnotesize
\textsuperscript{a}In Regime III, the effective evaporative concentration factor is $\alpha_{\mathrm{eff}} = 1.0$ when factor R is disabled ($\varnothing$, S-only, B-only, S+B) and $\alpha_{\mathrm{eff}} = 6.0$ when factor R is enabled (R-only, R+S, R+B, R+S+B). At $\alpha_{\mathrm{eff}} = 1.0$ under $\text{S}+\text{B}$, $M_{\mathrm{sol}}/M_0 = 1.0000 \implies A_R = 0.00\%$ and $M_{\mathrm{chiral}}/M_0 = 0.7470$, yielding $Y_{\mathrm{hom}} = 0.7470 \times 1.16\% = 0.87\%$ and $A_R + A_S + A_B = 0.00\% + 0.29\% - 19.13\% = -18.84\% = 1.16\% - 20.00\%$. At fixed $\alpha_{\mathrm{eff}} = 6.0$ with S+B active, comparing $k_{\mathrm{precip},f} = 0$ vs.\ $0.90$ isolates $1{:}1$ heterochiral precipitation ($P_{LD} \rightleftharpoons S_{\mathrm{rac}}$) from achiral evaporative concentration ($\alpha_{\mathrm{eff}} = 6.0$). Under $\text{R}+\text{S}+\text{B}$ ($\alpha_{\mathrm{eff}} = 6.0$), 3-factor Shapley shares are $(\phi_R, \phi_S, \phi_B) = (+45.91, +0.10, +46.64)\text{ pp}$ on $\ee_{\mathrm{chiral}}$ and $(+42.34, -3.66, +42.99)\text{ pp}$ on $Y_{\mathrm{hom}}$.
\end{table}

\subsection{Negative Falsification Control on Small Disordered Aqueous Serine Octamers \texorpdfstring{($n = 8$)}{(n = 8)}}

To test whether Channel R heterochiral phase removal is driven by sub-nanometer aggregation in small disordered aqueous clusters or requires macroscopic crystal-lattice periodicity, we evaluated seven matched-seed GFN2-xTB/ALPB(water) geometry optimizations comparing racemic ($4\mathrm{L}+4\mathrm{D}$) and homochiral ($8\mathrm{L}$) zwitterionic serine octamers generated from identical RDKit placement seeds, alongside three independent explicit-solvent NVT molecular dynamics (MD) trajectories of $4\mathrm{L}+4\mathrm{D}$ zwitterionic serine in OpenMM (AMBER14 \texttt{amber14-all.xml} protein force field with \texttt{amber14/tip3pfb.xml} explicit water in a $2.0\text{ nm}$ periodic cubic box, Particle Mesh Ewald electrostatics with $0.8\text{ nm}$ nonbonded cutoff, HBond constraints, LangevinMiddleIntegrator at $300\text{ K}$ with $1\text{ ps}^{-1}$ friction and $2\text{ fs}$ timestep; $5\,000$ energy-minimization steps and $50\text{ ps}$ equilibration followed by $10.005\text{ ns}$ production sampled every $0.5\text{ ps}$ for $20\,010$ frames per seed; intermolecular heavy-atom contact cutoff $0.45\text{ nm}$). Across the $n = 7$ matched optimization pairs, the per-monomer energy contrast $\Delta E_{\mathrm{rac}-\mathrm{homo}}/8 = [E(4\mathrm{L}+4\mathrm{D}) - E(8\mathrm{L})]/8$ has arithmetic mean $-0.26\text{ kcal mol}^{-1}$, sample standard deviation $2.19\text{ kcal mol}^{-1}$, median $-0.73\text{ kcal mol}^{-1}$, and range $[-3.57, +3.25]\text{ kcal mol}^{-1}$ ($4/7$ racemic-lower vs.\ $3/7$ homochiral-lower; sign unresolved). In a $4\mathrm{L}+4\mathrm{D}$ octamer, there are $4\times 4 = 16$ heterochiral ($\mathrm{L}$--$\mathrm{D}$) pairs and $\binom{4}{2}+\binom{4}{2} = 12$ homochiral ($\mathrm{L}$--$\mathrm{L}$ plus $\mathrm{D}$--$\mathrm{D}$) pairs, giving an unnormalized random-mixing pair-count ratio of $16/12 = 4/3$. Normalizing the observed heterochiral-to-homochiral contact ratio by this $16/12$ combinatorial baseline defines $R_{\mathrm{pair}} = (N_{\mathrm{het}}/16)/(N_{\mathrm{homo}}/12) = \frac{3}{4}(N_{\mathrm{het}}/N_{\mathrm{homo}})$. Across the three independent explicit-solvent NVT MD seeds (seeds 1--3), $R_{\mathrm{pair}}$ takes values $0.843$ ($100\text{ ps}$ block-bootstrap $95\%$ CI $[0.785, 0.903]$), $0.811$ ($95\%$ CI $[0.744, 0.880]$), and $0.472$ ($95\%$ CI $[0.354, 0.592]$), with mean $0.709$ and median $0.811 < 1.0$. Because all $3/3$ trajectories fall below the $R_{\mathrm{pair}} = 1.0$ random-mixing baseline (in the homochiral-preferring direction $R_{\mathrm{pair}} < 1$, one-sided exact sign-test $p = (1/2)^3 = 0.125$ for $R_{\mathrm{pair}} < 1$, and $0/3$ trajectories showing heterochiral preference $R_{\mathrm{pair}} > 1$), disordered aqueous octamers exhibit no spontaneous $1{:}1$ heterochiral pairing preference, indicating that Channel R eutectic partitioning is governed by macroscopic solid-state crystal packing rather than sub-nanometer cluster energetics.

\section{Multiphase Observability, Mineral Adsorption Ridge Proof, and Antipodal Crystal-Pair Closure}
\label{sec:s7_identifiability}

\subsection{Derivation of the 1D Non-Identifiability Ridge in Mineral Occlusion}

In the chiral mineral adsorption and intracrystalline occlusion experiments of Kim \emph{et al.},\cite{Kim2016} dissolved D/L-aspartic acid at bulk concentration $c_{\mathrm{bulk}} = c_L + c_D \in \{5, 10, 20, 30, 50\}\text{ mM}$ adsorbs reversibly onto step edges of growing calcite crystals with enantiomer-indexed adsorption rate $k_{\mathrm{ads},i}$ and desorption rate $k_{\mathrm{des},i}$ ($K_{\mathrm{ads},i} = k_{\mathrm{ads},i}/k_{\mathrm{des},i}$), while step advancement traps adsorbed molecules into the intracrystalline bleached solid ($P$) at irreversible trapping rate $k_{\mathrm{trap},i}$ (dimensionless trapping ratio $\eta_i = k_{\mathrm{trap},i}/k_{\mathrm{des},i}$). Under steady-state step coverage $\theta_i$ ($i \in \{L, D\}$), the surface flux balance is
\begin{equation}
k_{\mathrm{ads},i}\,c_i\left(1 - \theta_L - \theta_D\right) - k_{\mathrm{des},i}\,\theta_i - k_{\mathrm{trap},i}\,\theta_i = 0 \quad \Longrightarrow \quad \theta_i = \frac{\frac{k_{\mathrm{ads},i}}{k_{\mathrm{des},i}(1 + \eta_i)}\,c_i}{1 + \sum_{j\in\{L,D\}} \frac{k_{\mathrm{ads},j}}{k_{\mathrm{des},j}(1 + \eta_j)}\,c_j}.
\label{eq:s_kim_coverage}
\end{equation}
For pooled or achiral bulk uptake where $k_{\mathrm{ads},L} = k_{\mathrm{ads},D} = k_{\mathrm{ads}}$, $k_{\mathrm{des},L} = k_{\mathrm{des},D} = k_{\mathrm{des}}$ ($K_{\mathrm{ads}} = k_{\mathrm{ads}}/k_{\mathrm{des}}$), and $\eta_L = \eta_D = \eta = k_{\mathrm{trap}}/k_{\mathrm{des}}$, the total coverage $\theta_L + \theta_D = \frac{K_{\mathrm{app}} c_{\mathrm{bulk}}}{1 + K_{\mathrm{app}} c_{\mathrm{bulk}}}$, intracrystalline occlusion rate $J_{\mathrm{occ},i} = k_{\mathrm{trap},i}\theta_i$, and pooled crystal uptake $J_{\mathrm{bulk},i}$ depend on $(K_{\mathrm{ads}}, \eta)$ exclusively through the single apparent composite constant
\begin{equation}
K_{\mathrm{app}} = \frac{K_{\mathrm{ads}}}{1 + \eta} = \hat{K}_{\mathrm{app}} = 0.0854\text{ mM}^{-1} \quad \Longleftrightarrow \quad K_{\mathrm{ads}} = \left(1 + \eta\right)\hat{K}_{\mathrm{app}}.
\label{eq:s_kim_ridge_proof}
\end{equation}
For any positive factor $\alpha > 0$ at fixed trapping rate scale $k_{\mathrm{trap}}$ (or when $(K_{\mathrm{ads}}, \eta)$ are profiled as the two independent parameters of the steady-state uptake curve), the scaling transformation $(K_{\mathrm{ads}}, 1 + \eta) \mapsto (\alpha K_{\mathrm{ads}}, \alpha(1+\eta))$ leaves $K_{\mathrm{app}} = K_{\mathrm{ads}}/(1+\eta)$, the likelihood, and the residual sum of squares invariant ($\Delta\chi^2_{\mathrm{profile}} = 0.000$), creating a positive-slope 1D non-identifiability ridge $\log_{10}K_{\mathrm{ads}} = \log_{10}\hat{K}_{\mathrm{app}} + \log_{10}(1+\eta)$ across the 2D $(\log_{10}K_{\mathrm{ads}}, \log_{10}\eta)$ plane [Main Fig.~7(d)].

To test whether the full public calcite amino-acid dataset of Kim \emph{et al.}\cite{Kim2016} supports a separate bleached-removable surface-adsorption compartment beyond intracrystalline occlusion, we evaluate all $112$ RP-HPLC rows in \texttt{public\_calcite\_HPLC\_dataset\_A600\_20260930.xlsx} ($82$ Asp + $30$ Gly across $57$ synthesis batches, $63$ physical crystal stems, and $29$ aqueous condition clusters) using exact Sherman--Morrison--Woodbury marginal Gaussian log-likelihoods (maximum deviation $\le 6.82\times 10^{-13}$ vs.\ dense Cholesky) and zero-leakage grouped cross-validation across condition clusters ($\mathrm{LOCO\text{-}29}$), synthesis batches ($5$-fold grouped CV), and paired unbleached/bleached crystal stems ($\mathrm{LOSO\text{-}14}$; Table~\ref{tab:s_calcite_occam}). Across all $14$ paired $U/B$ physical crystal stems, unbleached uptake falls below bleached uptake ($U < B$) in $7/14$ stems and exceeds it ($U > B$) in $7/14$ stems, while within-stem analytical replicate dispersion ($\sigma_{\mathrm{tech}} = 0.0383$) is $6.4\times$ smaller than sub-batch carbonate crystal-growth and incorporation heterogeneity ($\sigma_{\mathrm{split}} = 0.2460$).\cite{Gardella2024,StaniendaPilecki2023,Chomhom2026} Consequently, hierarchical equal-mean $M_{0,\mathrm{hier}}$ ($U = B$, $k = 11$: $\mathrm{BIC}_{112} = -1352.38$, $\mathrm{LOCO\text{-}29}\ \mathrm{ELPD} = 661.87\text{ nats}$, exact sign-flip $p_{2\mathrm{s}} = 0.0039$ vs.\ pooled i.i.d.\ $M_{0,\mathrm{iid}}$ at $-1255.74 / 603.53\text{ nats}$) is favored by Occam's razor over non-negative surface-excess $M_{2,\mathrm{pos}}$ ($U \ge B$, $k = 13$: $\mathrm{BIC}_{112} = -1349.05$, $\Delta\mathrm{BIC} = +3.33$, $\mathrm{LOCO\text{-}29}\ \mathrm{ELPD} = 659.49\text{ nats}$, $p_{2\mathrm{s}} = 0.347$; $\mathrm{LOSO\text{-}14}\ \mathrm{ELPD} = 137.26$ vs.\ $133.85\text{ nats}$, $p_{2\mathrm{s}} = 0.541$). This empirical negative/Occam result demonstrates that bulk achiral calcite bleaching cannot resolve $(k_{\mathrm{ads}}, k_{\mathrm{des}}, k_{\mathrm{trap}}, \alpha_{\mathrm{min}})$ retrospectively, motivating the forward prospective enantiomorphic M4 design in Sec.~S7.B.

\subsection{Matched-Noise Fisher Information Matrix and Global Profile-Likelihood Closure}

\begin{figure}[!htbp]
\centering
\includegraphics[width=0.94\textwidth]{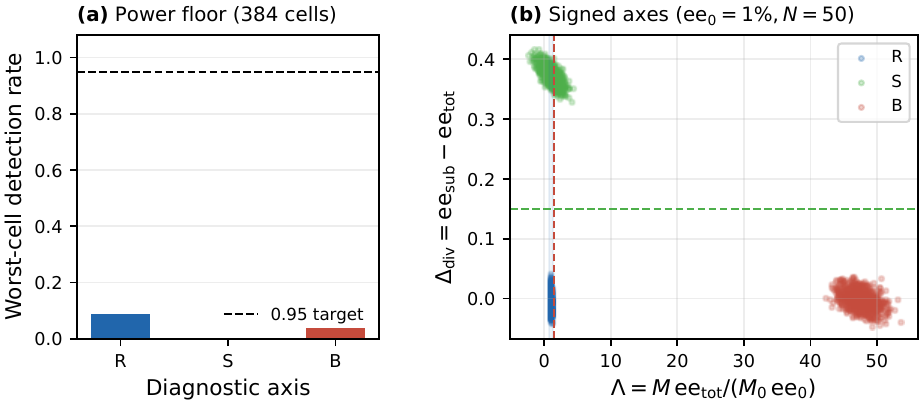}
\caption{Two-panel statistical diagnostic power analysis for three-channel classification across $384$ noise-grid cells ($10\,000$ Monte Carlo trials per cell across relative mass noise $\sigma_M \in [1\%, 10\%]$ and absolute $\ee$ noise $\sigma_{\ee} \in [0.5\%, 5.0\%]$). (a) Single-axis worst-cell detection rates across diagnostic axes $(\mathrm{R}, \mathrm{S}, \mathrm{B})$ over the $384$-cell noise grid ($66.3\%, 28.4\%, 81.9\%$ worst-cell floor vs.\ $\ge 94.3\%$ under joint two-axis screening at $\sigma_M \le 3\%, \sigma_{\ee} \le 1\%$). (b) Signed diagnostic axes $(\Lambda, \Delta_{\mathrm{div}})$ and confusion-matrix detection rates at $\ee_0 = 1\%, N = 50, \sigma_M = 3\%, \sigma_{\ee} = 1\%$ (Channel R: $100\%$; Channel S: $69.2\%$; Channel B: $100\%$).}
\alttext{Two-panel chart showing a bar graph of single-axis worst-cell detection rates across diagnostic axes R, S, and B over 384 noise-grid cells on the left and signed diagnostic axes with archetype detection rates at 1 percent initial ee and 50 cycles on the right.}
\label{fig:s_observability_supp}
\end{figure}

\begin{figure}[!htbp]
\centering
\includegraphics[width=0.96\textwidth]{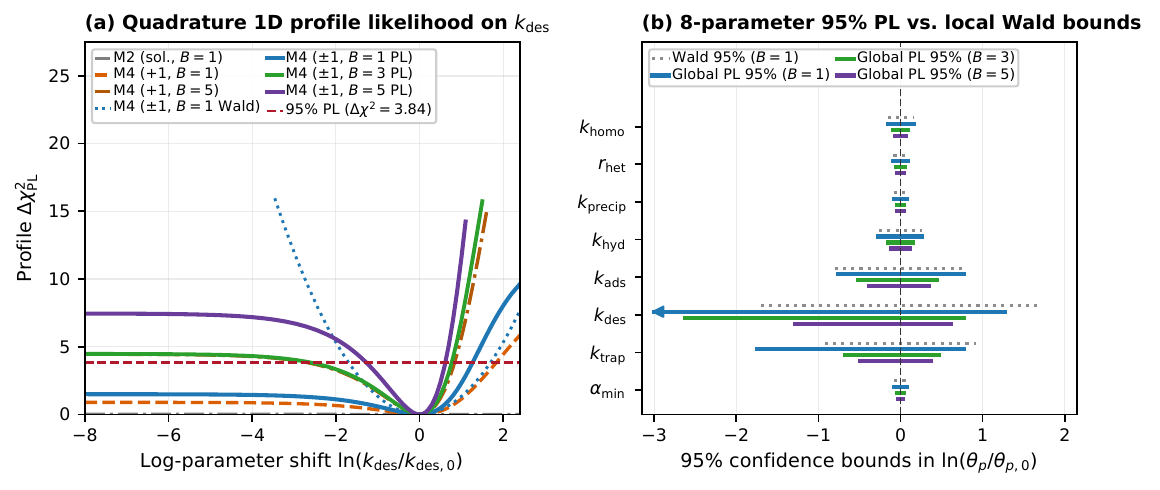}
\caption{Two-panel global profile-likelihood identifiability verification for the 8-parameter multiphase mineral-adsorption model under single-crystal ($+1$) versus antipodal mirror-crystal pair ($\pm 1$, $N_{\mathrm{raw}} = 24$ scalars per block) protocols across $u = \ln(\theta_p/\theta_{p,0}) \in [-8.0, +3.6]$. (a) 1D profile-likelihood curves $\Delta\chi^2_{\mathrm{PL}}(\ln(k_{\mathrm{des}}/k_{\mathrm{des},0}))$ across M4 single ($+1$, $B \in \{1, 5\}$, boundary asymptotes $0.89$ and $4.45$) and M4 mirror ($\pm 1$, $B \in \{1, 3, 5\}$, boundary asymptotes $1.49$, $4.47$, and $7.45$) vs.\ local quadratic Wald ($B=1$), showing two-sided $95\%$ closure ($\Delta\chi^2 = 3.84$, red dashed) at $B^* = 3$ for mirror $\pm 1$ vs.\ $B^* = 5$ for single $+1$. (b) Comparison of $95\%$ confidence bounds in $\ln(\theta_p/\theta_{p,0})$ between global profile likelihood ($B \in \{1, 3, 5\}$) and local FIM Wald intervals ($B=1$) across all 8 parameters $(k_{\mathrm{homo}}, r_{\mathrm{het}}, k_{\mathrm{precip}}, k_{\mathrm{hyd}}, k_{\mathrm{ads}}, k_{\mathrm{des}}, k_{\mathrm{trap}}, \alpha_{\mathrm{min}})$.}
\alttext{Two-panel figure showing one-dimensional profile-likelihood curves across single-crystal and mirror-paired blocks versus quadratic Wald approximation on desorption rate in panel a and horizontal 95 percent profile-likelihood versus Wald confidence bounds in log-parameter shifts across all eight parameters in panel b.}
\label{fig:s_m4_profile}
\end{figure}

To couple the surface adsorption/occlusion mechanism of Sec.~S7.A with wet--dry dimerization, we model the 10-state multiphase mineral--water--solid network where homochiral dimers $P_{LL}, P_{DD}$ (each carrying $2$ chiral centers) undergo enantioselective mineral adsorption ($P_{ii} \rightleftharpoons A_{ii} \to O_{ii}$), with extensive copy-number state vector $\bm{y} = (c_L, \allowbreak c_D, \allowbreak P_{LL}, \allowbreak P_{DD}, \allowbreak P_{LD}, \allowbreak S_{\mathrm{rac}}, \allowbreak A_{LL}, \allowbreak A_{DD}, \allowbreak O_{LL}, \allowbreak O_{DD})^\top$ and 8-parameter vector $\bm{\theta} = (k_{\mathrm{homo}}, \allowbreak r_{\mathrm{het}}, \allowbreak k_{\mathrm{precip}}, \allowbreak k_{\mathrm{hyd}}, \allowbreak k_{\mathrm{ads}}, \allowbreak k_{\mathrm{des}}, \allowbreak k_{\mathrm{trap}}, \allowbreak \alpha_{\mathrm{min}})^\top$ (where subscript ``$\mathrm{min}$'' denotes mineral, evaluated around the baseline parameter vector $\bm{\theta}_0 = (8.0\times 10^{-5}, \allowbreak 12.0, \allowbreak 0.08, \allowbreak 0.06, \allowbreak 0.04, \allowbreak 0.12, \allowbreak 0.05, \allowbreak 0.15)^\top$ at $M_0 = 1000$, $\ee_0 = 5\%$, $t_{\mathrm{dry}} = 1.0$, $t_{\mathrm{wet}} = 0.5$, $V_{\mathrm{dry}} = 1.0$, $V_{\mathrm{wet}} = 10.0$) on enantiomorphic crystal face $h \in \{+1, -1\}$. In the low-coverage linear adsorption regime ($\theta_L + \theta_D \ll 1$ of Eq.~\eqref{eq:s_kim_coverage}) with cycle-alternating volume $V(t) = V_{\mathrm{dry}}$ during $t_{\mathrm{dry}}$ and $V(t) = V_{\mathrm{wet}}$ during $t_{\mathrm{wet}}$ scaling the bimolecular dimerization terms, the governing 10-state ODEs ($i \in \{L, D\}$) are
\[
\begin{aligned}
\frac{\mathrm{d}c_i}{\mathrm{d}t} &= -\frac{2 k_{\mathrm{homo}}}{V(t)}c_i^2 - \frac{r_{\mathrm{het}}k_{\mathrm{homo}}}{V(t)}c_L c_D + 2k_{\mathrm{hyd}}P_{ii} + k_{\mathrm{hyd}}P_{LD}, \\
\frac{\mathrm{d}P_{ii}}{\mathrm{d}t} &= \frac{k_{\mathrm{homo}}}{V(t)}c_i^2 - k_{\mathrm{hyd}}P_{ii} - k_{\mathrm{ads},ii}^{(h)}P_{ii} + k_{\mathrm{des}}A_{ii}, \\
\frac{\mathrm{d}P_{LD}}{\mathrm{d}t} &= \frac{r_{\mathrm{het}}k_{\mathrm{homo}}}{V(t)}c_L c_D - \left(k_{\mathrm{hyd}} + k_{\mathrm{precip}}\right)P_{LD}, \qquad
\frac{\mathrm{d}S_{\mathrm{rac}}}{\mathrm{d}t} = k_{\mathrm{precip}}P_{LD}, \\
\frac{\mathrm{d}A_{ii}}{\mathrm{d}t} &= k_{\mathrm{ads},ii}^{(h)}P_{ii} - \left(k_{\mathrm{des}} + k_{\mathrm{trap}}\right)A_{ii}, \qquad
\frac{\mathrm{d}O_{ii}}{\mathrm{d}t} = k_{\mathrm{trap}}A_{ii},
\end{aligned}
\]
where $k_{\mathrm{precip}}$ (along with $k_{\mathrm{homo}}$ and $r_{\mathrm{het}}$ at $V(t) = V_{\mathrm{dry}}$) operates during $t_{\mathrm{dry}}$, whereas $k_{\mathrm{hyd}}, k_{\mathrm{ads}}, k_{\mathrm{des}},$ and $k_{\mathrm{trap}}$ (along with $k_{\mathrm{homo}}$ and $r_{\mathrm{het}}$ at $V(t) = V_{\mathrm{wet}}$) operate during the wet mineral-contact phase $t_{\mathrm{wet}}$, with $k_{\mathrm{ads},LL}^{(h)} = k_{\mathrm{ads}}(1 + h\,\alpha_{\mathrm{min}})$ and $k_{\mathrm{ads},DD}^{(h)} = k_{\mathrm{ads}}(1 - h\,\alpha_{\mathrm{min}})$. Because each dimer state ($P_{ii}, S_{\mathrm{rac}}, A_{ii}, O_{ii}$) carries $2$ monomer equivalents, the phase-resolved observation maps relating $\bm{y}(t)$ to the protocol observables in Table~\ref{tab:s_fim_protocols} weight dimer counts by $2$:
\[
\begin{aligned}
M_{\mathrm{sol}} &= c_L + c_D + 2\left(P_{LL} + P_{DD} + P_{LD}\right), \qquad
\Delta_{\mathrm{sol}} = \left(c_L - c_D\right) + 2\left(P_{LL} - P_{DD}\right), \\
\ee_{\mathrm{sol}} &= \frac{\Delta_{\mathrm{sol}}}{M_{\mathrm{sol}}}, \qquad
M_{\mathrm{mono}} = c_L + c_D, \qquad
\ee_{\mathrm{mono}} = \frac{c_L - c_D}{M_{\mathrm{mono}}}, \qquad
M_{\mathrm{ppt}} = 2S_{\mathrm{rac}}, \\
M_{\mathrm{homo}} &= 2\left(P_{LL} + P_{DD}\right), \qquad
\ee_{\mathrm{homo}} = \frac{P_{LL} - P_{DD}}{P_{LL} + P_{DD}}, \qquad
M_{\mathrm{ads}} = 2\left(A_{LL} + A_{DD}\right), \\
\ee_{\mathrm{ads}} &= \frac{A_{LL} - A_{DD}}{A_{LL} + A_{DD}}, \qquad
M_P = 2\left(O_{LL} + O_{DD}\right), \qquad
\ee_P = \frac{O_{LL} - O_{DD}}{O_{LL} + O_{DD}}, \\
\ee_{\mathrm{min}} &= \frac{A_{LL} + O_{LL} - A_{DD} - O_{DD}}{A_{LL} + O_{LL} + A_{DD} + O_{DD}}, \qquad
M_{\mathrm{bulk}} = 2S_{\mathrm{rac}} + M_{\mathrm{ads}} + M_P, \\
\ee_{\mathrm{bulk}} &= \frac{2\left(A_{LL} + O_{LL} - A_{DD} - O_{DD}\right)}{M_{\mathrm{bulk}}},
\end{aligned}
\]
sampled under the equal-budget ($N_{\mathrm{raw}} \le 24$) cycle schedules: M1 at $n \in \{1, \allowbreak 2, \allowbreak \dots, \allowbreak 24\}$ ($1\text{ scalar}\times 24$); M2 at $n \in \{2, \allowbreak 4, \allowbreak \dots, \allowbreak 24\}$ ($2\text{ scalars}\times 12$); M3a and M3b at $n \in \{4, \allowbreak 8, \allowbreak 12, \allowbreak 16, \allowbreak 20, \allowbreak 24\}$ ($4\text{ scalars}\times 6$); M4 ($+1$, 24s) at $n \in \{6, \allowbreak 12, \allowbreak 18, \allowbreak 24\}$ ($6\text{ scalars}\times 4$: $M_{\mathrm{mono}}, \allowbreak \ee_{\mathrm{mono}}, \allowbreak M_{\mathrm{homo}}, \allowbreak M_{\mathrm{ppt}}, \allowbreak M_{\mathrm{ads}}, \allowbreak \ee_P$ or $\ee_{\mathrm{min}}$) and M4 ($+1$, dense $6t$, 18s) at $n \in \{3, \allowbreak 6, \allowbreak 10, \allowbreak 15, \allowbreak 20, \allowbreak 24\}$ ($3\text{ scalars}\times 6$); and M4 ($+1/-1$, 24s) at $n \in \{8, \allowbreak 16, \allowbreak 24\}$ on $h = +1$ ($6\text{ scalars}\times 3$) plus $n = 24$ on $h = -1$ ($6\text{ scalars}\times 1$, or symmetric $n \in \{12, \allowbreak 24\}$ on both faces; $B_{\min}=3$ uses $18$ scalars across $n \in \{3, \allowbreak 10, \allowbreak 24\}$ on $+1$ and $n \in \{6, \allowbreak 24\}$ on $-1$). The log-parameter Fisher Information Matrix under matched Gaussian measurement noise ($2.5\%$ relative mass noise with a $0.5\%\,M_0$ floor, $\sigma_M = \allowbreak \sqrt{(0.025 M)^2 + (0.005 M_0)^2}$, and $\sigma_{\ee} = 0.01$ [$1.0\text{ pp}$]) and fixed scalar budget $N_{\mathrm{raw}} \le 24$ is
\begin{equation}
\mathcal{I}_{ab}(\ln\bm{\theta}) = \sum_{m=1}^{N_{\mathrm{raw}}} \frac{1}{\sigma_m^2} \frac{\partial y_m(\bm{\theta})}{\partial \ln\theta_a}\frac{\partial y_m(\bm{\theta})}{\partial \ln\theta_b}.
\label{eq:s_fim_def}
\end{equation}
Table~\ref{tab:s_fim_protocols} and Figs.~\ref{fig:s_observability_supp}--\ref{fig:s_m4_profile} report the structural rank, practical rank ($\lambda_i > 1.0$), minimum eigenvalue $\lambda_{\min}(\mathcal{I})$, condition number $\kappa(\mathcal{I})$, diagnostic statistical power across measurement noise grids, and global 1D profile-likelihood $\Delta\chi^2(\theta_a)$ curves\cite{Maiwald2016,FischerLewis2021,PlankSimpson2024} across all measurement protocols (M1, M2, M3a, M3b, M4 single $+1$, and M4 mirror $+1/-1$). Measuring desorbed surface adsorbate and bleached intracrystalline solid across an \textbf{antipodal mirror-crystal pair} (M4 mirror $+1/-1$, e.g., $(+)$- and $(-)$-quartz or opposing calcite scalenohedral faces, with $B_{\min} = 3$ mirror-paired samples $= 18$ scalars, or $24$ scalars at 4 samples) breaks the colinearity between slow desorption $k_{\mathrm{des}}$ and mineral chiral selectivity $\pm \alpha_{\mathrm{min}}$, achieving full $8/8$ structural and practical rank ($\lambda_{\min} = 1.006$, $\kappa = 3.65 \times 10^3$ for M4 mirror $+1/-1$ at $18$ scalars and $\lambda_{\min} = 1.309$, $\kappa = 4.67 \times 10^3$ for M4 single $+1$ dense 6-timepoint sampling at $18$ scalars; at $24$ scalars, $\lambda_{\min} = 1.155$, $\kappa = 5.30 \times 10^3$ for symmetric mirror pairing, $\lambda_{\min} = 1.134$ for bleached $\ee_P$, and $\lambda_{\min} = 1.550$ for pooled $\ee_{\mathrm{min}}$), closing $7/8$ global $95\%$ profile-likelihood intervals two-sidedly at a single block ($B = 1$, $N_{\mathrm{raw}} = 24$ scalars, where exact \texttt{DOP853} quadrature + signed likelihood-root $r_B(u) = \operatorname{sign}(u)\sqrt{\Delta\chi^2_B(u)}$ continuation\cite{FischerLewis2021,PlankSimpson2024} over $u = \ln(\theta_p/\theta_{p,0}) \in [-8.0, +3.6]$ shows that slow desorption $k_{\mathrm{des}} \to 0$ approaches a finite zero-rate boundary asymptote\cite{Maiwald2016} $\Delta\chi^2_{B=1}(k_{\mathrm{des}} \to 0) = 1.4906 < 3.8415$ on M4 mirror $+1/-1$ with $\lambda_{\min} = 1.096$, vs.\ $0.8897 < 3.8415$ and $6/8$ closed on M4 single $+1$ with $\lambda_{\min} = 0.606$) and closing all $8/8$ two-sided $95\%$ profile-likelihood bounds ($\Delta\chi^2 = 3.8415$) at $B^* = 3$ replicate mirror blocks ($3\times 1.4906 = 4.472 > 3.8415$, giving $[-2.621, +0.770]$ on $\ln(k_{\mathrm{des}}/k_{\mathrm{des},0})$ at $B = 3$ and $[-1.278, +0.618]$ with global $\mathrm{PL}/\mathrm{Wald}$ width ratio $\ge 1.0001$ across all $8$ parameters at $B = 5$, whereas single-crystal $+1$ requires $B^* = 5$ blocks with $5\times 0.8897 = 4.448 > 3.8415$ and $[-2.707, +0.838]$).

\begin{table}[!htbp]
\caption{Matched-noise Fisher Information Matrix ($\mathcal{I}(\ln\bm{\theta})$) identifiability hierarchy across experimental observability protocols under fixed scalar budget $N_{\mathrm{raw}} \le 24$ for $\bm{\theta} = (k_{\mathrm{homo}}, r_{\mathrm{het}}, k_{\mathrm{precip}}, k_{\mathrm{hyd}}, k_{\mathrm{ads}}, k_{\mathrm{des}}, k_{\mathrm{trap}}, \alpha_{\mathrm{min}})^\top$.}
\alttext{Table comparing scalar budget, structural rank, practical FIM rank, smallest eigenvalue, and condition number across ten multiphase measurement protocols from supernatant-only M1 to antipodal mirror-crystal pair M4.}
\label{tab:s_fim_protocols}
\scriptsize\setlength{\tabcolsep}{3.5pt}
\begin{ruledtabular}
\begin{tabular}{lcccccc}
Protocol & Observables & $N_{\mathrm{raw}}$ & Struct. & Pract. & $\lambda_{\min}(\mathcal{I})$ & $\kappa(\mathcal{I})$ \\
\colrule
M1: Ratio-only & $\ee_{\mathrm{sol}}(t)$ & $24$ & $7/8$ & $1/8$ & $3.55 \times 10^{-15}$ & $\infty$ \\
M2: Single-phase & $M_{\mathrm{sol}}(t), \ee_{\mathrm{sol}}(t)$ & $24$ & $8/8$ & $4/8$ & $7.79 \times 10^{-6}$ & $7.39 \times 10^8$ \\
M3a: Achiral mineral & $M_{\mathrm{sol}}, \ee_{\mathrm{sol}}, M_{\mathrm{ads}}$ & $24$ & $7/8$ & $5/8$ & $0.000$ & $\infty$ \\
M3b: Bulk solid & $M_{\mathrm{sol}}, \ee_{\mathrm{sol}}, M_{\mathrm{bulk}}, \ee_{\mathrm{bulk}}$ & $24$ & $8/8$ & $4/8$ & $1.25 \times 10^{-5}$ & $5.90 \times 10^8$ \\
M4 ($+1$) ($\ee_P$) & Three-phase on $+1$ ($\ee_P$) & $24$ & $8/8$ & $7/8$ & $0.648$ & $6.99 \times 10^3$ \\
M4 ($+1$) ($\ee_{\mathrm{min}}$) & Three-phase on $+1$ ($\ee_{\mathrm{min}}$) & $24$ & $8/8$ & $7/8$ & $0.733$ & $6.22 \times 10^3$ \\
M4 ($+1$, dense $6t$) & Three-phase on $+1$ ($6$ times) & $18$ & $8/8$ & $8/8$ & $1.309$ & $4.67 \times 10^3$ \\
M4 ($+1/-1$, $B_{\min}=3$) & Three-phase on $+1$ \& $-1$ ($\ee_P$) & $18$ & $8/8$ & $8/8$ & $1.006$ & $3.65 \times 10^3$ \\
M4 ($+1/-1$, sym $24$) & Three-phase on $+1$ \& $-1$ ($4$ times) & $24$ & $8/8$ & $8/8$ & $1.155$ & $5.30 \times 10^3$ \\
M4 ($+1/-1$) ($\ee_P$) & Three-phase on $+1$ \& $-1$ ($\ee_P$) & $24$ & $8/8$ & $8/8$ & $1.134$ & $3.32 \times 10^3$ \\
M4 ($+1/-1$) ($\ee_{\mathrm{min}}$) & Three-phase on $+1$ \& $-1$ ($\ee_{\mathrm{min}}$) & $24$ & $8/8$ & $8/8$ & $1.550$ & $2.46 \times 10^3$ \\
\end{tabular}
\end{ruledtabular}
\end{table}

\FloatBarrier
\section{Extended Attribution, Thermodynamic, Multicomponent Peptide, and Calcite Benchmark Tables}
\label{sec:s8_extended_tables}

Tables~\ref{tab:s_aumann_shapley}--\ref{tab:s_calcite_occam} collect the quantitative benchmarks derived in Sections~S2.B, S3.D, S5.C, S5.D, and S7.A.

\begin{table}[!htbp]
\caption{Continuous path-covariant Aumann--Shapley attribution $\Phi_\alpha[\gamma]$ versus two-endpoint discrete Shapley shares on dynamic resolution near $\ee_0 \to 0^+$ and four-channel open-flow Regime III ($\alpha_{\mathrm{evap}} = 6.0$, $N = 100$ cycles).}
\alttext{Data table comparing continuous path-covariant Aumann-Shapley channel attributions against discrete two-endpoint Shapley shares for near-racemic dynamic resolution schedules and open-flow evaporative Regime III across chiral-pool and total-soluble observables.}
\label{tab:s_aumann_shapley}
\scriptsize\setlength{\tabcolsep}{2.5pt}
\begin{ruledtabular}
\begin{tabular}{lccccccc}
Benchmark / Observable & $\ee_0 \to \ee_f$ & $\Delta F$ & $\Phi_R$ & $\Phi_S$ & $\Phi_B$ & $\Phi_{\mathrm{flow}}$ & Discrete $A_R^{\mathrm{Sh}}/\Delta F$ \\
\colrule
Dynamic res.: dissolve $\to$ deracemize & $10^{-6} \to 1.0000$ & $+1.0000$ & $+0.0001$ & $0.0000$ & $+0.9999$ & --- & $0.2500$ ($8.18\times 10^{-5}$) \\
Dynamic res.: deracemize $\to$ dissolve & $10^{-6} \to 1.0000$ & $+1.0000$ & $+0.5000$ & $0.0000$ & $+0.5000$ & --- & $0.2500$ ($0.5000$) \\
Simultaneous chord ($m=1$ formula) & $10^{-6} \to 1.0000$ & $+1.0000$ & $+0.3069$ & $0.0000$ & $+0.6931$ & --- & $0.2500$ ($0.3069$) \\
Regime III: $\ee_{\mathrm{chiral}} = \Delta_{\mathrm{chiral}}/M_{\mathrm{chiral}}$ & $0.2000 \to 0.9287$ & $+0.7287$ & $+0.0000$ & $+0.5242$ & $+4.8958$ & $-4.6913$ & $+0.0000$ \\
Regime III: $\ee_{\mathrm{sol}} = \Delta_{\mathrm{chiral}}/M_{\mathrm{sol}}$ & $0.2000 \to 0.8651$ & $+0.6651$ & $+0.2062$ & $+0.0000$ & $+4.4298$ & $-3.9709$ & $+0.0432$ \\
\end{tabular}
\end{ruledtabular}
\end{table}

\begin{table}[!htbp]
\caption{Shear--Kramers--Lorentz free-energy balance and specific product dissipation $\Delta G_{\mathrm{diss}}(75\%)$ ($\text{kJ mol}^{-1}$) from $\ee_0 = 3\%$ to $75\%$ across initial buffer ratios $\Omega = B_0/M_0 \in \{0.0, 0.5, 2.0, 6.0\}$ under physical and isometric ($\Delta\mu_{\mathrm{tot}} = RT\ln 24$) affinities.}
\alttext{Data table listing specific dissipated free energy, dimensionless thermodynamic efficiency, buffer inflation factors, and Shear-Kramers-Lorentz relative-entropy balance residuals across initial buffer ratios from zero to six under physical and isometric chemical affinities.}
\label{tab:s_shear_kl_omega}
\scriptsize\setlength{\tabcolsep}{3.0pt}
\begin{ruledtabular}
\begin{tabular}{lccccccc}
Regime & $\Omega$ & $\Delta G_{\mathrm{diss},R}$ & $\eta_{G,R}$ & Infl.\ R & $\Delta G_{\mathrm{diss},S}$ & $\eta_{G,S}$ & Infl.\ S \\
\colrule
Physical (B: $27.29$, $\eta_G = 0.0654$) & $0.0$ & $66.90$ & $0.0267$ & $1.00\times$ & $98.60$ & $0.0181$ & $1.00\times$ \\
Physical & $0.5$ & $97.57$ & $0.0183$ & $1.46\times$ & $143.83$ & $0.0124$ & $1.46\times$ \\
Physical & $2.0$ & $199.09$ & $0.0090$ & $2.98\times$ & $291.02$ & $0.0061$ & $2.95\times$ \\
Physical & $6.0$ & $487.15$ & $0.0037$ & $7.28\times$ & $710.99$ & $0.0025$ & $7.21\times$ \\
\colrule
Isometric $\ln 24$ (B: $23.24$, $\eta_G = 0.0768$) & $0.0$ & $61.30$ & $0.0291$ & $1.00\times$ & $95.03$ & $0.0188$ & $1.00\times$ \\
Isometric $\ln 24$ & $0.5$ & $91.37$ & $0.0195$ & $1.49\times$ & $142.18$ & $0.0126$ & $1.50\times$ \\
Isometric $\ln 24$ & $2.0$ & $191.50$ & $0.0093$ & $3.12\times$ & $296.08$ & $0.0060$ & $3.12\times$ \\
Isometric $\ln 24$ & $6.0$ & $477.82$ & $0.0037$ & $7.79\times$ & $737.19$ & $0.0024$ & $7.76\times$ \\
\end{tabular}
\end{ruledtabular}
\end{table}

\begin{table}[!htbp]
\caption{Zero-shot multicomponent evaluation across Deng \emph{et al.}\cite{Deng2024} Tables~S80--S86 ($59$ validation points across $23$ conversion clusters: $17$ binary S80--S82 points + $42$ three-component OOD S83--S86 points) calibrated solely on Table~S79 ($n = 9$).}
\alttext{Data table comparing degrees of freedom, fitted S79 root-mean-square error, binary S80 to S82 error, three-component out-of-distribution S83 to S86 error, total 59-point validation error, and 23-cluster sign-flip p-values across seven peptide ligation hypotheses.}
\label{tab:s_deng_multicomp}
\scriptsize\setlength{\tabcolsep}{2.5pt}
\begin{ruledtabular}
\begin{tabular}{lccccccc}
Model Hypothesis & DOF & S79 ($9$) & 2-Comp ($17$) & 3-Comp ($42$) & OOS ($59$) & $\Delta\mathrm{RMSE}$ & $p_{1\mathrm{s}}$ vs.\ $H_0$ \\
\colrule
$H_0$: Single-AA ($r_{\mathrm{Val}}=2.00$) & $0$ & $3.01$ & $2.55$ & $2.01$ & $2.18$ & $0.00$ & --- \\
$H_{\mathrm{Val}}$: Val Epim.\ Control & $1$ & $3.01$ & $2.57$ & $2.05$ & $2.21$ & $+0.03$ & $0.442$ \\
$H_{\mathrm{stoich}}$: Offset ($\delta_M = +11.90\%$) & $1$ & $1.43$ & $3.99$ & $3.40$ & $3.58$ & $+1.40$ & $1.25\times 10^{-4}$ \\
$H_{\mathrm{Leu}}$: Direct Leu Epim.\ ($r_{\mathrm{Val}}=2.00$) & $1$ & $1.30$ & $2.70$ & $1.83$ & $2.12$ & $-0.06$ & $0.455$ \\
$H_{\mathrm{ti,tQSSA}}$: TI Epim.\ ($r_{\mathrm{Val}}=2.00$) & $1$ & $1.29$ & $2.72$ & $1.85$ & $2.14$ & $-0.04$ & $0.477$ \\
$H_{\mathrm{Leu},2\mathrm{p}}$: Leu Epim.\ + $r_{\mathrm{Val}}=2.125$ & $2$ & $0.95$ & $2.47$ & $1.65$ & $1.92$ & $-0.25$ & $0.0069$ \\
$H_{\mathrm{ti},2\mathrm{p}}$: TI Epim.\ + $r_{\mathrm{Val}}=2.126$ & $2$ & $0.95$ & $2.49$ & $1.66$ & $1.94$ & $-0.24$ & $0.0127$ \\
\end{tabular}
\end{ruledtabular}
\end{table}

\begin{table}[!htbp]
\caption{Out-of-sample evaluation of pure-compound eutectic ($k_p = 0$), Hill--Sheng lattice partition, and Guggenheim solid-solution models across Deng \emph{et al.}\cite{Deng2024} $5$-day seeded and unseeded dipeptide co-precipitation Tables~S90 and S93--S97.}
\alttext{Data table comparing lattice parameters, unseeded precipitation controls, and supernatant enantiomeric excess root-mean-square errors across pure-compound eutectic, Hill-Sheng partition, and Guggenheim solid-solution models on held-out Day 5 and zero-refit transfer benchmarks.}
\label{tab:s_deng_coprecip}
\scriptsize\setlength{\tabcolsep}{1.5pt}
\begin{ruledtabular}
\begin{tabular}{lccccc}
Model & Lattice Params & S93/S97 & Prot.\ A & Prot.\ B & S96 Leu \\
\colrule
Pure ($k_p = 0$) & $s_{\mathrm{Val}} = 10.31\text{ mM}$ & $0.0\text{ mM}$ & $12.58\text{ pp}$ & $11.06\text{ pp}$ & $15.03\text{ pp}$ \\
Hill--Sheng (\texttt{HS-Kp}) & $s_{\mathrm{Val}} = 10.31$, $k_p = 0.184, 0.758$ & $0.0\text{ mM}$ & $2.20\text{ pp}$ & $3.24\text{ pp}$ & $1.49\text{ pp}$ \\
Guggenheim ($G^{\mathrm{E}} \le 0$) & $s_{\mathrm{Val}} = 12.24$, $a_{0,\mathrm{Leu}} = -1.378$ & $0.0\text{ mM}$ & $3.85\text{ pp}$ & $2.40\text{ pp}$ & $0.95\text{ pp}$ \\
\end{tabular}
\end{ruledtabular}
\end{table}

\begin{table}[!htbp]
\caption{Hierarchical Occam model comparison and grouped out-of-sample cross-validation across all $112$ RP-HPLC rows ($82$ Asp + $30$ Gly, $57$ batches, $63$ crystal stems, $29$ aqueous conditions, $14$ paired $U/B$ stems) of Kim \emph{et al.}\cite{Kim2016}}
\alttext{Data table comparing parameter count, exact marginal log-likelihood, AIC, BIC, condition-grouped LOCO-29 ELPD, five-fold batch-grouped ELPD, and stem-grouped LOSO-14 ELPD across hierarchical equal-mean and surface-excess calcite models.}
\label{tab:s_calcite_occam}
\scriptsize\setlength{\tabcolsep}{3.0pt}
\begin{ruledtabular}
\begin{tabular}{lccccccc}
Calcite Uptake Model & $k$ & $\ln\mathcal{L}_{112}$ & $\mathrm{AIC}_{112}$ & $\mathrm{BIC}_{112}$ & $\mathrm{LOCO\text{-}29}$ & $5$-Fold Batch & $\mathrm{LOSO\text{-}14}$ \\
\colrule
$M_{0,\mathrm{iid}}$: Pooled i.i.d.\ Langmuir ($U = B$) & $9$ & $+649.10$ & $-1280.21$ & $-1255.74$ & $603.53$ & $613.77$ & $140.86$ \\
$M_{0,\mathrm{corr}}$: Correlated $U/B$ stem ($U = B$) & $10$ & $+666.00$ & $-1312.01$ & $-1284.82$ & $602.27$ & $610.84$ & $107.37$ \\
$M_{0,\mathrm{hier}}$: Stem + split + tech ($U = B$) & $11$ & $+702.14$ & $-1382.28$ & $-1352.38$ & $661.87$ & $666.06$ & $133.85$ \\
$M_{0,\mathrm{batch}}$: Batch + split + tech ($U = B$) & $11$ & $+702.17$ & $-1382.34$ & $-1352.43$ & $663.76$ & $675.10$ & $137.01$ \\
$M_{1,\mathrm{iid}}$: Pooled additive excess ($U \ge B$) & $11$ & $+652.15$ & $-1282.30$ & $-1252.39$ & $602.21$ & $614.85$ & $141.47$ \\
$M_{1,\mathrm{corr}}$: Correlated surface excess ($U \ge B$) & $12$ & $+674.05$ & $-1324.09$ & $-1291.47$ & $594.09$ & $615.27$ & $100.25$ \\
$M_{2,\mathrm{pos}}$: Hierarchical surface excess ($U \ge B$) & $13$ & $+705.20$ & $-1384.39$ & $-1349.05$ & $659.49$ & $671.09$ & $137.26$ \\
$M_{2b,\mathrm{biv}}$: Bivariate Henry $U/B$ ($U \ge B$) & $17$ & $+709.72$ & $-1385.43$ & $-1339.22$ & $554.96$ & $634.98$ & $124.68$ \\
\end{tabular}
\end{ruledtabular}
\end{table}

\FloatBarrier

\end{document}